\documentclass[aps,prb,twocolumn,showpacs,superscriptaddress,citeautoscript,longbibliography]{revtex4-1}
\usepackage{graphicx} 
\usepackage{amsmath} 
\usepackage{bm} 
\usepackage{natbib}
\usepackage{natmove}
\usepackage[colorlinks=true, linkcolor=blue, citecolor=blue, urlcolor=blue, linktoc=page, bookmarks=false, pdfstartview={FitH}, pdfborder={0 0 0.0 [3 3]}]{hyperref} 
\usepackage{color} 
\usepackage{dcolumn} 
\newcolumntype{.}{D{.}{.}{2.1}}
\newcolumntype{-}{D{.}{.}{4.0}}
\usepackage{makecell}
\usepackage{multirow}
\usepackage{cleveref}
\usepackage[T1]{fontenc}  
\crefname{figure}{Fig.}{Figs}
\crefname{table}{Table}{Tables}
\crefname{section}{Sec.}{Section}
\renewcommand{\today}{\number\day \space \ifcase \month \or January\or February\or March\or April\or May \or June\or July\or August\or September\or October\or November\or December\fi \space \number\year} 
\newcommand{\mor}[1]{\multicolumn{1}{r}{#1}}
\newcommand{\moc}[1]{\multicolumn{1}{c}{#1}}

\begin{document}

\title{DFT study of itinerant ferromagnetism in {\textit p}-doped monolayers of MoS$_\text{2}$}
\author{Yuqiang Gao}
\email[Email: ]{Y.Gao@utwente.nl}
\affiliation{Faculty of Science and Technology and MESA$^+$ Institute for Nanotechnology, University of Twente, P.O.~Box~217, 7500~AE Enschede, The Netherlands}
\affiliation{Department of Applied Physics, Northwestern Polytechnical University, Xi'an, China}

\author{Nirmal Ganguli}
\email[Email: ]{nganguli@iiserb.ac.in}
\altaffiliation{Present address: Department of Physics, Indian Institute of Science Education and Research Bhopal, Bhauri, Bhopal 462066, India}
\affiliation{Faculty of Science and Technology and MESA$^+$ Institute for Nanotechnology, University of Twente, P.O.~Box~217, 7500~AE Enschede, The Netherlands}

\author{Paul J. Kelly\thanks{corresponding author}}
\email[Email: ]{P.J.Kelly@utwente.nl}
\affiliation{Faculty of Science and Technology and MESA$^+$ Institute for Nanotechnology, University of Twente, P.O.~Box~217, 7500~AE Enschede, The Netherlands}
\affiliation{The Center for Advanced Quantum Studies and Department of Physics, Beijing Normal University, 100875 Beijing, China}
\date{\today}
\begin{abstract}
We use density functional theory to explore the possibility of making the semiconducting transition-metal dichalcogenide MoS$_2$ ferromagnetic by introducing holes into the narrow Mo $d$ band that forms the top of the valence band. In the single impurity limit, the repulsive Coulomb potential of an acceptor atom and intervalley scattering lead to a twofold orbitally degenerate effective-mass like $e'$ state being formed from Mo $d_{x^2-y^2}$ and $d_{xy}$ states, bound to the K and K$'$ valence band maxima. It also leads to a singly degenerate $a'_1$ state with Mo $d_{3z^2-r^2}$ character bound to the slightly lower lying valence band maximum at $\Gamma$. Within the accuracy of our calculations, these $e'$ and $a'_1$ states are degenerate for MoS$_2$ and accommodate the hole that polarizes fully in the local spin density approximation in the impurity limit. With spin-orbit coupling included, we find a single ion magnetic anisotropy of $\sim 5\,$meV favouring out-of-plane orientation of the magnetic moment. Pairs of such hole states introduced by V, Nb or Ta doping are found to couple ferromagnetically unless the dopant atoms are too close in which case the magnetic moments are quenched by the formation of spin singlets. Combining these exchange interactions with Monte Carlo calculations allows us to estimate ordering temperatures as a function of $x$. For $x \sim 9\%$, Curie temperatures as high as 100K for Nb and Ta and in excess of 160K for V doping are predicted. Factors limiting the ordering temperature are identified and suggestions made to circumvent these limitations.
\end{abstract}

\maketitle

\section{Introduction}
\label{sec:intro}

\vspace{-4mm}
The discovery of ferromagnetism in (In,Mn)As \cite{Ohno:prl92} and (Ga,Mn)As \cite{Ohno:apl96} and predictions for achieving room temperature ordering \cite{Dietl:sc00} sparked a huge effort to realize a dilute magnetic semiconductor (DMS) that might lead to a semiconductor-based spin electronics (``Spintronics''). After twenty-five years of intensive research, the maximum ordering temperature has stagnated at values too low for extensive applications \cite{Dietl:natm10}. The number of material systems being considered has proliferated but it is not clear what the fundamental limit is to the ordering temperature achievable in any particular material system. There are many reasons for the low ordering temperatures \cite{Jungwirth:rmp06, Sato:rmp10} but the essential dilemma is that the open $d$ shell states of magnetic impurities like Mn are quite localized. While this favours the onsite exchange interaction that is the origin of the Hund's-rule spin alignment and makes the ionic moment insensitive to temperature, it leads to weaker exchange interactions between pairs of impurity ions that determine the Curie temperature $T_C$, the ferromagnetic ordering temperature. To increase $T_C$, the concentration of impurity atoms has to be increased. This is accompanied by a variety of adverse effects such as a nonuniform distribution of magnetic impurities or the formation of antisite defects that are electron donors which counter the intended increase in the concentration of holes. In many semiconductors, transition metal ions introduce ``deep levels'', tightly bound partially occupied states in the fundamental gap of the semiconductor. At high dopant concentrations, these form deep impurity bands that dominate the (transport) properties of a material that is no longer a semiconductor and from the electronic structure point of view, is an entirely new material.

In a quite different context, it was long believed that long-range magnetic ordering would not be possible in two-dimensional (2D) materials \cite{Mermin:prl66, Hohenberg:pr67}. However the observation of ferromagnetism in ultrathin epitaxial layers of e.g., Fe on Au substrates demonstrated that the Mermin-Wagner theorem is not watertight, violation of the proof usually being attributed to magnetocrystalline anisotropy \cite{UMS}. The recent observation of ferromagnetic ordering in two different chromium-based 2D crystalline materials Cr$_2$Ge$_2$Te$_6$ \cite{Gong:nat17} and CrI$_3$ \cite{Huang:nat17} nonetheless attracted considerable attention \cite{Samarth:nat17}. One reason was because of the general interest in 2D materials, triggered by spectacular observations on graphene \cite{Novoselov:sc04, Novoselov:pnas05, Novoselov:nat12}. This interest was reinforced by the realization that the properties of semiconductors like MoS$_2$ could also be importantly different in few- and mono-layer form \cite{Splendiani:nanol10, Mak:prl10, Radisavljevic:natn11} and was compounded by the desirability of stacking layers of 2D materials with different properties \cite{Geim:nat13} whereby the lack of a ferromagnetic material in a vast profusion of 2D materials was a striking lacuna \cite{Gibertini:natn19}. Because the Curie temperatures of monolayers of the chromium based materials \cite{Gong:nat17, Huang:nat17} is low, $\lesssim 50\,$K, the very recent reports that the transition metal dichalcogenide VSe$_2$ \cite{Bonilla:natn18} and Fe$_3$GeTe$_2$ \cite{Deng:nat18} exhibit ferromagnetism at room temperature acquires huge significance.

\begin{figure}[t]
\includegraphics[scale = 0.35]{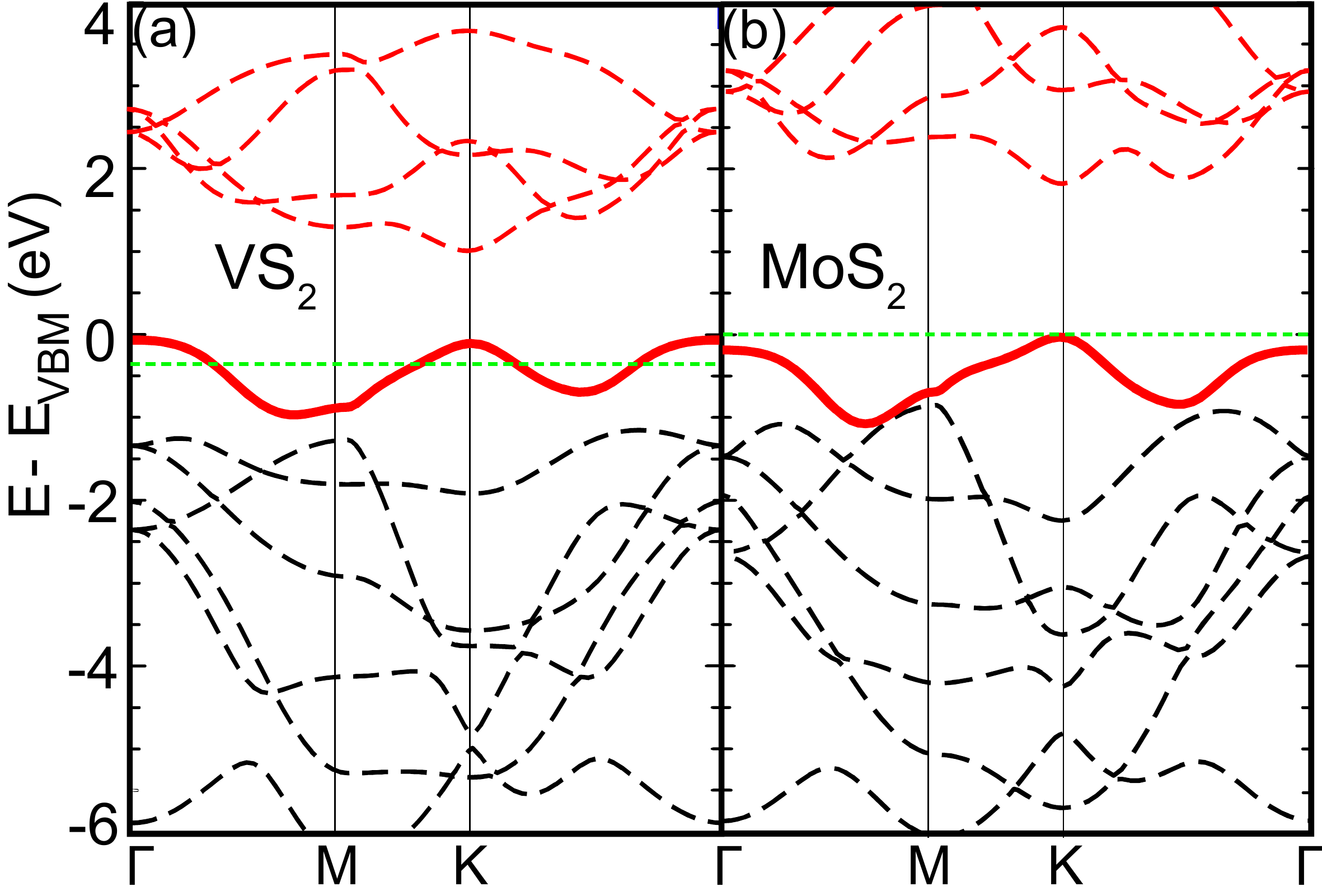}
\caption{Non spin-polarized band structures of monolayers of trigonal prismatic 1H VS$_2$ (a) and MoS$_2$ (b). The Fermi level is indicated by a horizontal dashed green line. The valence band maximum (VBM) at the K and K$'$ points has mixed $d_{x^2-y^2}$ and $d_{xy}$ character. The slightly lower-lying valence band maximum at the $\Gamma$ point has $d_{3z^2-r^2}$ character.
}
\label{figA}
\end{figure}

The ferromagnetism of VS$_2$ and VSe$_2$ was predicted with the aid of density functional theory (DFT) calculations \cite{footnote2}. The driving force behind the magnetic ordering can be understood in terms of the band structure of the nonmagnetic 1H phase shown in \cref{figA}(a) that is very similar to that of the isostructural MoS$_2$ shown in \cref{figA}(b) but with one valence electron per formula unit less so that it is metallic with the Fermi level situated in the middle of the solid red band.
Bulk multilayered MoS$_2$ is a non-magnetic semiconductor with an indirect bandgap of about 1~eV. In monolayer form it was predicted to have a larger, direct gap \cite{Li:jpcc07} and this was confirmed experimentally where direct gaps of $\sim 1.8\,$eV have been reported \cite{Splendiani:nanol10, Mak:prl10}. In the figure, the ``nominal'' Mo $4d$ bands are indicated in red, the black bands are sulphur-derived $3p$ bands. The interaction of the Mo-$d$ and S-$p$ states is such that a large covalent bonding-antibonding gap is formed leaving a single Mo-$d$ band (solid red line) with mixed $\{d_{x^2-y^2}, d_{xy}, d_{3z^2-r^2}\}$ character in the fundamental band gap \cite{Bromley:jpc72, Mattheiss:prb73}. For MoX$_2$, this band is completely filled but for VX$_2$ it is only half full. The dispersion of only about 1~eV leads to a high average density of states of $\sim 2\,$states/eV and the gain in energy achieved by exchange-splitting this narrow band more than offsets the kinetic energy cost. The bandwidth reduction in 2D that leads to larger band gaps is favourable for itinerant ferromagnetism because of the higher average densities of states (DoS) than in three dimensions. Likewise $3d$ elements are more favourable than $4d$ and $4d$ more favourable than $5d$ because of the greater localization of the $d$ electrons and concomittant smaller bandwidth as the principal quantum number decreases.

A number of intrinsic defects have been found to form local moments \cite{Ataca:jpcc11b, Lu:nrl14, Hong:natc15, Li:prb16, Khan:jpcm18} in MX$_2$ materials and suggestions have been made to make the MX$_2$ materials magnetic by adsorption of impurity atoms \cite{He:apl10, Ataca:jpcc11b, Dolui:prb13, Lu:nrl14}, or by substituting M or X atoms with impurity atoms \cite{Karthikeyan:nanol19, Cheng:prb13, Yue:pla13, Ramasubramaniam:prb13, Mishra:prb13, Dolui:prb13, Yun:pccp14, Qi:jpcm14, Gil:jpcm14, Andriotis:prb14, Lu:nrl14, Zhang:jac15, Miao:jms16, Zhao:jac16, Zhao:rsca16, Fan:nrl16, Robertson:acsn16, Singh:am17, Miao:ass18, Mekonnen:ijmpb18}. Even though the Mermin-Wagner theorem \cite{Mermin:prl66, Hohenberg:pr67} tells us that there is no long range ordering in two dimensions for isotropic Heisenberg exchange, few attempts have been made to determine the exchange coupling between magnetic impurities \cite{Ramasubramaniam:prb13, Mishra:prb13, Dolui:prb13, Qi:jpcm14, Gil:jpcm14, Fan:nrl16, Mekonnen:ijmpb18} and it was only very recently that the magnetic anistropy of a defect was calculated, for an antisite defect in MoS$_2$ \cite{Khan:jpcm18}.
Replacing some of the M atoms with Hund's-rule coupled transition metal atoms like Mn or Fe gives rise to deep impurity levels in the semiconductor gap. Where attempts have been made to estimate the Curie temperature, the predicted values are either very low or the concentration of transition metal dopant is so high that the doped material is no longer a semiconductor \cite{Cheng:prb13, Ramasubramaniam:prb13, Mishra:prb13, Qi:jpcm14, Gil:jpcm14, Miao:ass18}.
Based upon the electronic structure shown in Fig.~\ref{figA}(b), we explore a different approach to making MoS$_2$ ferromagnetic in this manuscript
\cite{[A short summary of this work appeared in ]Gao:prb19}.


\begin{figure}[b]
\includegraphics[scale = 0.54]{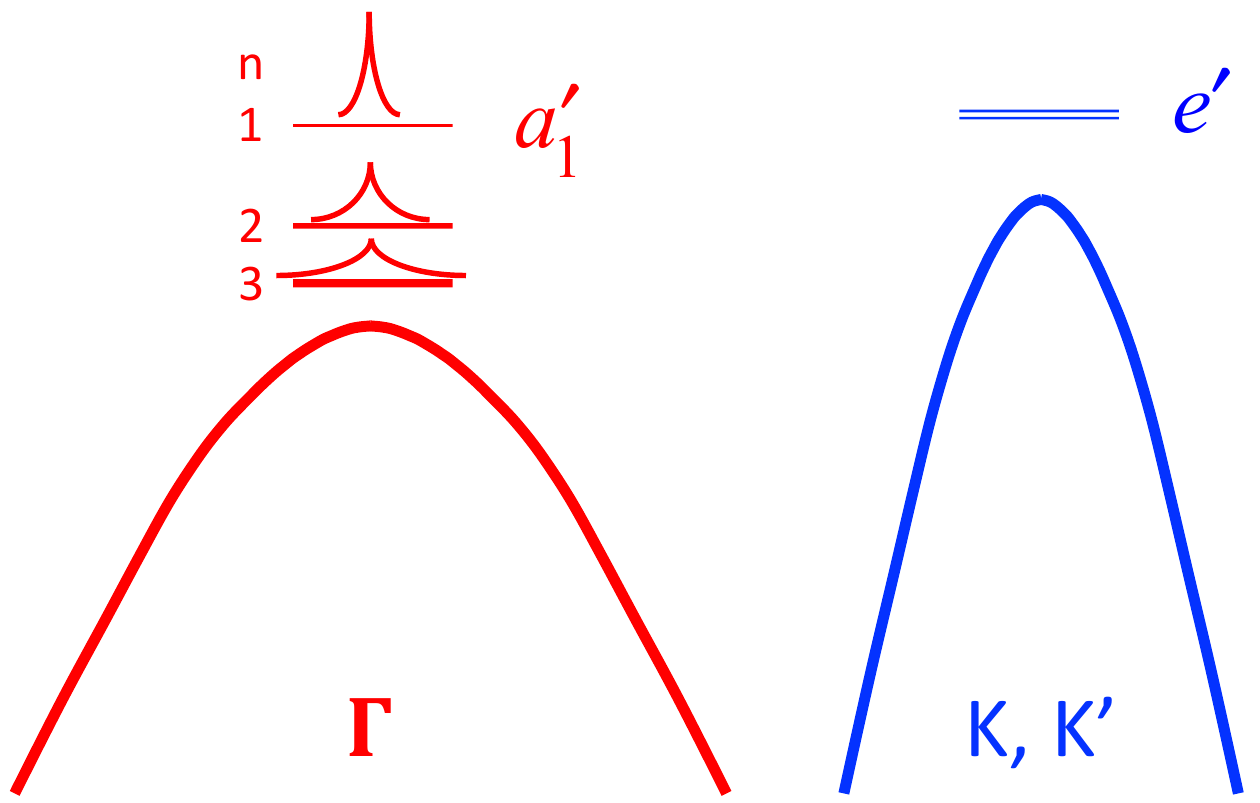}
\caption{Schematic of the effective mass acceptor states bound to the valence band maxima (VBM): an $e'$ state bound to the K-K$'$ VBM and an $a'_1$ state to the $\Gamma$ VBM. $n$ is the principal quantum number and only $n=1, 2, 3$ levels of the Rydberg series are sketched at the $\Gamma$ point. }
\label{figB}
\end{figure}

Group VIB Mo has a 4$d^5$5$s^1$ electronic configuration and, in a dichalcogenide like MoS$_2$, is nominally Mo$^{4+}$ with one up-spin and one down-spin $d$ electron so it is nonmagnetic as seen in \cref{figA}(b). When a Mo atom is substituted by a group VB atom like V, Nb or Ta, then the dopant atom e.g. V$^{4+}$, has a single unpaired $d$ electron and a single hole is thereby introduced into the narrow Mo 4$d$ band; substitution of a group IVB atom (Ti, Zr, Hf) will introduce two holes per dopant atom. In the impurity limit, the asymptotic Coulomb potential leads to a series of hydrogenic states bound to the top of the valence band; to the maxima at the K and K$'$ points with mixed Mo $d_{x^2-y^2}$ and $d_{xy}$ character and to the slightly lower valence band maximum at the $\Gamma$ point with Mo $d_{3z^2-r^2}$ character and a large effective mass \cite{Peelaers:prb12}.

The aim of this paper \cite{Gao:prb19} is to determine if there are dopant atoms whose potential is sufficiently similar to that of the host Mo atom that only weakly bound, effective-mass like states are formed above the valence band edge, \cref{figB}. At low concentrations these bound states should polarize and form impurity bands that have such a high density of states that they remain exchange split \cite{Edwards:jpcm06}. At finite temperatures these polarized bound holes will be excited into the valence band giving rise to a DMS. The key objectives of this paper are to determine (i) whether single acceptor dopant atoms give rise to polarized effective-mass like states in the MoS$_2$ host system and to determine the position of these states with respect to the valence band edge; (ii) whether the interaction between pairs of dopant atoms favours ferromagnetic or antiferromagnetic alignment and to identify the nature of the interaction, Zener $p$-$d$ type, double exchange etc. \cite{Jungwirth:rmp06, Sato:rmp10} and understand the factors determining it; (iii) the magnetic anisotropy of single impurities, the so-called single ion anisotropy (SIA); (iv) the ordering temperature and express it in terms of parameterized models that describe the dopant-induced states and their interactions in order to identify the most promising regions of parameter space to realize a room temperature DMS.

To do this we use density functional theory total energy calculations to determine ground state energies of single acceptor impurities. We outline the methods used and give some technical details specific to the present work in Sec.~\ref{sec:CD}.
Our results are presented in Sec.~\ref{sec:results} beginning with a study of the single impurity limit of a substitutional vanadium atom in Sec.~\ref{ssec:SILV} including the effects of spin polarization and local atomic relaxation. The binding of pairs of V dopants is considered in Sec.~\ref{sec:BIP} and their magnetic ``exchange'' interaction in Sec.~\ref{sec:MI} with special attention being devoted to understanding the quenching of the magnetic moments of close pairs of impurity ions. In Sec.~\ref{ssec:NbTa} we briefly compare V with Nb and Ta.
Sec.~\ref{sec:MO} is concerned with the question of magnetic ordering and begins with a study of the single ion anisotropy of V impurities in Sec.~\ref{ssec:SIA} to justify using an Ising spin model with the exchange interactions from Sec.~\ref{sec:MI} and the Monte Carlo techniques described in Sec.~\ref{ssec:MC} to estimate ordering temperatures in Sec.~\ref{ssec:CT}.
A comparison of our findings with other calculations in \cref{sec:Comp} leads us to consider how using the generalized gradient approximation (GGA) would alter our local density approximation (LDA) results. After a brief discussion in Sec.~\ref{sec:Discussion} some conclusions are drawn in \cref{sec:Conclusions}.

\vspace{-2mm}
\section{Computational Details}
\label{sec:CD}

Calculations of the total energy and structural optimizations were carried out within the framework of density functional theory (DFT) using the projector augmented wave (PAW) method \cite{Blochl:prb94b} and a plane-wave basis set with a cut-off energy of 400 eV as implemented in the {\sc vasp} code \cite{Kresse:prb93b, Kresse:prb96, Kresse:prb99}. Monolayers of MX$_2$ periodically repeated in the $c$ direction were separated by more than 20~\AA\ of vacuum to avoid spurious interaction.

\begin{table}[t]
\caption{In-plane lattice constant $a$, distance between sulphur atoms $d_{\rm SS}$ (thickness of an MoS$_2$ monolayer), Mo-S bond length $d_{\rm MoS}$, energy gap $\Delta\varepsilon_g$, and energy difference between the valence band maxima (VBM) at the K and $\Gamma$ points $\Delta_{\rm K\Gamma}=\varepsilon_{\rm K}-\varepsilon_{\Gamma}$ in LDA and GGA for bulk and monolayer (ML) MoS$_2$.
A van der Waals functional should be used to obtain a reasonable interlayer separation for bulk layered MoS$_2$. Because we are only interested in monolayers of MoS$_2$ in this paper, we have used the experimental value of $c$ to obtain the bulk results shown here.
}
\label{tab:A}
\begin{ruledtabular}
\begin{tabular}{llllllc}
     &     & $a$(\AA)  & $d_{\rm SS}$(\AA)
                                    & $d_{\rm MoS}$(\AA)
                                            & $\Delta\varepsilon_g$(eV)
                                                     & $\Delta_{\rm K\Gamma}$  \\
\hline
Bulk & GGA & 3.183     & 3.127      & 2.42      & 0.885      & -0.640  \\
     & LDA & 3.125     & 3.115      & 2.38      & 0.748      & -0.640  \\
     & Exp & 3.160$^a$ & 3.172$^a$  & 2.41$^a$  & 1.290$^c$  & -0.600$^b$  \\
ML   & GGA & 3.185     & 3.130      & 2.42      & 1.650      &  0.012  \\
     & LDA & 3.120     & 3.115      & 2.38      & 1.860      &  0.150  \\
     & Exp & 3.160     & 3.172      & 2.41      & 1.900$^c$  &  0.140$^b$  \\
  \end{tabular}
  \end{ruledtabular}
$^a$Ref.\onlinecite{Boker:prb01},
$^b$Ref.\onlinecite{Jin:prl13}
$^c$Ref.\onlinecite{Mak:prl10}
\end{table}

\begin{figure}[b]
\includegraphics[scale = 0.75]{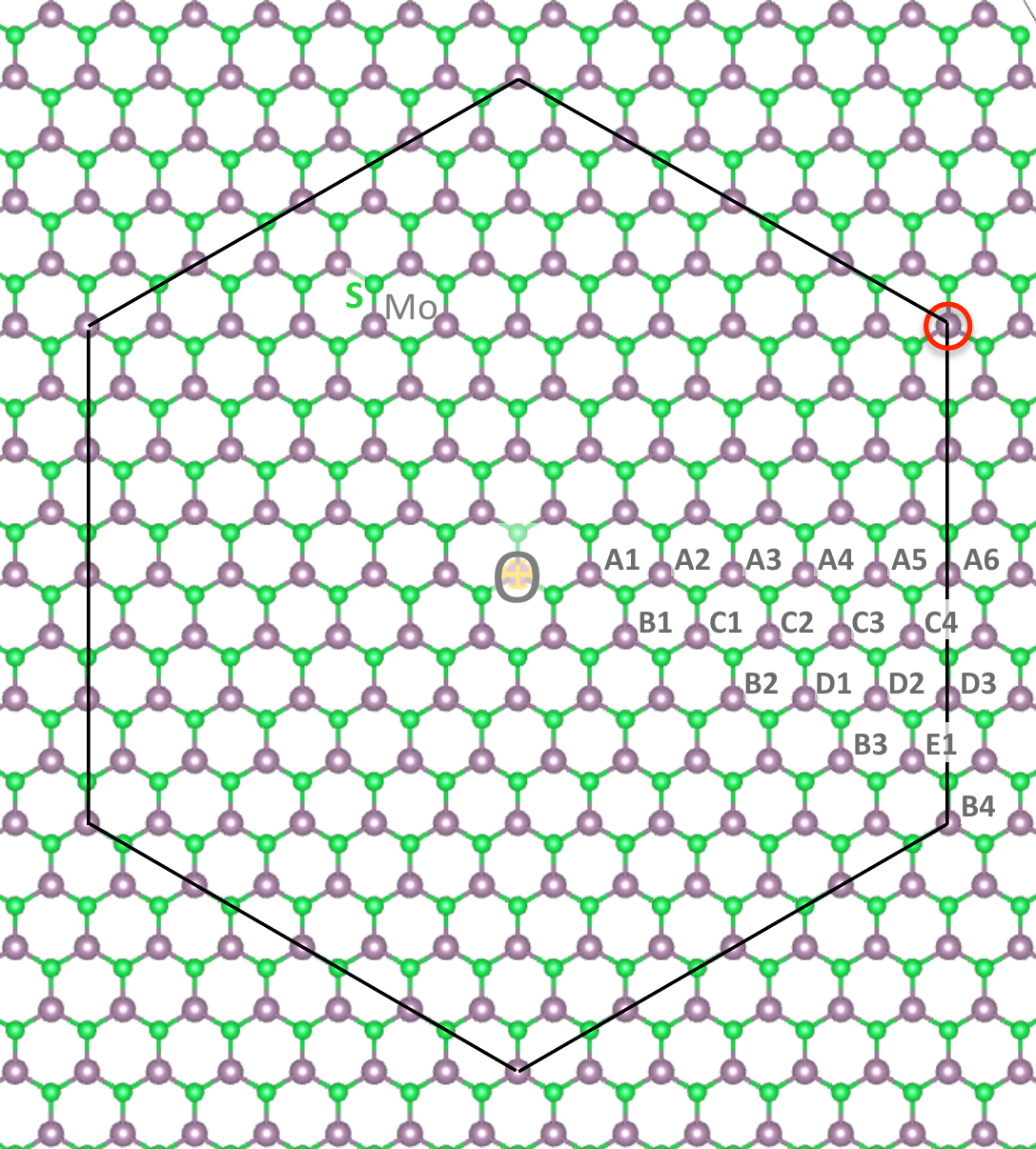}
\caption{Sketch of a 12$\times$12 MoS$_2$ supercell with a substitutional atom at the origin, O. Shown is the more symmetric Wigner-Seitz cell. The potential on the Mo atom indicated with a red circle that is furthest from this atom will be used to identify the host valence band maximum (VBM). Mo atoms at various distances from the central atom are labelled A1-A6, B1-B4, C1-C4, D1-D3 and E1 for later reference.
}
\label{figC}
\end{figure}

\begin{figure*}[t]
\includegraphics[scale = 0.34]{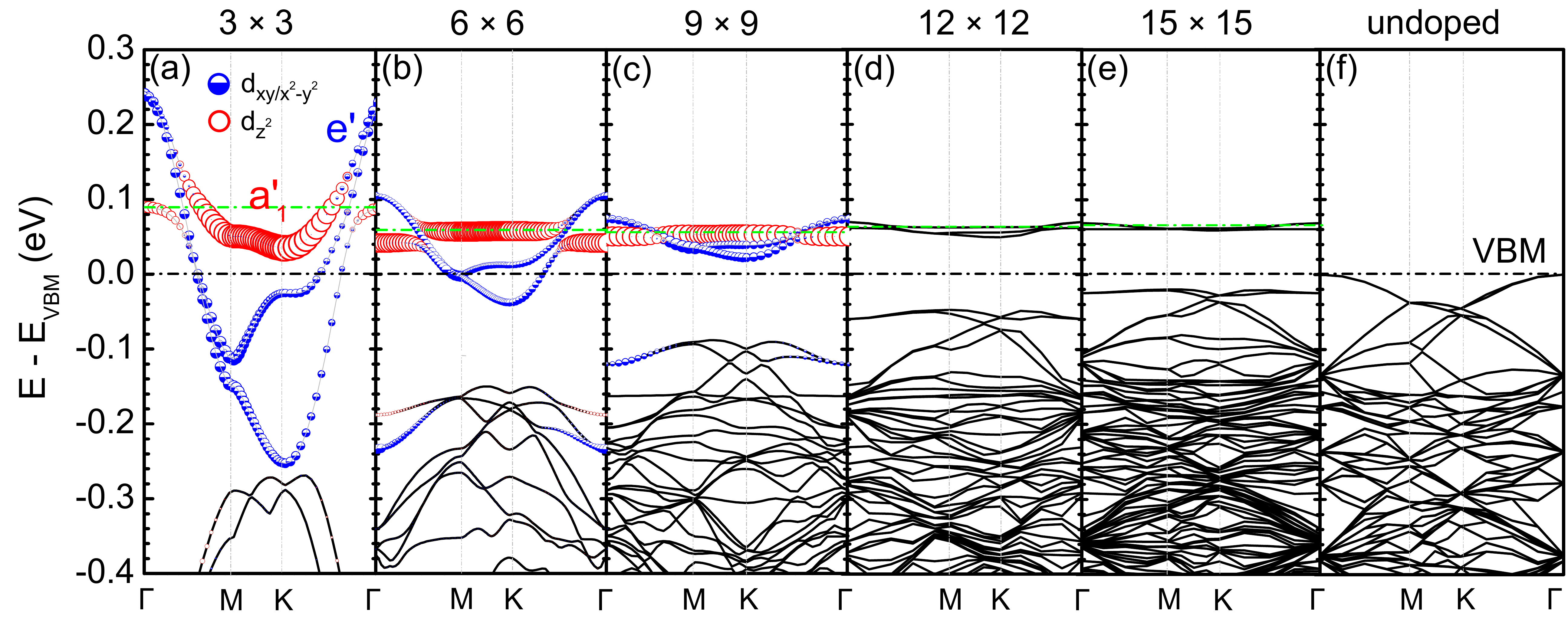}
\caption{The band structure of an MoS$_2$ monolayer with a single Mo atom replaced by V without relaxation in 3$\times$3, 6$\times$6, 9$\times $9, 12$\times$12 and 15$\times$15 supercells (a-e) aligned with respect to the valence band maximum determined with respect to semicore level states on the Mo atom furthest from the origin (horizontal black dot-dashed line at energy zero) and of an undoped MoS$_2$ monolayer (f). The Fermi energy is indicated by the horizontal green dot-dashed line. The contribution from vanadium $d_{3z^2-r^2}$ and \{$d_{x^2-y^2},d_{xy}$\} orbitals are show as open red circles and half-filled blue circles, respectively in (a-c) where the symbol size is proportional to the population of the corresponding state.
}
\label{figD}
\end{figure*}

The equilibrium structural parameters for bulk and monolayer MoS$_2$ were calculated in both the LDA \cite{Perdew:prb81} and GGA \cite{Perdew:prl96} and are given in Table~\ref{tab:A}. It can be seen that the GGA slightly overestimates lattice constants and bond lengths compared to experiment \cite{Boker:prb01}. The LDA underestimates them by more than the GGA overestimates them, a result found for many materials. In the present case, the agreement with experiment is still very reasonable for both LDA and GGA. However, we see that for an MoS$_2$ monolayer the LDA gives a better description of the energy levels near the valence band maximum (VBM) than does the GGA, in particular the important quantity $\Delta_{{\rm K} \Gamma}=\varepsilon_{\rm K}-\varepsilon_{\Gamma}$, the position of the VBM at the $\Gamma$ point, $\varepsilon_{\Gamma}$, relative to the top of the valence band at the K point, $\varepsilon_{\rm K}$ \cite{Jin:prl13}. To describe acceptor states accurately, it is important to have a good description of the host band structure in the vicinity of the VBM so we will describe exchange and correlation effects in this paper using the local spin density approximation LSDA as parameterized by Perdew and Zunger \cite{Perdew:prb81}. Results obtained with the GGA are considered in \cref{sec:Comp}.

We model substitutional impurities and impurity pairs in $N \times N$ in-plane supercells with $N$ as large as 15 using the calculated equilibrium lattice constant for the pure monolayer (ML) host, \cref{figC}. Local geometries are first relaxed using $N=6$ and only a small differential relaxation needs to be performed in the larger supercells. Interactions between pairs of impurities were studied in 12$\times$12 supercells. The atomic positions were relaxed using a 2$\times$2$\times$1 $\Gamma$-centered $k$-point mesh until the forces on each ion were smaller than 0.01 eV/\AA. Spin-polarized calculations were performed with a denser mesh corresponding to 4$\times$4 $k$-points for a 12$ \times$12 unit cell.

\section{Results}
\label{sec:results}

Impurity states in semiconductors are usually described in one of two limits: (i) in effective mass theory (EMT) where the main emphasis is on the Rydberg series of bound states tied to the conduction band minima or valence band maxima formed in response to a Coulomb potential or (ii) in the tight-binding limit where the main emphasis is on the local chemical binding, atomic relaxation and impurity states formed deep in the fundamental bandgap associated with an impurity potential very different to the host atomic potential \cite{Pantelides:rmp78, Lannoo:81, Altarelli:82}. Because there is no consensus of how best to combine both aspects \cite{Smith:sr17}, we consider the behaviour of shallow acceptor states in a periodic supercell geometry in some detail in the following section.

\subsection{Single impurity limit: V in MoS$_2$}
\label{ssec:SILV}

\begin{figure}[bthp]
\includegraphics[scale = 0.25]{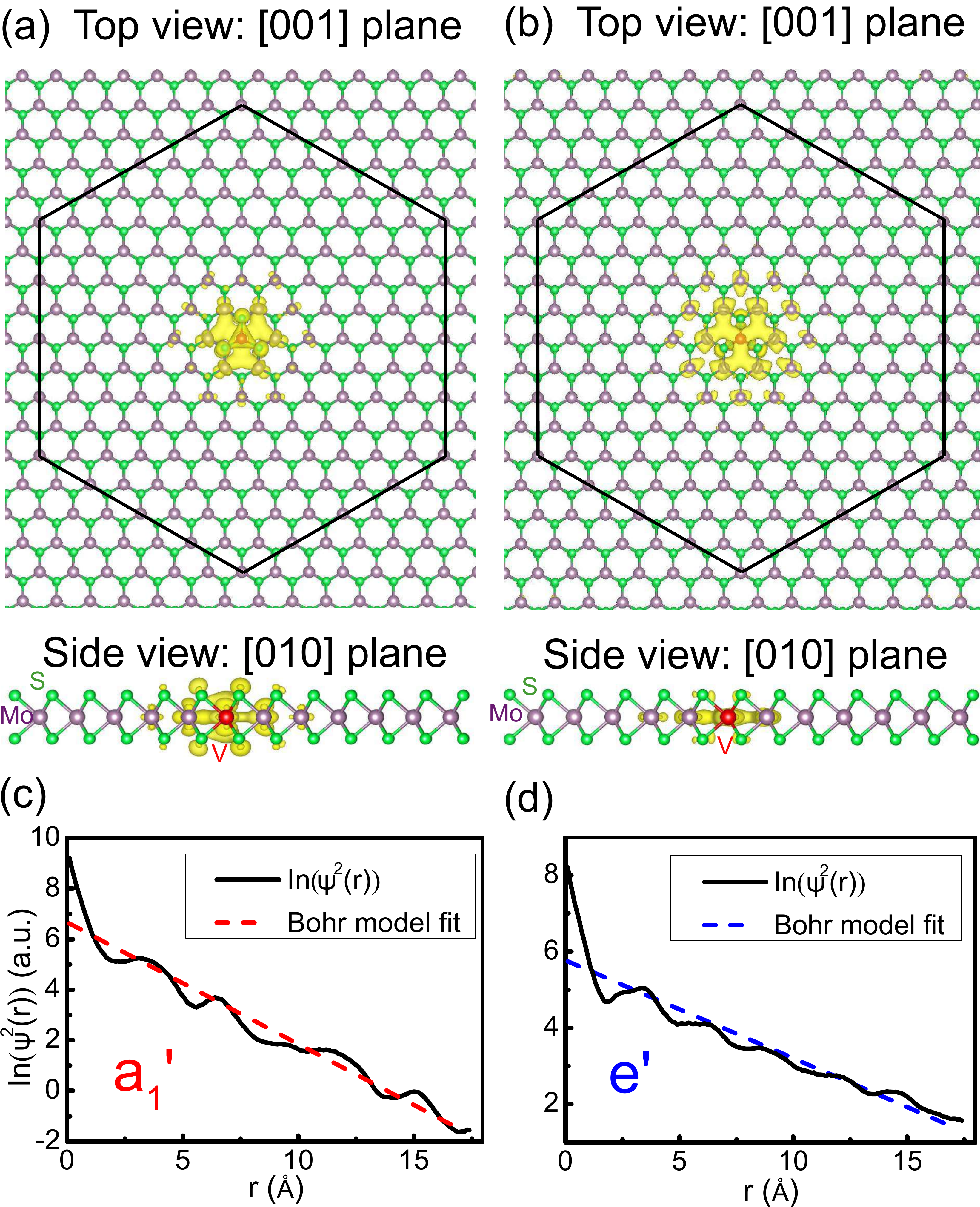}
\caption{Charge density plots for the $a'_1$ (lhs) and $e'$ (rhs) states at the $\Gamma$ point in \cref{figD}(d) in the central [001] plane through the Mo atoms (top view) and [010] plane (side view) for a 12$\times$12 supercell. The isosurface levels are 0.001 $e/\mbox{\AA}^3$. (c) and (d): circularly averaged charge densities fitted with a Bohr model. }
\label{figE}
\end{figure}

We begin by replacing a single Mo atom in an $N \times N$ MoS$_2$ supercell with a V atom with one valence electron less, \cref{figC}. To more easily identify the downfolded host bands we choose $N$ to be a multiple of three whereby the K and K$'$ points fold down to the $\Gamma$ point of the reduced BZ. The energy bands for this supercell before relaxing the local geometry are shown in \cref{figD} for $N=3$, 6, 9, 12 and 15. The repulsive (for electrons; attractive for holes) impurity potential is seen to push not one but three impurity states out of the valence band to form localized states labeled $a'_1$ and $e'$ under the local $D_{3h}$ symmetry, \cref{figB}. By projecting the corresponding wavefunctions at the $\Gamma$ point onto spherical harmonics on the ${\rm V_{Mo}}$ site, we find that the singly degenerate $a'_1$ state has V $d_{3z^2-r^2}$ character while the $e'$ state that is doubly degenerate at the center of the BZ has V \{$d_{x^2-y^2}, d_{xy}$\} character. The corresponding partial charge density plots are shown on the left- respectively right-hand sides (lhs, rhs) of \cref{figE}. By fitting the wave functions of the impurity states to a hydrogenic wave function $\psi(r)$=A$\exp(-r/a_0^*)$, we find effective Bohr radii $a_0^*$ of 4.2 \AA\  and 8.0 \AA\  for the $a'_1$ and $e'$ states, respectively in \cref{figE}(c) and \cref{figE}(d).

We identify these $a'_1$ and $e'$ states with the most tightly bound (effective mass like) acceptor states formed when a screened Coulomb potential is introduced by substitution of a Mo atom by V (Nb or Ta). In the single impurity limit, intervalley scattering leads to a twofold orbitally degenerate effective mass like state formed from Mo \{$d_{x^2-y^2}, d_{xy}$\} states bound to the K and K$'$ valence band maxima in \cref{figA} and a singly degenerate state with Mo $d_{3z^2-r^2}$ character bound to the slightly lower lying valence band maximum at $\Gamma$ in \cref{figA}. Within the accuracy of our calculations, these $e'$ and $a'_1$ states are (accidentally) degenerate for MoS$_2$ and accommodate the hole that we will see polarizes fully in the local spin density approximation \cite{Gunnarsson:prb74}. The shape of the dispersion of the impurity states is essentially independent of the supercell size so the bands can be described with a single effective hopping parameter. The $a'_1$ state exhibits very little dispersion consistent with the out-of-plane $d_{3z^2-r^2}$ orbital character at $\Gamma$ where the weak dispersion of the host MoS$_2$ bands is described by a large effective mass \cite{Aghajanian:sr18}. In the language of effective mass theory (EMT), the binding energy of the $a'_1$ state is dominated by the central cell correction \cite{Pantelides:rmp78}.

In the rightmost panel of \cref{figD}, we show the band structure of an undoped monolayer calculated in a $15 \times 15$ supercell so the K point VBM is downfolded onto $\Gamma$. If we compare this with the impurity supercell bands on the left, we see that even for $N=15$, \cref{figD}(e), the interaction of the impurity bands and the VBM still suppresses the VBM quite noticeably, by more than 20~meV.

\subsubsection{Screened impurity potential}
\label{sssec:SIP}

\begin{figure}[b]
\includegraphics[scale = 0.33]{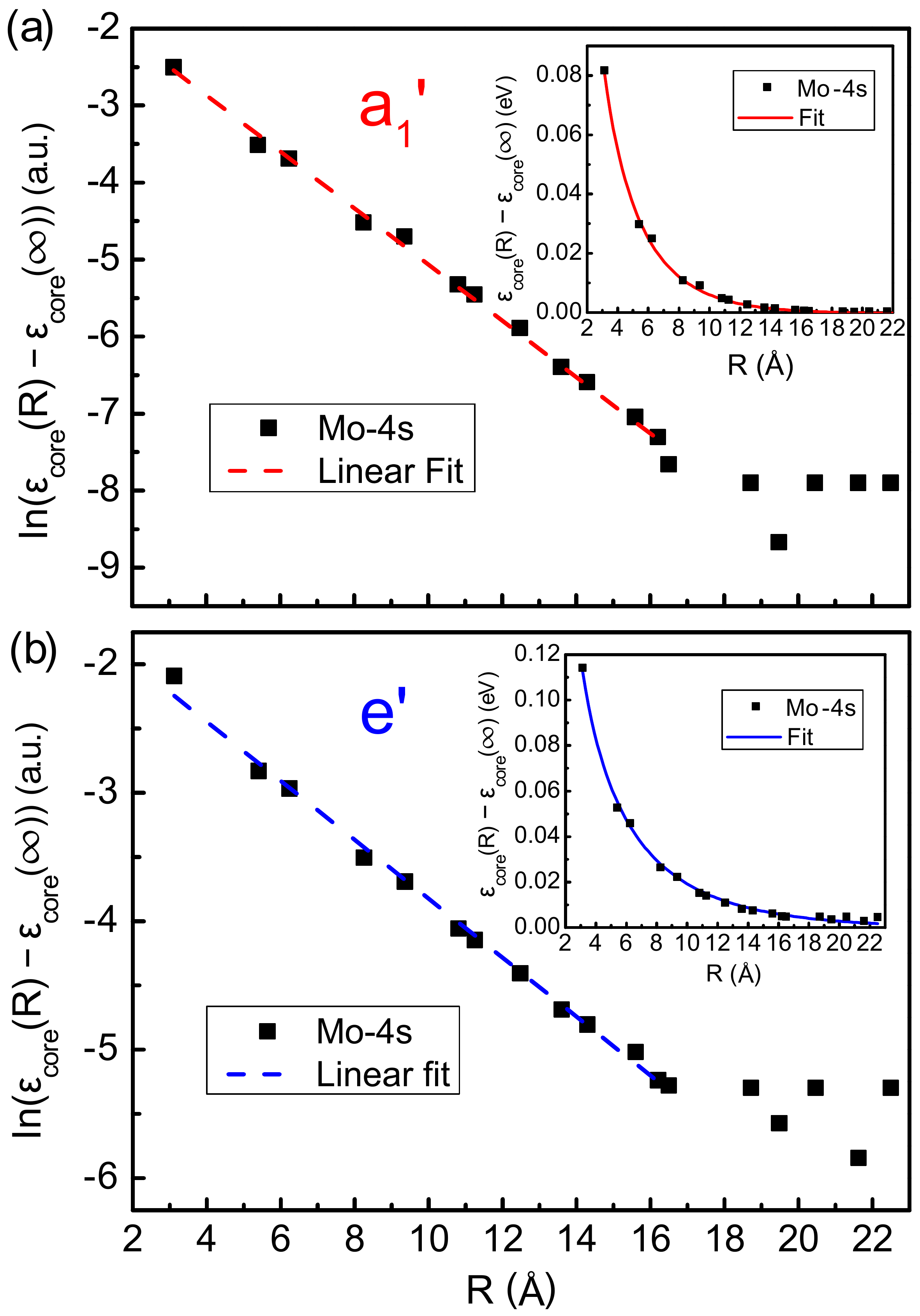}
\caption{Dependence of the Mo $4s$ semicore level on the separation from the ${\rm V_{Mo}}$ dopant ion. The Coulomb potential of the V dopant is screened by the host valence electrons and by the $a'_1$ hole (upper panel) respectively by the $e'$ hole (lower panel). The 18 data points refer to the 18 inequivalent Mo atoms labelled in \cref{figC}. The asymptotic value $\varepsilon_{\rm core}(\infty)$ was determined by fitting the calculated data points in the insets to an exponential wave function and using this fit (red and blue curves) to extrapolate to $R=\infty$.
}
\label{figF}
\end{figure}

Identifying the valence band maximum (VBM) in an impurity supercell calculation is complicated by the Rydberg series of effective mass like states associated with the single impurity whose wavefunctions will overlap with their periodic images and form bands that overlap and hybridize with the ``true'' valence band states, \cref{figB}. To disentangle those effects, we first determine the position of the VBM with respect to Mo 4$s$ semicore states, $\varepsilon_{4s}^{\rm Mo}$, for an undoped monolayer of MoS$_2$; $\varepsilon$ denotes a Kohn-Sham eigenvalue. For a sufficiently large impurity supercell, the position of $\varepsilon_{4s}^{\rm Mo}$ for the Mo atom furthest from the impurity (indicated with a red circle in \cref{figC}) relative to the VBM should be asymptotically equal to the corresponding energy separation for an undoped monolayer of MoS$_2$ because the impurity potential far from the dopant center will be completely screened by the bound charge of the neutral impurity in an $a'_1$ or $e'$ bound state. To test this hypothesis quantitatively, we plot $\varepsilon^{\rm Mo}_{4s}$  with respect to its asymptotic value as a function of the separation of Mo from the impurity V ion in the insets of \cref{figF} for $N=12$ supercells (symbols). The corresponding results for the S semicore 2$s$ state $\varepsilon^{\rm S}_{2p}$ are shown in Appendix~\ref{sec:SRA} and yield similar conclusions.

In an impurity supercell calculation, a localized electron in a (semi)core level on a Mo atom a distance $R$ from the impurity atom will see a screened $1/\epsilon_r R$ repulsive potential that is partially compensated by the charge of the bound hole, $n_{\rm hole}(r)$. Here $\epsilon_r$ is the relative static dielectric constant. We can ``measure'' this screened Coulomb potential by studying how Mo (and S) semicore levels behave as a function of their separation from the central V atom. The perturbing electrostatic potential seen by the core electrons has the form
\begin{subequations}
\begin{align}
\!\!\! \varepsilon_{\rm core}(R)-\varepsilon_{\rm core}(\infty)
 &= \frac{1}{\epsilon_r R}\Big(1 - \int_0^R n_{\rm hole}(r) 2\pi r dr \Big) \label{eq:sha} \\
 &= \frac{e^\frac{-2R}{a^*_0}(\frac{2R}{a^*_0}+1)}{\epsilon_r R}
\label{eq:shb}
\end{align}
\end{subequations}
where in \eqref{eq:sha} $ n_{\rm hole}(r) = \int_{-\infty}^{\infty} n_{\rm hole}(r,z) dz$ and $n_{\rm hole}(r,z)$ is obtained by integrating $|\psi_i(r,\theta,z)|^2 $ over $\theta$ for $i=a'_1$ or $e'$. In \eqref{eq:shb}, we assume that $\psi_i(r,\theta,z)$ is the solution of a strictly two dimensional hydrogenic problem \cite{Yang:pra91}. If we take the natural logarithm of \eqref{eq:shb}, the slope is $-2/a^*_0$ for large values of $R$ and we can extract $a^*_0$ from \cref{figF}.

In the supercell band structures shown in \cref{figD}, the $a'_1$ and $e'$ derived states overlap and $n_{\rm hole}(r)$ is a mixture of these two states with different masses $m_h$. To circumvent this complication, we calculate the electronic structure at the K (or M) point where the lowest unoccupied state has $a'_1$ character. By using a sufficiently small temperature broadening we can obtain the corresponding charge density and obtain the result shown in \cref{figF} (upper panel). Alternatively, we calculate the electronic structure at the $\Gamma$ point where the lowest unoccupied state has $e'$ character to obtain \cref{figF} (lower panel). The ab-initio values of $\varepsilon_{\rm core}(R)$ and $\ln[\varepsilon_{\rm core}(R)-\varepsilon_{\rm core}(\infty)]$ are fit quite well with \eqref{eq:shb} with an effective Bohr radius of $a^*_0\sim 5.4$~\AA\ and $\epsilon_r=20$ for the $a'_1$ hole and  $a^*_0 \sim 8.7$~\AA\ and $\epsilon_r=13$ for the $e'$ hole. The values of $\epsilon_r$ should be compared to recent calculations for the in-plane ``macroscopic'' dielectric constant where $\epsilon_r=15$ was found for monolayers of MoS$_2$ as well as for bulk MoS$_2$ with negligible ionic contribution to the screening \cite{Laturia:tdma18}. The deviation of $\epsilon_r=20$ from the macroscopic value is not very surprising in view of the localization of the $a'_1$ hole that does not ``see'' many unit cells of MoS$_2$. The value for the $e'$ hole is reasonable.

At large values of $R$ in \cref{figF}, the potential felt by the core states is seen not to decay but to oscillate. We  attribute this to the accumulation of the residual hole charge at the supercell boundary that is a consequence of charge neutrality. The data points in \cref{figF} that deviate from the trend line are to be found outside the circle inscribed in the hexagonal WS cell.

In the effective mass approximation the effective Bohr radius $a^*_0= \epsilon_r/m_h \times 0.529$ \AA\ and the ground state binding energy with respect to the appropriate VBM is $\varepsilon_b =m_h/\epsilon_r^2 \times 13.606\,$eV. From the band structure in \cref{figA}, the effective mass in units of the free electron mass $m_0$ is $m_h \sim 0.56$ at the K point VBM and 3.42 at the $\Gamma$ point VBM, consistent with a previous calculation \cite{Yun:prb12}. Combining these masses with $\epsilon_r=15$  \cite{Laturia:tdma18} leads to values of $a^*_0 \sim 14$~\AA\ and $\varepsilon_b \sim 34\,$meV for $e'$ holes and $a^*_0 \sim 2.3$~\AA\ and $\varepsilon_b \sim 207\,$meV for $a'_1$ holes, respectively. At best the EMT is indicative but is clearly not quantitative for the most strongly bound acceptor states - a conclusion that is not especially surprising in view of the expected central cell correction for ground states \cite{Pantelides:rmp78} as well as the strong localization of both states.

The screening of the impurity potential by (i) the MoS$_2$ valence electrons and (ii) by the bound impurity hole means that the residual perturbation measured by the core states decreases rapidly with $R$ allowing us to estimate the position of the reference core state far from the impurity and therefore of the VBM to an accuracy of a few meV for $N=12$. This procedure was used to estimate the position of the VBM and of the impurity states with respect to it for each supercell size shown in \cref{figD} (dot-dashed line).

The same results can be obtained more simply by noting that the repulsive potential that binds a Rydberg series to the top of the valence band has little effect on the conduction band edge. Since we know the value of the band gap, the VBM can be determined from the conduction band minimum. We verified that this leads to the same results as the more elaborate procedure discussed in the foregoing.

\subsubsection{Hydrogenic perturbation model}
\label{sssec:HPM}

\begin{figure}[t]
\includegraphics[scale = 0.35]{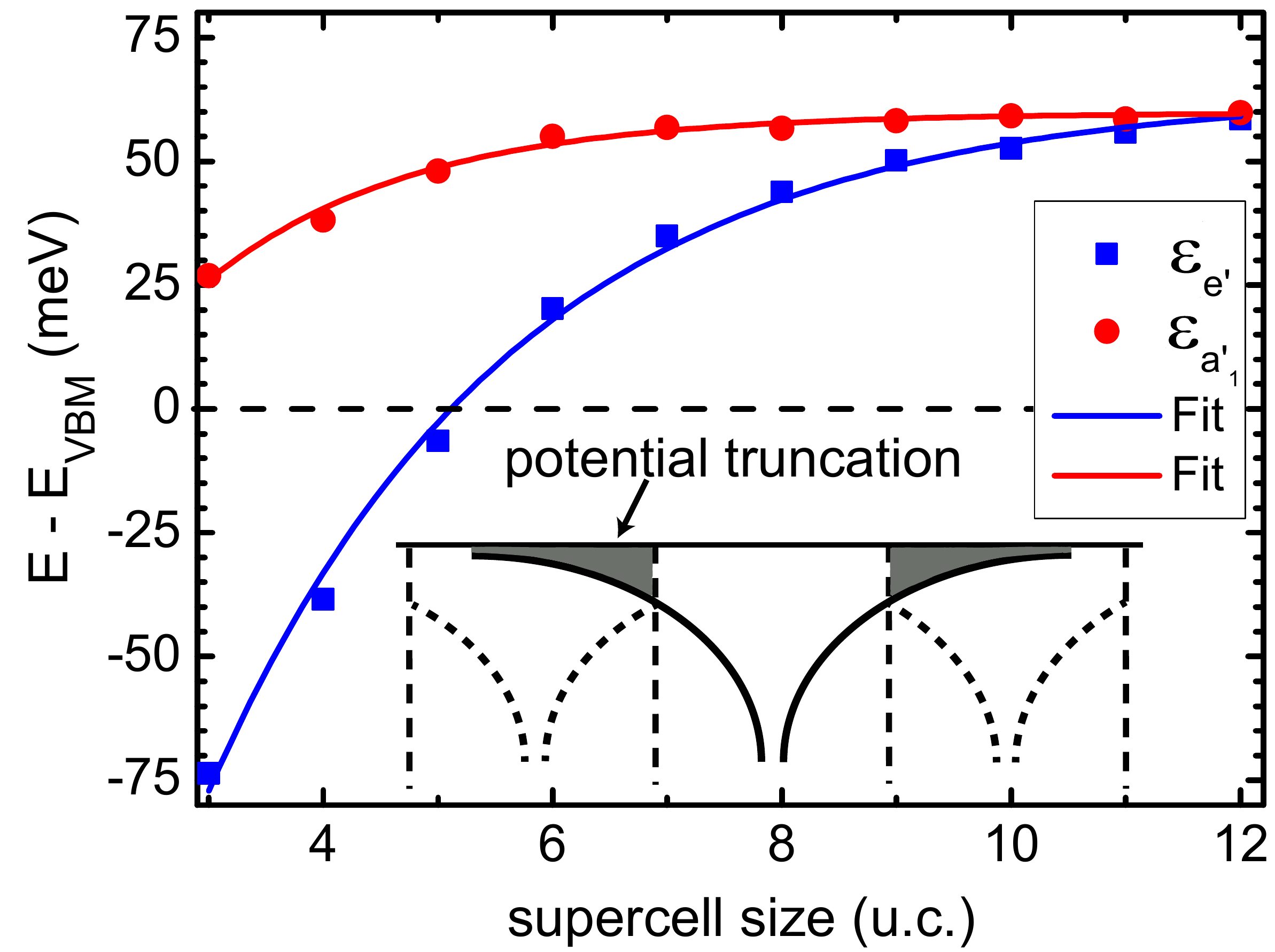}
\caption{Dependence on the supercell size $N$ of the (spin-degenerate) $a'_1$ and $e'$ impurity levels induced by a substitutional vanadium atom ${\rm V_{Mo}}$ with respect to the valence band maximum. The solid lines are fits to the data points using a model that includes the tail of the Coulomb potential in first order perturbation theory. (Inset) Truncation of the Coulomb potential in a supercell calculation.
}
\label{figG}
\end{figure}

Now that we have established procedures for determining the position of the VBM, we see that the $a'_1$ and $e'$ impurity bands in \cref{figD} not only narrow as the supercell size $N$ is increased but rise with respect to the VBM. To make this clearer, we plot their centers of gravity
\begin{equation}
\label{average}
  \bar{\varepsilon}_i = \frac{\sum_{n{\bf k}} f_{in}({\bf k}) \varepsilon_n({\bf k})}
                             {\sum_{n{\bf k}}f_{in}({\bf k})}
\end{equation}
with respect to the VBM in \cref{figG}; both levels are seen to rise as a function of $N$. The probability $f_{in}({\bf k})$ is the $i$ character of the wavefunction $\psi_{n{\bf k}}$ obtained by projecting $\psi_{n{\bf k}}$ onto site centered orbitals $\beta_i$ and $i \equiv Rlm$ is a composite site, angular momentum index. Here we have chosen $i$ to be the $d_{x^2-y^2}$ and $d_{xy}$ Kubic harmonics on the ${\rm V_{Mo}}$ atom for the $e'$ state and $d_{3z^2-r^2}$ on ${\rm V_{Mo}}$ for the $a'_1$ state and the summation is carried out over the entire Brillouin zone and over the three split off impurity bands in \cref{figD}.  Since the position of the impurity levels introduced by ${\rm V_{Mo}}$ atoms will play an important role in determining the magnetic moment and exchange interaction between impurities, we wish to understand this increase.

A Coulomb potential in a semiconductor gives rise to a Rydberg series of bound states. A finite supercell cannot describe the asymptotic form of the potential correctly but will truncate it on the supercell boundary. In a self-consistent calculation, the requirement of charge neutrality will lead to the charge in the tail of the hydrogenic state accumulating on the supercell boundary. As the supercell size is increased, more of the ``tail'' of the (repulsive) Coulomb potential is described correctly, leading to the rise of the impurity levels seen in Figs.~\ref{figD} and \ref{figG}.

The effect of truncating the Coulomb potential can be estimated using a simple two dimensional (2D) hydrogenic \cite{Yang:pra91} model and first order perturbation theory. For simplicity we assume a circular geometry and replace the 2D Wigner-Seitz cell with a circle of radius $S$ with the same area $\pi S^2 = A_{\rm WS}$. The correction to the ground state energy of a hydrogen atom in 2D is
\begin{align}
\label{equation2D}
  \int_{S}^{\infty} R^2(r) \frac{e^2}{4 \pi\epsilon_r r} 2 \pi r \,dr =
  \frac{e^2}{ 2 \pi\epsilon_r a_0^*} e^{-2S/a_0^*}
\end{align}
where a$_0^*$ is the effective Bohr radius, $\epsilon_r$ is the relative dielectric constant and $R(r)$ is the radial part of the 2D hydrogenic wave function \cite{Yang:pra91} for a screened Coulomb potential. Taking the top of the valence band $\varepsilon_{\rm VBM}$ of an ideal MoS$_2$ monolayer, as estimated in the previous subsection, to be zero, we fit the ab-initio calculated data points with the solid curves shown in \cref{figG}. The fit is very good and deviations can be attributed to local screening effects in the ``real'' inhomogeneous crystal as modelled in DFT.

The $a'_1$ and $e'$ impurity levels increase in energy with increasing supercell size and converge to a (coincidentally) common value of $\sim 62\,$meV in the single impurity limit ($N \rightarrow \infty$). From the fitting, we obtain another estimate of the effective Bohr radius of 8.3 \AA\ for the $e'$ state, of 5.5 \AA\ for the $a'_1$ state and of $\epsilon_r \sim 10.0$ for the in-plane dielectric constant.
These values should be compared to the EMT predictions of
$a_0^* \sim 14$~\AA\ and $\varepsilon_b \sim 34\,$meV with respect to the K point VBM  for the $e'$ holes and
$a_0^* \sim 2.3$~\AA\ and $\varepsilon_b \sim 207\,$meV with respect to the $\Gamma$ point VBM for the $a'_1$ holes. Taking the LDA value of $\Delta_{{\rm K}\Gamma}\sim 150\,$meV from \cref{tab:A} into account, we would expect to find the $a'_1$ ground state at $207-150=57\,$meV above the K point VBM.

\begin{figure}[b]
\includegraphics[scale = 0.35]{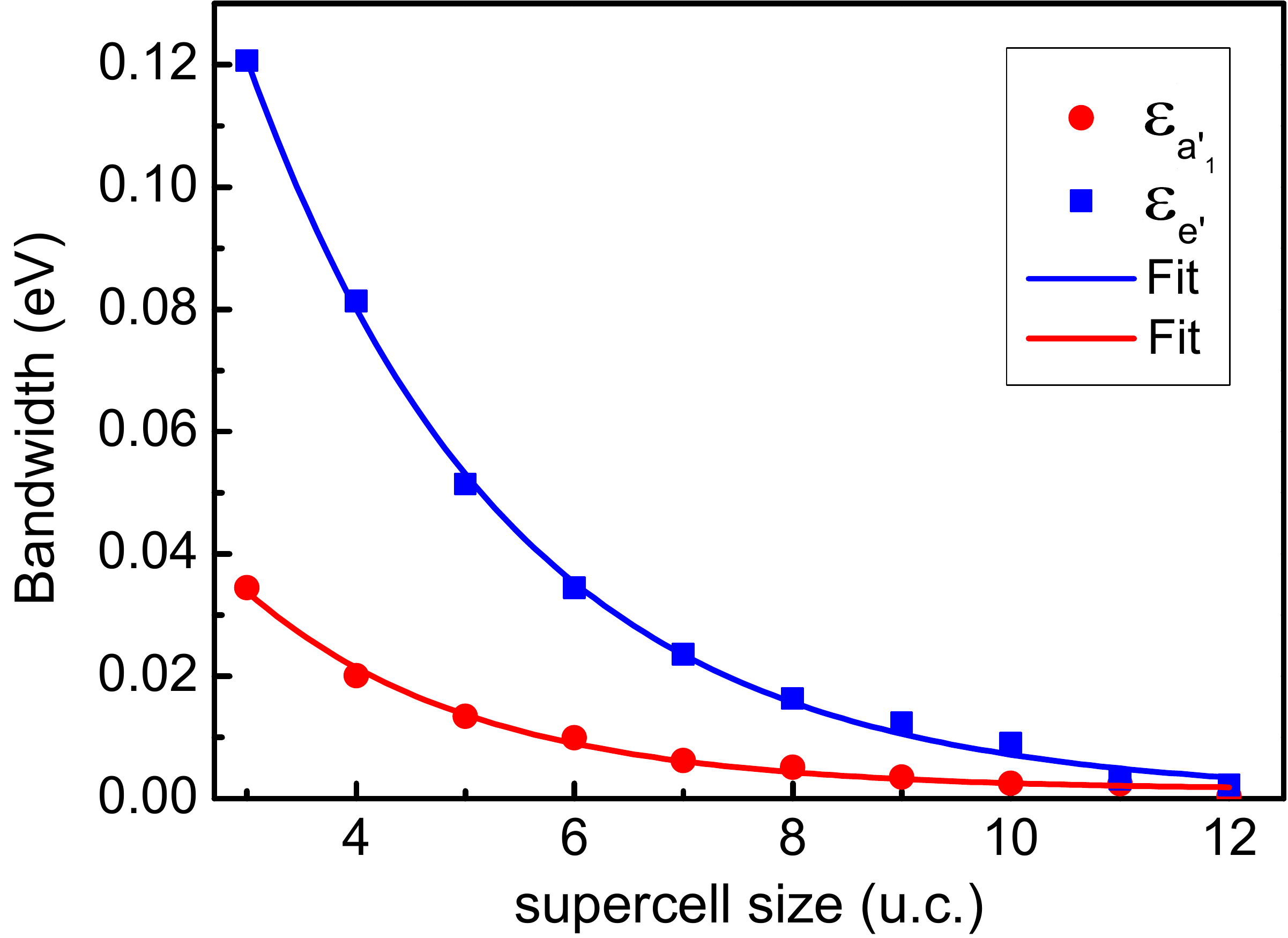}
\caption{Dependence of the band width of (spin-degenerate) impurity bands induced by substitutional V impurities on the supercell size $N$. The data points calculated using \eqref{bandwidth} are fit using the hydrogen model discussed in the text.}
\label{figH}
\end{figure}

According to \cref{figG}, the highest $e'$ impurity states emerge from the valence band when the supercell size is larger than 5$\times$5. This is consistent with the effective Bohr radius of the impurity levels deduced in \cref{figE}. For impurity states with higher principal quantum numbers, the Bohr radii are at least twice as large. These states are not sufficiently localized in the Coulomb potential to appear above the valence band maximum for the largest supercells we have studied.

We can also determine effective Bohr radii from the dependence of the widths of the impurity bands, shown in \cref{figD}, on the supercell size $N$ because of the dependence of the bandwidth on the overlap of impurity wavefunctions in neighboring supercells. In \cref{figH} we plot the second moment of the impurity bands
\begin{align}
\label{bandwidth}
  \sqrt{\frac{\sum_{n{\bf k}} f_{in}({\bf k}) (\varepsilon_n({\bf k})-\bar{\varepsilon}_i)^2}
                             {\sum_{n{\bf k}}f_{in}({\bf k})}} \propto e^{-R/a_0^*}
\end{align}
as a function of $N$ where a$_0^*$ is the effective Bohr radius and $R$ is the distance between dopants in neighboring supercells. From the fitting, we get effective Bohr radii of 7.8 \AA\ for the $e'$ state and 5.2 \AA\ for the $a'_1$ state which are consistent with our earlier results summarized in \cref{tabB}.

\begin{table}[t]
\caption{Summary of the $a'_1$ and $e'$ bound state effective Bohr radii $a_0^*$~(\AA) derived in different ways without relaxation.
}
\label{tabB}
\begin{ruledtabular}
\begin{tabular}{llrrrrrr}
            &        & \multicolumn{2}{c}{V}
                                      & \multicolumn{2}{c}{Nb}
                                                   & \multicolumn{2}{c}{Ta} \\
\cline{3-4} \cline {5-6} \cline {7-8}
            &        & $a'_1 (\Gamma) $
                              & \moc{$e'$(K)}
                                      & $a'_1 (\Gamma) $
                                            & \moc{$e'$(K)}
                                                   & $a'_1 (\Gamma) $
                                                     & \moc{$e'$(K)} \\
\hline
\cref{figE}  &                      & 4.2   &  8.0  & 5.3 & 10.0 & 5.2 & 10.3 \\
\cref{figF}, & Eq.~\ref{eq:shb} Mo  & 5.4   &  8.7  & 6.2 & 9.6 & 5.8 &  9.7  \\
\cref{figgC},& Eq.~\ref{eq:shb} S   & 6.5   &  8.8  & 6.8 & 10.0 & 6.5 & 10.0  \\
\cref{figG}, & Eq.~\ref{equation2D} & 5.5   &  8.3  & 5.9 & 10.0 & 6.2 & 10.4 \\
\cref{figH}, & Eq.~\ref{bandwidth}  & 5.2   &  7.8  & 5.2 &  9.8 & 5.4 & 10.1 \\
EMT          &($\epsilon_r=15$)     & 2.3   & 14.0  & 2.3 & 14.0 & 2.3 & 14.0 \\
\end{tabular}
\end{ruledtabular}
\end{table}

\subsubsection{Effect of relaxation}
\label{sssec:SILV-rel}

\begin{figure}[b]
\includegraphics[scale = 0.35]{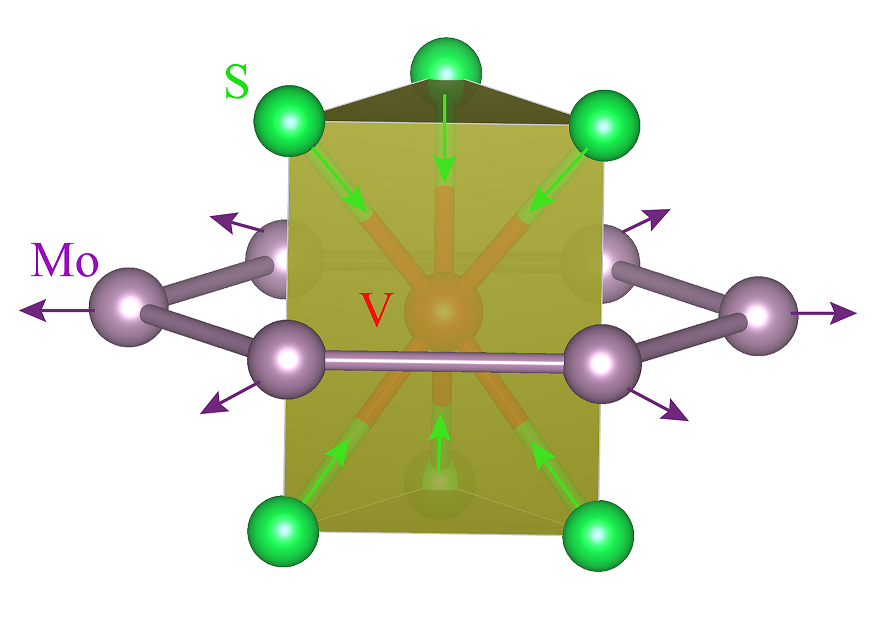}
\caption{(color online). Schematic of the relaxation about a vanadium atom on a substitutional Mo site, ${\rm V_{Mo}}$, in  MoS$_2$.
}
\label{figI}
\end{figure}

\begin{figure*}[t]
\includegraphics[scale = 1.23]{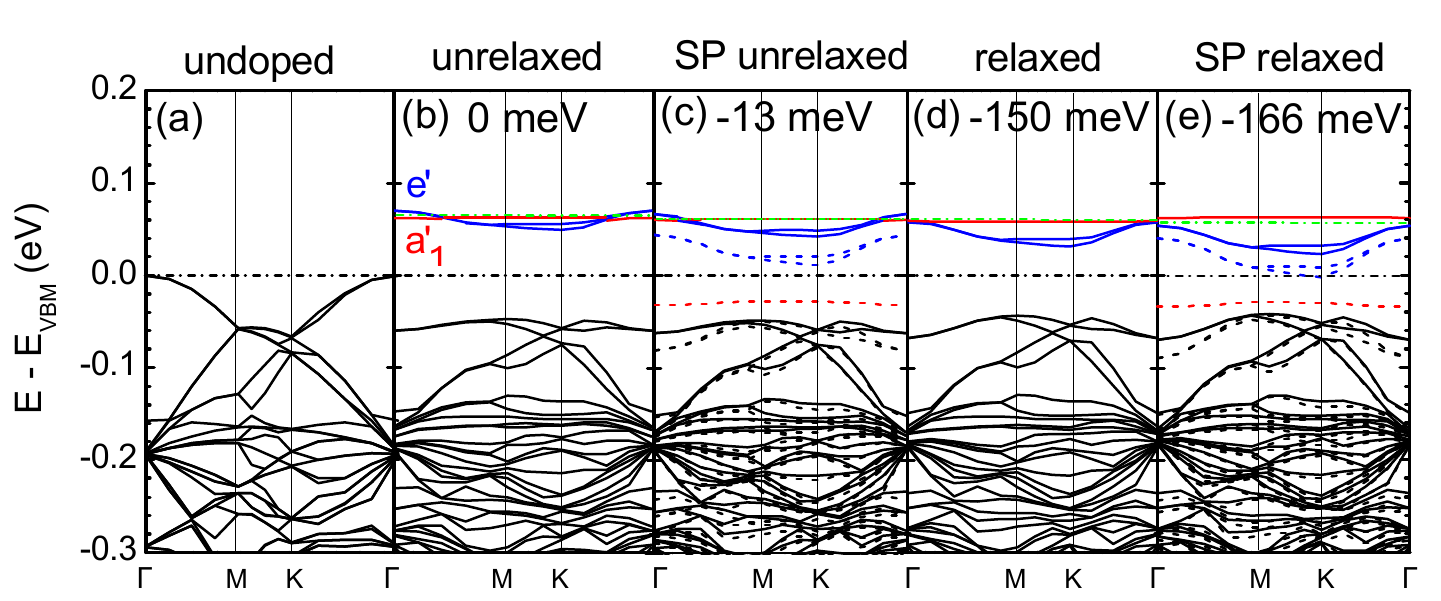}
\caption{Effect of relaxation and spin-polarization on the electronic structure of a 12$\times$12 supercell for an MoS$_2$ monolayer with a single Mo atom replaced by V.
(a) Reference bands for an undoped MoS$_2$ monolayer, energy bands for a single substitutional V impurity (b) without relaxation, (c) with spin polarization (SP) and without relaxation, (d) with relaxation, without spin polarization, (e) spin polarized and relaxed. The $a'_1$ level is red, the $e'$ states are blue. In (c) and (e), the solid (dashed) lines indicate minority (majority) spin states. The energy gain with respect to the unrelaxed case is given in each panel in meV. The zero of energy is the VBM and the Fermi level is indicated by a green dot-dashed line.
}
\label{figJ}
\end{figure*}

One of the most attractive and useful features of a plane wave basis is the ease with which Hellmann-Feynman forces can be calculated. This makes it simple to  determine how the host MoS$_2$ crystal relaxes locally in response to substituting a Mo atom with V, \cref{figI}.  According to the electronic structure \cref{figD}(d) for the unrelaxed geometry, shown enlarged in \cref{figJ}(b), the Fermi level is essentially pinned in the orbitally nondegenerate $a'_1$ state and the system does not undergo a Jahn-Teller (JT) distortion; if we begin geometry optimisation from a JT distorted configuration, the system relaxes back to a symmetric one. Consistent with this, we find only symmetry-conserving (``breathing mode'') relaxation about the vanadium ion with the six nearest neighbour sulphur atoms relaxing towards the V atom and the six in-plane neighbouring Mo atoms relaxing radially away, shown in \cref{figI}. The displacements converge rapidly with supercell size to $\Delta d_{\rm V-S}=0.060\,$\AA\ and $\Delta d_{\rm V-Mo}=0.008\,$\AA\ as seen in \cref{tabC}. The total energy gain from relaxation is 150~meV.

The band structures of the unrelaxed and relaxed 12$\times$12 impurity supercells are compared in panels (b) and (d) of \cref{figJ}. The band structures are aligned on the VBM, located at the $\Gamma$ point as seen in \cref{figJ}(a) for an undoped monolayer, using the $4s$ semicore level shift of the ``B4'' Mo atom on the boundary of the Wigner-Seitz cell furthest from the dopant V atom, see \cref{figC}.
The main effect of relaxation is to lift the quasidegeneracy of the $a'_1$ and $e'$ impurity states, \cref{figJ}(d). The increased V-Mo bond length leads to a lowering of the center of gravity of the $e'$ state with respect to the $a'_1$ state that is antibonding with respect to the neighbouring S $p$ states. As a consequence, the hole state acquires essentially pure $a'_1$ character.

\begin{table}[t]
\caption{Relaxation of nearest neighbour S and Mo shells about a substitutional vanadium atom as a function of the $N \times N$ supercell size $N$. Atomic displacmenents in \AA.
}
\label{tabC}
\begin{ruledtabular}
\begin{tabular}{lrrrrr}
$N$                    & \moc{3} & \moc{6} & \moc{9} & \moc{12} & \moc{15}\\
\hline
$\Delta d_{\rm V-S}$   & -0.060  & -0.060  & -0.060  & -0.060   & -0.060  \\
$\Delta d_{\rm V-Mo}$  &  0.003  &  0.005  &  0.008  &  0.008   &  0.008  \\
\end{tabular}
\end{ruledtabular}
\end{table}

\subsubsection{Spin polarization}
\label{sssec:SILV-SP}

The electronic structure of an unpaired spin in an orbitally nondegenerate $a'_1$ impurity state resembles that of a free hydrogen-like atom and, like a free atom, its total energy can be lowered by allowing the electron to polarize in the local spin density approximation \cite{Gunnarsson:prb74, Gunnarsson:prb76}. The result of doing so in the dilute limit is shown in Figs.~\ref{figJ}(c) and (e). Without relaxation, the localized and dispersionless $a'_1$ state splits by 91~meV leaving the hole with mixed $a'_1$--$e'$ character and a magnetic moment of $m = 1 \mu_B$. Expressing the exchange splitting in terms of an effective Stoner parameter $I_{\rm xc}$ as $\Delta \varepsilon = mI_{\rm xc}$ results in a value of $I_{\rm xc}$ of 91~meV that is substantially less than the free atom value of $I_{\rm xc} \sim 0.7\,$eV \cite{Janak:prb77}.
 In the local density approximation, it is the local electron density that drives the exchange splitting and the small exchange splitting can be understood in terms of the much lower spin density of the impurity state compared to that of a free atom. Consistent with this picture is the even smaller exchange splitting of the more delocalized $e'$ state that is only $\sim 25 \,$meV. Before relaxation, the partial occupation of $a'_1$ and $e'$ states allows the $e'$ state to ``freeload'' on the much more localized $a'_1$ electron density enhancing its spin polarization and exchange splitting (24.7 meV) which decreases to 15.2 meV after relaxation, \cref{tabF}.
The mixing of the $e'$ impurity state with what will eventually become the top of the valence band is clearly seen in the 12$\times$12 supercell in terms of the large exchange splitting of the uppermost host valence band state (black solid and dashed bands) at the $\Gamma$ point. The total energy gain from spin polarization of 13 meV without relaxation or 16 meV with relaxation is dwarfed by the 150~meV energy gain from relaxation.

\begin{table}[b]
\caption{Calculated magnetic moment (in $\mu_B$) for an MoS$_2$ monolayer supercell doped with V, Nb and Ta as a function of the $N \times N$ supercell size without (Un) and with (Re) relaxation. For V the effect of an onsite Coulomb repulsion parameter $U=1\,$eV was examined for the relaxed case. The reciprocal space sampling density is constant.
}
\label{tabD}
\begin{ruledtabular}
\begin{tabular}{ll...........}
$N$ &  & \mor{3} & \mor{4} & \mor{5} & \mor{6} & \mor{7} & \mor{8} & \mor{9} & \mor{10} & \mor{11} & \mor{12} \\
\hline
V  & \text{Un}   & 0.00 & 0.00 & 0.61 & 0.82 & 0.97 & 1.00 & 1.00 & 1.00 & 1.00 & 1.00 \\
   & \text{Re}   & 0.00 & 0.00 & 0.81 & 1.00 & 1.00 & 1.00 & 1.00 & 1.00 & 1.00 & 1.00 \\
   & \text{U}    & 0.54 & 0.68 & 1.00 & 1.00 & 1.00 & 1.00 & 1.00 & 1.00 & 1.00 & 1.00 \\
\hline
Nb & \text{Un}   & 0.00 & 0.00 & 0.00 & 0.34 & 0.54 & 0.68 & 0.87 & 0.93 & 1.00 & 1.00 \\
   & \text{Re}   & 0.00 & 0.00 & 0.00 & 0.66 & 0.75 & 0.89 & 1.00 & 1.00 & 1.00 & 1.00 \\
\hline
Ta & \text{Un}   & 0.00 & 0.00 & 0.00 & 0.32 & 0.42 & 0.57 & 0.86 & 0.95 & 1.00 & 1.00 \\
   & \text{Re}   & 0.00 & 0.00 & 0.00 & 0.68 & 0.83 & 0.92 & 1.00 & 1.00 & 1.00 & 1.00 \\
\end{tabular}
\end{ruledtabular}
\end{table}

For smaller supercell sizes, the dispersion of the $e'$ state increases until it overlaps the unoccupied $a'_1$ level and begins to quench the spin polarization for $N < 8$, \cref{tabD}. Reducing the supercell size further increases the quenching and when, in addition, the impurity potential fails to pull the impurity levels above the VBM for supercell sizes smaller than $5 \times 5$, the magnetic moment disappears. Relaxation enhances the magnetic moment by reducing the overlap of the $a'_1$ and $e'$ states in spite of the unfavourable increase of the $e'$ state dispersion. Even a very small value of the Coulomb repulsion parameter $U=1\,$eV \cite{Dudarev:prb98} can lead to a $3\times3$ supercell becoming polarized. Most of the discrepancies in the literature can be explained in terms of the supercell size, k-point sampling, exchange-correlation potential, $U$ etc. \cite{Dolui:prb13, Yue:pla13, Yun:pccp14, Andriotis:prb14, Lu:nrl14, Miao:jms16, Singh:am17, Miao:ass18, Wu:pla18}

\subsubsection{Formation energies}
\label{sssec:SILV-fe}

\begin{table}[b]
\caption{Formation energies in eV of substitutional V, Nb and Ta impurities in an MoS$_2$ monolayer for a 12$\times$12 supercell. }
\label{tabE}
\begin{ruledtabular}
\begin{tabular}{llll}
           &  V   &   Nb   &   Ta \\
\hline
 Unrelaxed & 0.40 &  0.01  & -0.12  \\
 Relaxed   & 0.25 & -0.15  & -0.23  \\
\end{tabular}
\end{ruledtabular}
\end{table}

The formation energy of a substitutional dopant ${\rm X_{Mo}}$ is defined as
\begin{equation}
E_{\rm form}[{\rm X_{Mo}}]
=E_{\rm tot}[{\rm MoS_2\!:\!X}]
-E_{\rm tot}[{\rm MoS_2}]+\mu_{\rm Mo}-\mu_{\rm X}
\label{eq:form3}
\end{equation}
where E$_{\rm tot}$[{\rm MoS$_2$:X}] is the total energy of an MoS$_2$ monolayer with one Mo atom replaced by one X atom, E$_{\rm tot}$[{\rm MoS$_2$}] is the total energy of a pristine MoS$_2$ monolayer and $\mu_{\rm Mo}$ and $\mu_{\rm X}$ are the total energies per atom of Mo and X in their bulk metallic bcc phases, respectively.
Taking the (spin-polarized) S$_2$ molecule as the reference chemical potential for S, the heat of formation of a MoS$_2$ monolayer, $E_{\rm form}[\rm MoS_2]$, was calculated to be -5.31 eV/formula unit.   The formation energy of V$_{\rm Mo}$ is small and those of Nb$_{\rm Mo}$ and Ta$_{\rm Mo}$ actually become negative when relaxed indicating that doping MoS$_2$ with these group V elements should be experimentally feasible, \cref{tabE}.

\subsection{Binding of V impurity pairs}
\label{sec:BIP}

Two substitutional dopant V atoms will have a negligible interaction energy when sufficiently far apart. This energy can be calculated as follows. First define a reference energy $E_{\rm V}^N$ for a single V dopant atom substituting a Mo atom in MoS$_2$ as
\begin{equation}
\label{form}
  E_{\rm V}^N = E_{\rm tot}^N[{\rm V_{Mo}}] -E_{\rm tot}^N[{\rm MoS}_2]
\end{equation}
where $E_{\rm tot}^N[{\rm MoS}_2]$ is the total energy of an $N \times N$ supercell of MoS$_2$ in equilibrium and $ E_{\rm tot}^N[{\rm V_{Mo}}] $ is the total energy of the same supercell with one Mo atom replaced with a V atom. To calculate absolute formation energies, suitable chemical potentials would need to be included to take account of where the V atom came from and where the Mo atom went to; we will not be concerned with those here. The binding energy $E_b$ is then
\begin{equation}
  E_b^N(R) = E_{\rm tot}^N[{\rm V}_2(R)]
          -E_{\rm tot}^N[{\rm MoS}_2] -2 E_{\rm V}^N
\label{eq:bind1}
\end{equation}
where $E_{\rm tot}^N[{\rm V}_2(R)]$ is the total energy of a supercell with two Mo atoms a distance $R$ apart substituted with V atoms and the last two terms on the right do not depend on $R$. We consider the two cases where relaxation is (Re) and is not (Un) included.

\begin{figure}[t]
\includegraphics[scale = 0.35]{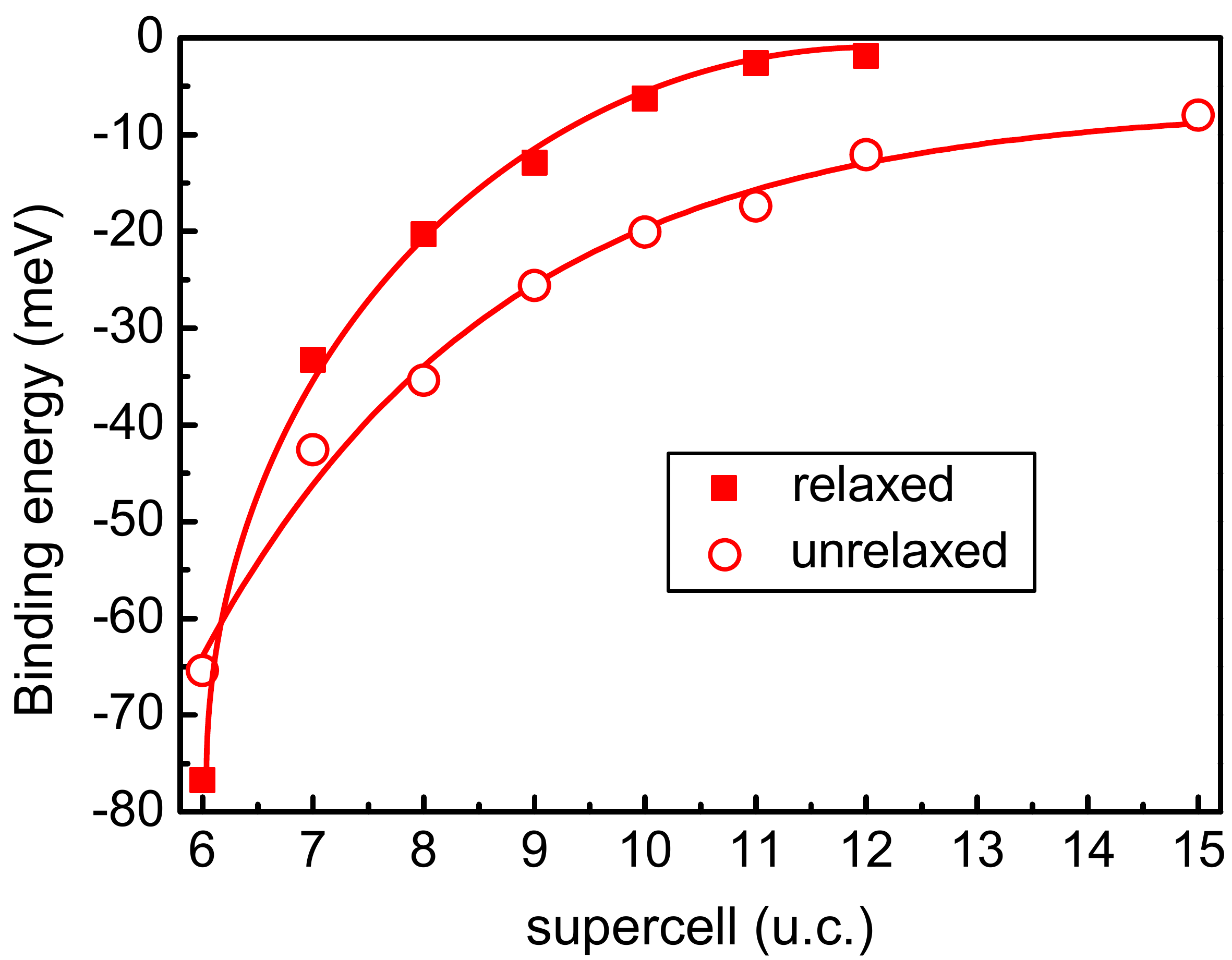}
\caption{$E_b^N(R_{\rm max})$ as a function of the supercell size $N$ for spin-polarized unrelaxed (open red circles) and relaxed (filled red squares) geometries. The red lines are a guide for the eye.
}
\label{figK}
\end{figure}

For supercells containing two substitutional V atoms as far apart as possible ($R_{\rm max}$), with one V atom at the origin and the second at the corner site in \cref{figC}, the binding energy $E_b^N(R_{\rm max})$ is shown as a function of the supercell size $N$ in \cref{figK} with spin polarization included. Although $E_b^N(R_{\rm max})$ does not change much for $N \ge 12$, there is still a surprisingly large binding energy of $\sim 8$~meV for $N=15$ in the unrelaxed case. We can trace this to the near degeneracy of the minority-spin $e'$ and $a'_1$ related bands shown in \cref{figJ}(c) for the unrelaxed V$_{\rm Mo}$ as well as the relatively long range of the $e'$ holes. When relaxation is included, the hole becomes localized in the dispersionless minority-spin $a'_1$ band, \cref{figJ}(e), it becomes much easier to converge the total energy (with respect to BZ sampling and self-consistency) and $E_b^{12}(R_{\rm max})$ decreases fast to $\sim 2$~meV for $N=12$. In general, when there is a gap between occupied and unoccupied states, total energies can be converged better. Since the problem has to do with the (separation independent) reference energy $E_{\rm V}^N$, it turns out to be better to consider
\begin{equation}
  E_b^N(R) = E_{\rm tot}^N [{\rm V}_2(R)]
            -E_{\rm tot}^N [{\rm V}_2(R=\infty)]
\label{eq:bind2}
\end{equation}
and approximate $E_{\rm tot}^N [{\rm V}_2(R=\infty)] \sim E_{\rm tot}^N [{\rm V}_2(R_{\rm max)}]$

\begin{figure}[b]
\includegraphics[scale = 0.32]{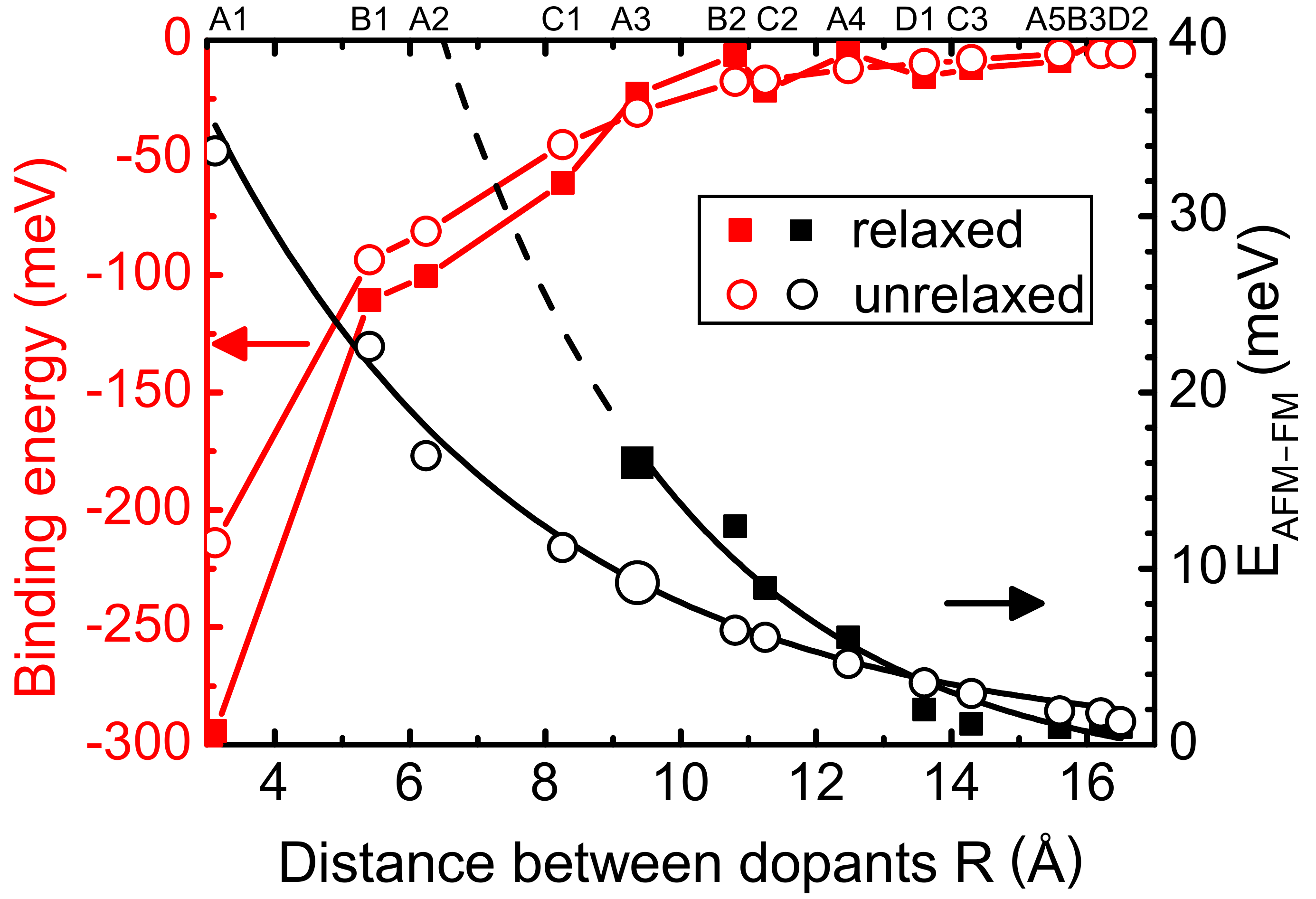}
\caption{Red symbols: interaction energies of dopant ${\rm V_{Mo}}$ atoms in a 12$\times$12 MoS$_2$ supercell as a function of their separation. The binding energy was calculated using \eqref{eq:bind2} for unrelaxed (open red circles) and relaxed (filled red squares) geometries. The lines are a guide to the eye.
Black symbols: total energy differences between parallel and antiparallel aligned spins on the V dopant atoms without (open black circles) and with (filled black squares) relaxation. The lines are fits to an exponentially decaying function. The dashed black line extrapolates the relaxed exchange interaction to separations where it is found to be quenched. The labels along the top of the figure indicate the sites in \cref{figC} and the large symbols refer to the A3 configuration discussed in the text and \cref{figP}.
}
\label{figL}
\end{figure}

\begin{figure*}[t]
\includegraphics[scale = 1.60]{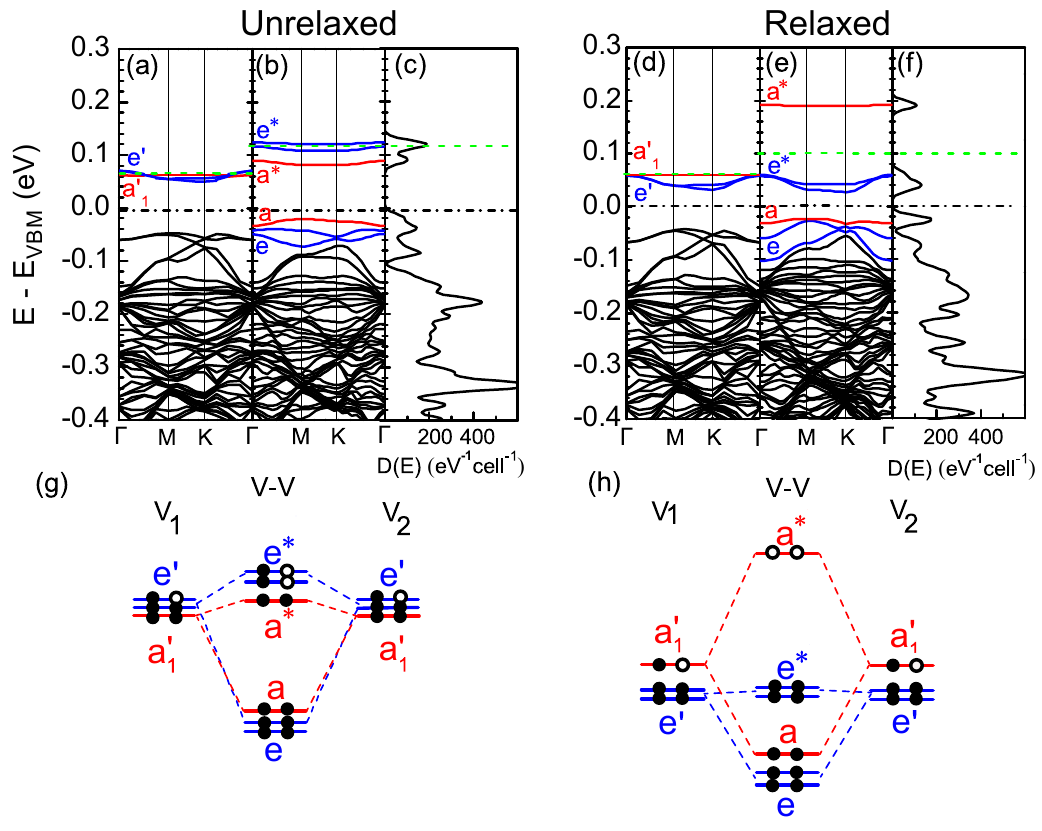}
\caption{Spin unpolarized electronic structure (bands and DoS) for a 12$\times$12 MoS$_2$ supercell with two V atoms in an A1 configuration on nearest neighbour Mo sites (b,c,e,f) and, for comparision, for a single ${\rm V_{Mo}}$ atom (a,d) corresponding to \cref{figJ}((b,d). Without (a,b,c,g) and with (d,e,f,h) atomic relaxation.
Schematic of the coupling mechanism between two V dopants with two holes without (g) and with (h) atomic relaxation. Impurity bands are highlighted in red and blue. The K point VBM is set to be zero (dot-dashed black line) and the Fermi level is shown as a dashed green line.
}
\label{figM}
\end{figure*}

Using a 12$\times$12 supercell and \eqref{eq:bind2} we explore the pair binding energy $E_b^{12}(R)$ for ${\rm V_{Mo}}$ dopants as a function of their separation $R$ in \cref{figL} where one dopant atom is assumed at the site marked 0 in \cref{figC}.
Without relaxation, the (absolute value of the) binding energy decreases monotonically from a value of $\sim 220$~meV for V atoms on neighbouring Mo sites to a value of $\sim 0\,$meV at the maximum separation in a 12$\times$12 supercell (open red circles, left axis). Because these energies are so small, we will later assume that dopant atoms are randomly distributed in real materials that are not in full thermodynamic equilibrium.

With relaxation (filled symbols), the magnitude of the binding energy increases for separations $R$ smaller than a critical separation, $R_c \sim 8.5\,$\AA, and does not change for separations larger than this. This behaviour is intimately related to quenching of the magnetic moments for V dopant atoms closer than $R_c$ and for these separations, $R < R_c$, an exchange interaction cannot be determined. We proceed to consider the magnetic interactions.

\subsection{Magnetic Interaction of impurity pairs}
\label{sec:MI}

We estimate the exchange interaction $\Delta E(R)$ between pairs of dopant atoms as the energy difference between configurations with the V magnetic moments aligned parallel (``ferromagnetically'', FM) and antiparallel (``antiferromagnetically'', AFM)
\begin{equation}
 \Delta E(R) = E_{\rm tot}\left[{\rm V}_2^{\rm AFM}(R)\right]
             - E_{\rm tot}\left[{\rm V}_2^{\rm FM}(R)\right]
\label{eq:form2}
\end{equation}
in 12$\times$12 supercells so that the interaction between periodic images is acceptably small. Because the spin-polarized calculations are computationally expensive, care is taken to construct suitable starting V$_2^{\rm FM}$ configurations using ``superpositions'' of relaxed, spin-polarized local atomic configurations for single ${\rm V_{Mo}}$. Starting V$_2^{\rm AFM}$ configurations are constructed from relaxed V$_2^{\rm FM}$ configurations so only the much smaller differential relaxation needs to be calculated.

The energy difference between antiferromagnetic and ferromagnetic ordering without (open black circles) and with (filled black squares) relaxation is shown on the right axis of \cref{figL}. Before relaxation, neighbouring V dopant atoms show FM coupling with a total moment of $2\,\mu_B$ or $1\,\mu_B$ per V for all separations. The interaction decreases monotonically and exponentially from a maximum of $\sim 33\,$meV for nearest neighbours with a decay length of $\sim 5.3$ \AA.
With relaxation included the magnetic moments are quenched for separations $R$ smaller than a critical separation, $R_c \sim 8.5\,$\AA, and for these separations an exchange interaction cannot be determined. For separations $R > R_c$, the coupling remains ferromagnetic and is enhanced. Because the maximum value of the magnetic ordering temperature will depend strongly on this relaxation-induced behaviour, we need to understand its origin.

\begin{figure}[b]
\includegraphics[scale=0.36]{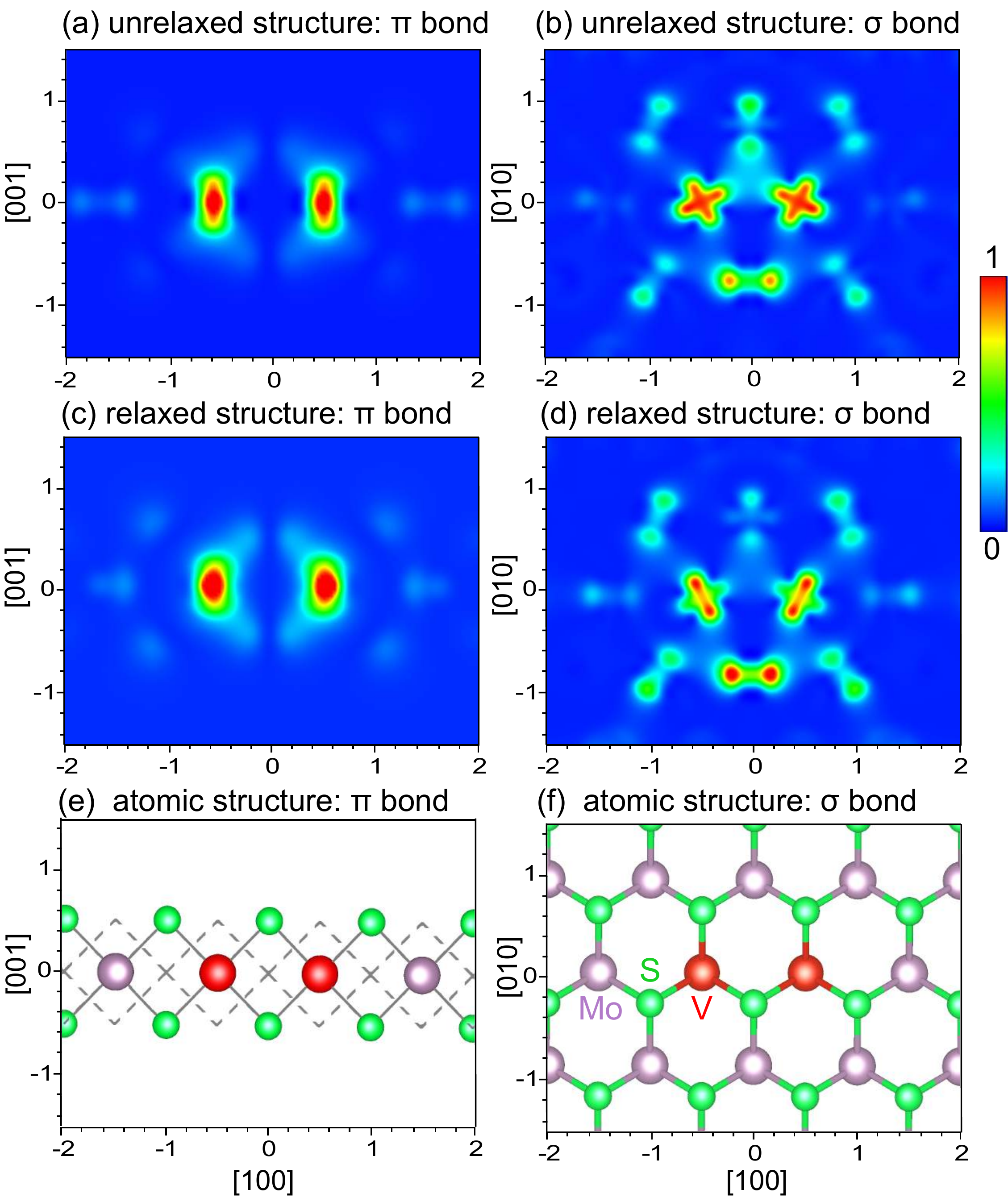}
\caption{Side view of the partial charge distributions of the $\pi$ bond (left panels) and top view of the $\sigma$ bond (right panels) without (a, b) and with (c, d) atomic relaxation. The corresponding atomic structures are shown in (e,f).
}
\label{figN}
\end{figure}

\subsubsection{Quenching of moments for  $R < R_c$}
\label{sssec:QM}

To do so, we consider a 12$\times$12 supercell for an MoS$_2$ monolayer with a pair of Mo atoms on neighbouring sites substituted with V. The unpolarized supercell electronic structures and DoS are shown in \cref{figM} without (left) and with relaxation (right). The corresponding band structures for a single V impurity are included in the left panels for reference. In the spirit of a defect molecule model, \cref{figM}(a-c) suggests that the $e'$ orbitals form $\sigma$ bonding-antibonding $e$-$e^*$ pairs scarcely lifting the degeneracy of the $e$ states while the $a'_1$ states interact less strongly to form a $\pi$ bonding-antibonding $a$-$a^*$ pair. Without relaxation, the strength of the $\pi$ bond between the $a'_1$ orbitals is not strong enough to raise the $a^*$ level above the $e^*$ level and the two holes reside on the fourfold orbitally and spin degenerate $e^*$ states as sketched in \cref{figM}(g). This leads to a DoS peak at the Fermi level that is unstable with respect to exchange splitting.

Relaxation results in a structure where the neighbouring S atoms move closer to the V atoms, the two V atoms move apart and the $a'_1$ levels on individual ${\rm V_{Mo}}$ atoms are lifted clear of the $e'$ levels. The reduced V-S separation strengthens the $\pi$ bond [\cref{figN}(a) versus \cref{figN}(c)] through hybridization between vanadium $d_{3z^2-r^2}$ and sulphur $p_x$ and $p_y$ orbitals, while the $\sigma$ bonds formed by vanadium $\{d_{xy},d_{x^2-y^2}\}$ orbitals are weakened by the increased V-V separation [\cref{figN}(b) versus \cref{figN}(d)]. This makes the $\pi$ bond the dominant bonding interaction between dopants. The $a$-$a^*$ splitting is increased so much by relaxation that the $a^*$ level is lifted well above the $e^*$ level and the two holes are accommodated in an orbitally nondegenerate state (rhs of \cref{figM}).

\begin{figure}[t]
\includegraphics[scale = 0.35]{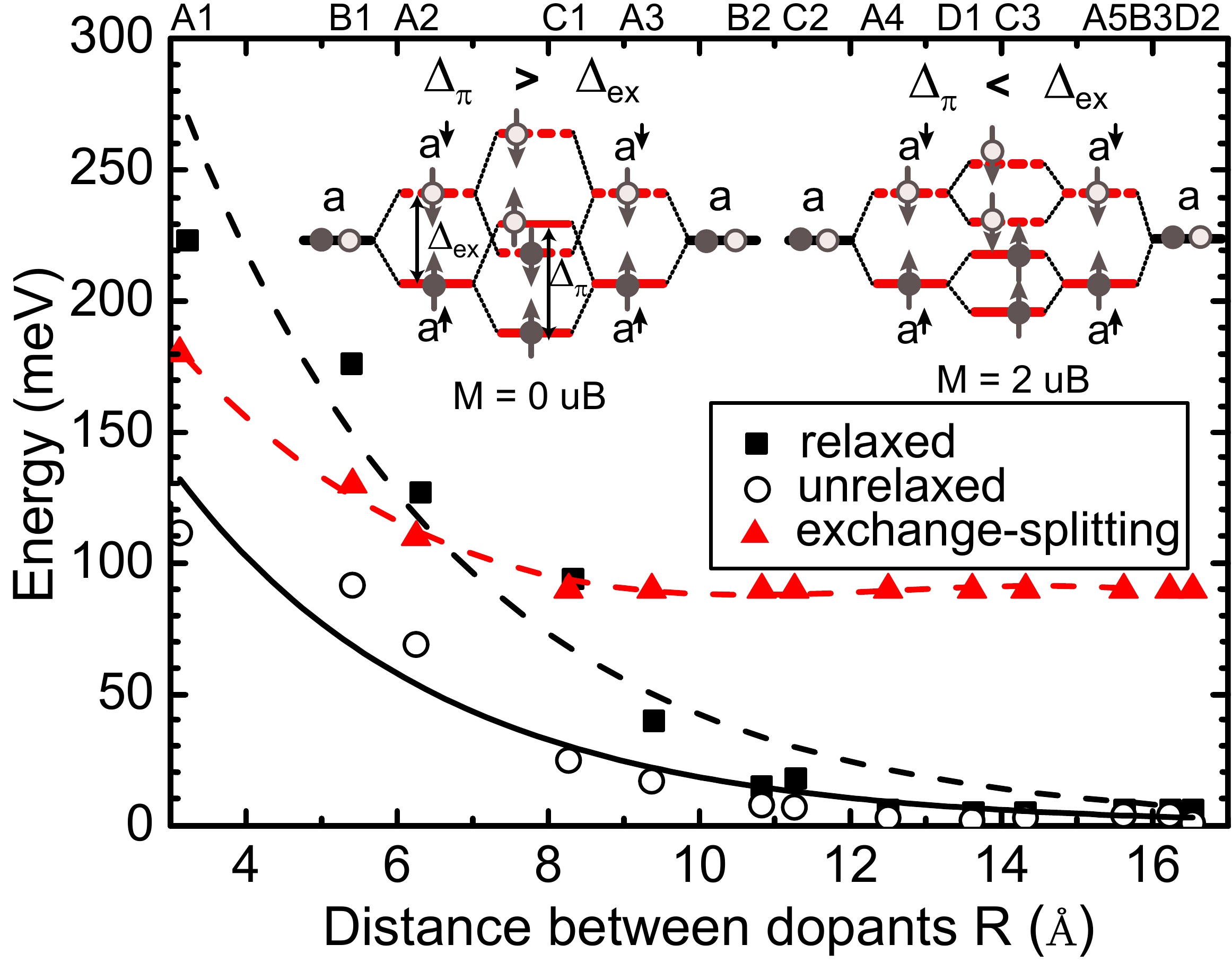}
\caption{Unpolarized $a-a^*$ level separation with (filled black squares) and without (open black circles) atomic relaxation plotted as a function of the separation between the substitutional dopant atoms in a 12$\times$12 MoS$_2$ supercell. The $a^*$ exchange splitting for different (unrelaxed) configurations is indicated by red triangles. The insets show the energy level schemes with spin polarization included for relaxed configurations with impurity separations below (lhs) and above (rhs) the critical separation of 8.5 \AA, respectively.
}
\label{figO}	
\end{figure}

To demonstrate how competition between bonding and exchange interactions of the V $d_{3z^2-r^2}$ orbitals leads to the quenching of the magnetic moments, we examine how these interactions depend on the separation between the dopant atoms. We define the bond strength $\Delta_\pi$ of the $\pi$ bond to be the $a^*$--$a$ bonding-antibonding splitting. \cref{figO} shows how $\Delta_\pi$ depends on the impurity separation $R$ with (filled black squares) and without (open black circles) structural relaxation. As $R$ increases, $\Delta_\pi$ decreases because of the decreasing wavefunction overlap. Without relaxation (open circles), $\Delta_\pi$ is smaller than the exchange splitting (red triangles, dashed red line) for all separations and a triplet state would form as indicated in the rhs inset of \cref{figO}. With relaxation (filled squares), $\Delta_\pi$ is larger and exceeds the exchange splitting at distances smaller than $\sim 7\,$\AA. A singlet state is formed to gain bonding energy, as sketched in the lhs inset of \cref{figO}, and this leads to the quenching of the magnetic moment. The critical quenching separation is twice the effective Bohr radius $a_0^*=4.2\,$\AA\ for the $a'_1$ state, implying the formation of a $\pi$ bond. In general, to quench the magnetic moment, the $\pi$ bonding interaction should be strong enough to make the $a^*$ state the highest lying state.

Lastly, we note a significant enhancement of the exchange splitting when two vanadium atoms are close, \cref{figO}. From a value of 92~meV for single V dopants, the increase in hole density at short separations doubles the exchange splitting to $\sim 180\,$meV for (unrelaxed) V dopants on neighbouring Mo sites.

\subsubsection{Enhancement of Exchange Interaction for  $R > R_c$}
\label{sssec:xyz}

\begin{figure}[t]
\includegraphics[scale = 0.50]{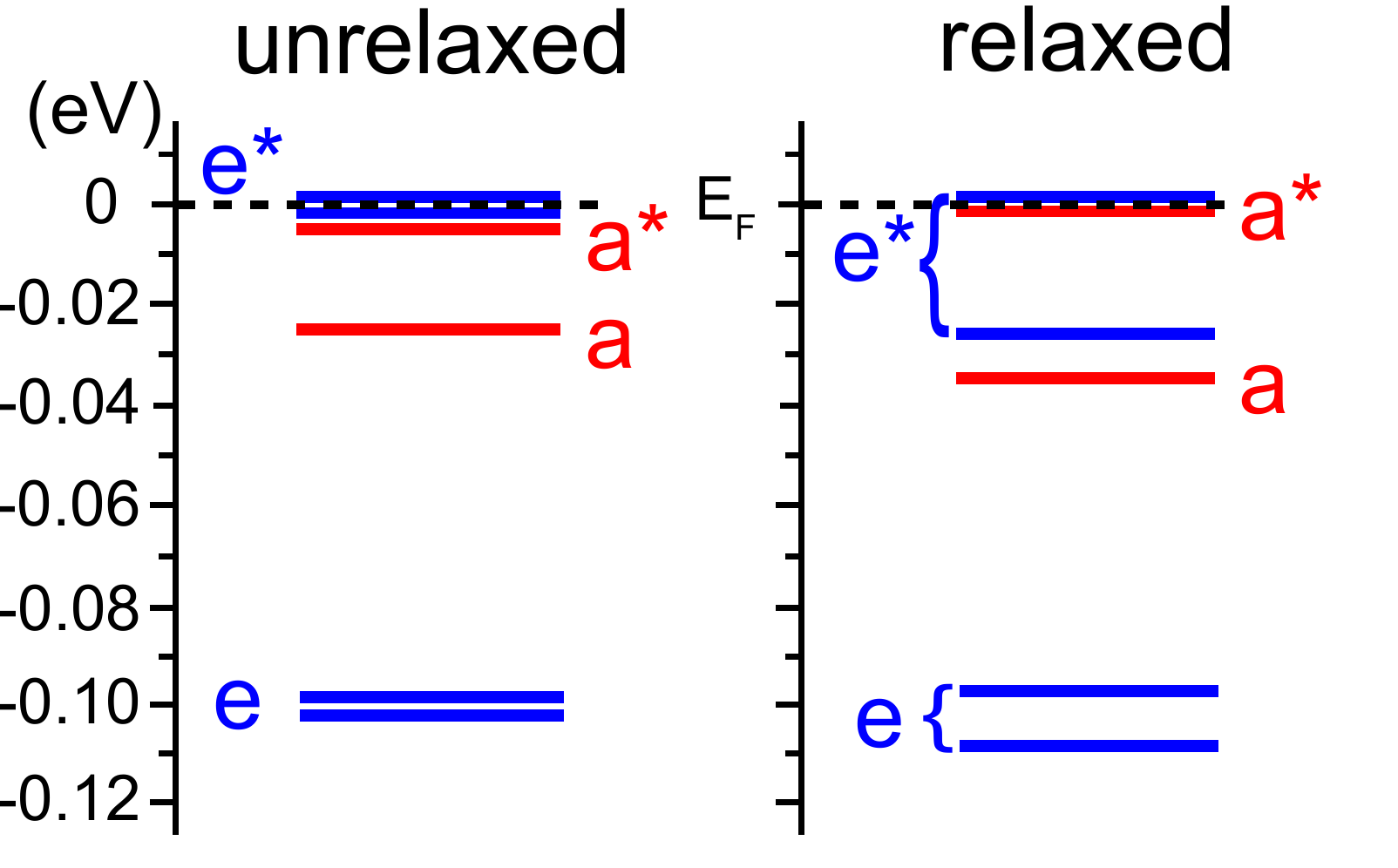}
\caption{Defect level structure calculated without spin polarisation for an A3 configuration of two V$_{\rm Mo}$ atoms. (lhs) unrelaxed and (rhs) relaxed. The levels are calculated from the appropriate weighted average of the $\Gamma$, K/K$'$ and M eigenvalues of the corresponding bands. $a$ and $a^*$ (red), $e$ and $e^*$ (blue) levels originate in the $a'_1$ and $e'$ levels for single V$_{\rm Mo}$ dopants. The zero of energy is the Fermi level indicated by a black dashed horizontal line.}
\label{figP}
\end{figure}

For separations greater than $R_c$, the exchange interaction is strongly enhanced by relaxation before decaying more strongly than the unrelaxed case till it eventually becomes smaller when $R\sim 13$\AA, \cref{figL}. We can understand the enhancement by considering in \cref{figP} the defect levels associated with the A3 configuration, \cref{figC}. For the unrelaxed structure (lhs), the breaking of the local $D_{3h}$ symmetry is negligible and the bonding $e$ and antibonding $e^*$ states remain doubly degenerate. At this separation of 9.55~\AA, the bonding interaction of the $a'_1$ states is much less that of the $e'$ states so that the $e^*$ level is the lowest unoccupied level to which both holes gravitate. Because it is degenerate, the $e^*$ level can exchange split with both holes aligned to form a triplet spin state. The exchange splitting is weak because of the delocalisation of the $e$ levels.

Relaxation reduces the V-S bond length while increasing the V-V bond length and breaks the local $D_{3h}$ symmetry leading to a significant splitting of the degenerate $e$ and $e^*$ levels as well as a small increase in the bonding interaction between the $a'_1$ levels (\cref{figP}, rhs). The net result is that the $a^*$ level and highest $e^*$ level become degenerate and accommodate the two holes. Because of the greater localization of the $a$ levels, this leads to an enhancement of the exchange splitting for the parallel (FM) configuration of the two V$_{\rm Mo}$ dopants (compared to the unrelaxed case) and a reduction for the antiparallel (AFM) configuration with a corresponding increase of the $E_{\rm AFM}-E_{\rm FM}$ energy difference (large symbols in \cref{figL}). To a good approximation the interaction strength only depends on the separation and decays exponentially more rapidly than the unrelaxed case with a much reduced decay length of 3.6~\AA\ reflecting the greater localization of the $a'_1$ holes.

\subsection{Nb and Ta in MoS$_2$}
\label{ssec:NbTa}

We expect Nb and Ta to more closely resemble Mo than V with a weaker central cell potential leading to less localized impurity states than in the case of V. Nb and Ta will turn out to have very similar effective Bohr radii and binding energies that lead to virtually indistinguishable magnetic properties.

\subsubsection{Single impurity limit: Nb and Ta in MoS$_2$}
\label{ssec:SILNbTa}

\begin{figure}[b]
\includegraphics[scale = 0.35]{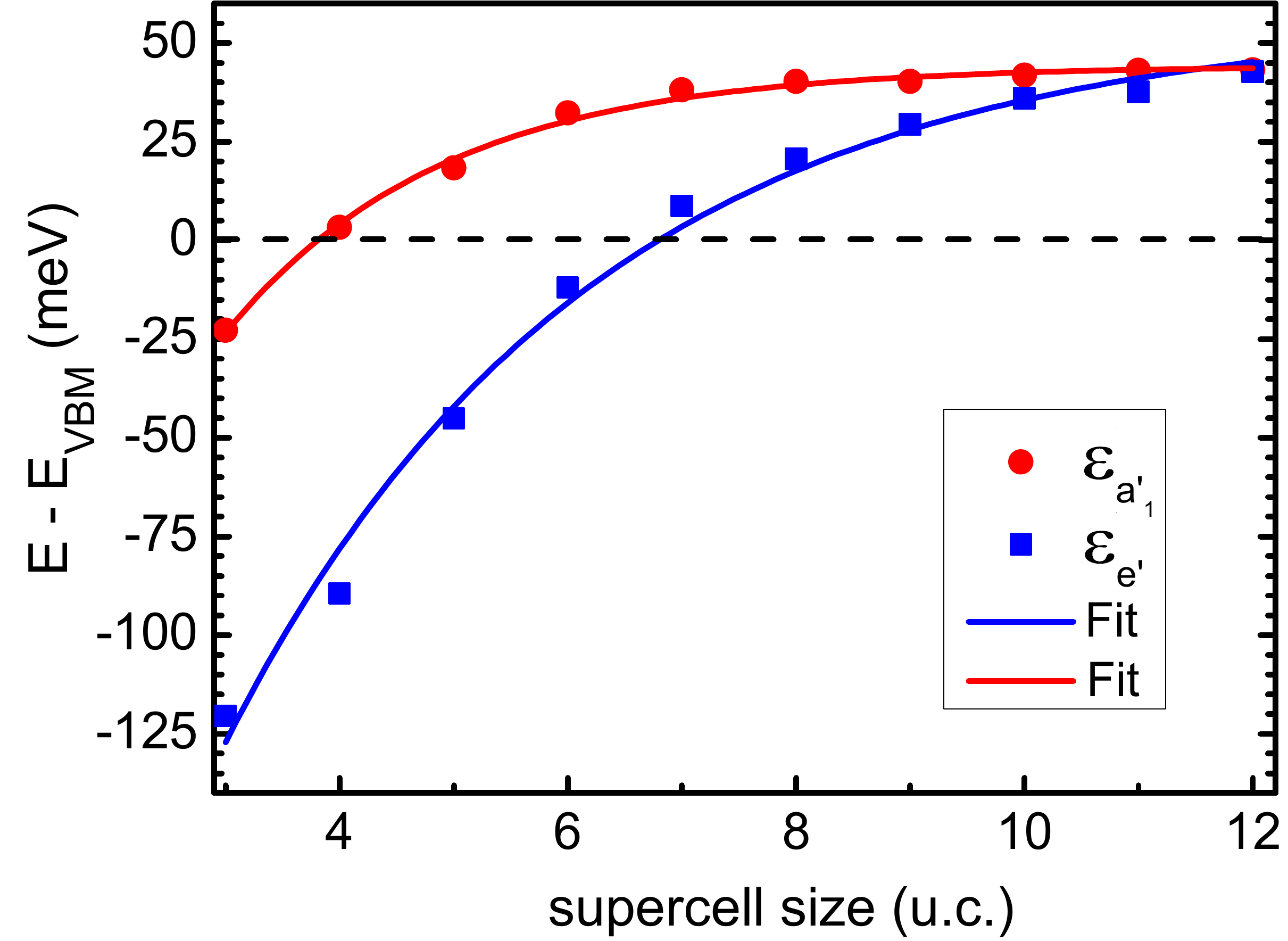}
\caption{Same as \cref{figG} but for Nb instead of V. The corresponding results for Ta are virtually indistinguishable.
}
\label{figQ}
\end{figure}

For an unrelaxed Nb (Ta) substitutional impurity, we find an effective Bohr radius of 10.0 (10.3) \AA\ for the $e'$ state and 5.3 (5.2) \AA\ for the $a'_1$ state by fitting the circularly averaged wave function in a 12$\times$12 supercell to be compared to values of 8~\AA\ and 4.2~\AA, respectively, for V found in \cref{ssec:SILV}, \cref{tabB}. The binding energies of these $e'$ and $a'_1$ states converge to a common value of $\sim 45\,$meV in the large supercell, single impurity limit \cref{figQ}; the results for Nb and Ta are virtually indistinguishable and only those for Nb are shown. The smaller binding energies and larger effective Bohr radii make the Nb (Ta) impurity states more sensitive to the supercell truncation of the impurity potential compared to V. Increasing delocalization of the holes from V$\rightarrow$Ta leads to a reduction of the exchange splittings, \cref{tabF}, and a magnetic moment is found to develop only when $N > 5$. Total polarization only occurs for $N > 10$, see \cref{tabD}.

After relaxation, the Nb-Mo (Ta-Mo) bond length increases by 0.040 (0.036)~\AA, leading to a  lowering of the $e'$ state. The hole then goes into the $a'_1$ state whose exchange splitting increases while that of the $e'$ state decreases because of the reduced overlap in space of the $e'$ and $a'_1$ partial electron densities. As we already saw for V in \cref{tabD}, relaxation enhances the magnetic moments.

\begin{table}[t]
\caption{Summary of the exchange splitting $\Delta_{\rm ex}$ in meV of $a'_1$ and $e'$ bound states in unrelaxed (Un) and relaxed (Re) structures in 12$\times$12 supercells.}
\label{tabF}
\begin{ruledtabular}
\begin{tabular}{cccccccc}
    & \multicolumn{2}{c}{V}
                     & \multicolumn{2}{c}{Nb}
                                  & \multicolumn{2}{c}{Ta} \\
\cline{2-3} \cline {4-5} \cline {6-7}
$\Delta_{\rm ex}$
    & $a'_1 (\Gamma) $
            & \moc{$e'$(K)}
                     & $a'_1 (\Gamma) $
                            & \moc{$e'$(K)}
                                  & $a'_1 (\Gamma) $
                                         & \moc{$e'$(K)} \\
\hline
Un  & 91.3  &  24.7  & 32.4 & 10.9 & 35.6 & 10.8 \\
Re  & 96.1  &  15.2  & 53.2 & 9.8 & 52.0 & 9.7 \\
\end{tabular}
\end{ruledtabular}
\end{table}

\subsubsection{Interaction of Nb (Ta) impurity pairs}

\begin{figure}[t]
\includegraphics[scale = 0.32]{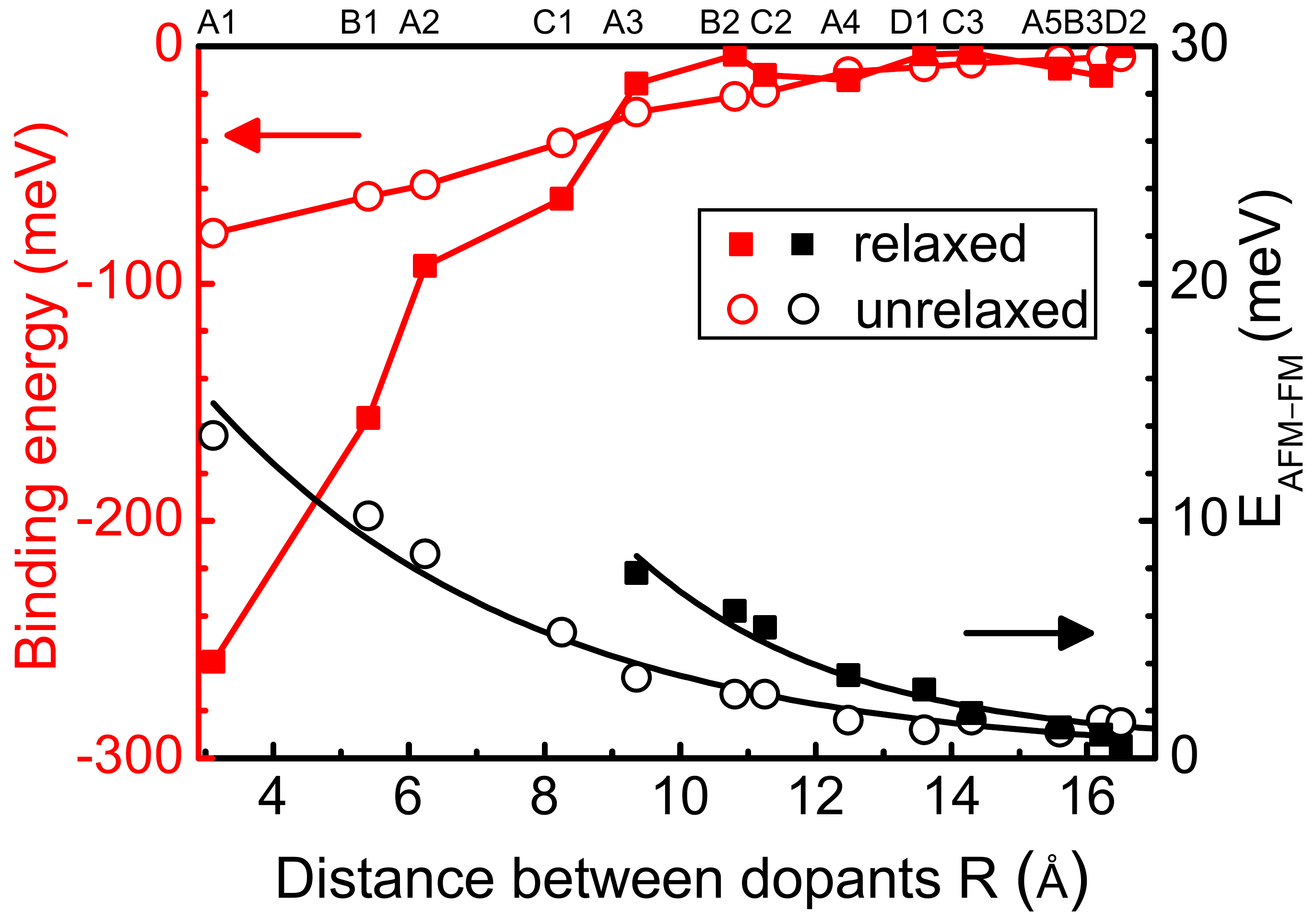}
\caption{Same as \cref{figL} but for Nb instead of V. The corresponding results for Ta are virtually indistinguishable.
}
\label{figR}
\end{figure}

Because of the smaller bound state energies and weaker spin polarization, the exchange interaction between pairs of Nb (Ta) dopant atoms is weaker than that between pairs of V atoms, \cref{figL}, and decays more slowly with increasing separation, \cref{figR}. To a good approximation the interaction strength only depends on the separation $R$ and decays exponentially with $R$ with a decay length of 5.2 (5.8)~\AA\ versus 3.6 for V when relaxation is included.

The exchange splitting of the $a'_1$ and $e'$ levels and their relative hole occupations determine the strength of their different exchange interactions. Relaxation raises the $a^*$ level to become degenerate with the upper $e^*$ level, enhancing the exchange splitting of the $a$ level while reducing that of the $e$ level (see \cref{tabF}) and changes the character of the holes from $e$-like to $a$+$e$-like. This increases the strength of the exchange interaction but leads to a faster decay as we saw in \cref{figL,figR} for vanadium.

In summary, before relaxation, the long-range weak $e'$ ferromagnetic interaction dominates while the strong short-range $a'_1$ interaction dominates after relaxation, the near-degeneracy of the upper $e^*$ and $a^*$ levels making it possible to form a triplet without violating the Pauli exclusion principle.

\section{Magnetic Ordering}
\label{sec:MO}

The Ising spin model in two dimensions undergoes a phase transition to long-range magnetic order at a finite temperature \cite{Onsager:pr44, Yang:pr52}. For a Heisenberg model with isotropic exchange interactions, thermal fluctuations destroy long-range magnetic ordering in two dimensions at any finite temperature  \cite{Mermin:prl66, Hohenberg:pr67}. The Ising spin model, with spin dimensionality $n=1$, is recovered by assuming a generalized Heisenberg spin Hamiltonian with isotropic exchange and strong perpendicular anisotropy. Though the predictions of such generalized Heisenberg models are not identical to those of the Ising spin model, the consensus is that for ferromagnetism to exist in two-dimensional systems, magnetic anisotropy is essential. We therefore begin this section on magnetic ordering by studying the magnetic anisotropy of a single substitutional dopant, the so-called single ion anisotropy (SIA).

\begin{figure*}[t]
\includegraphics[scale = 0.38]{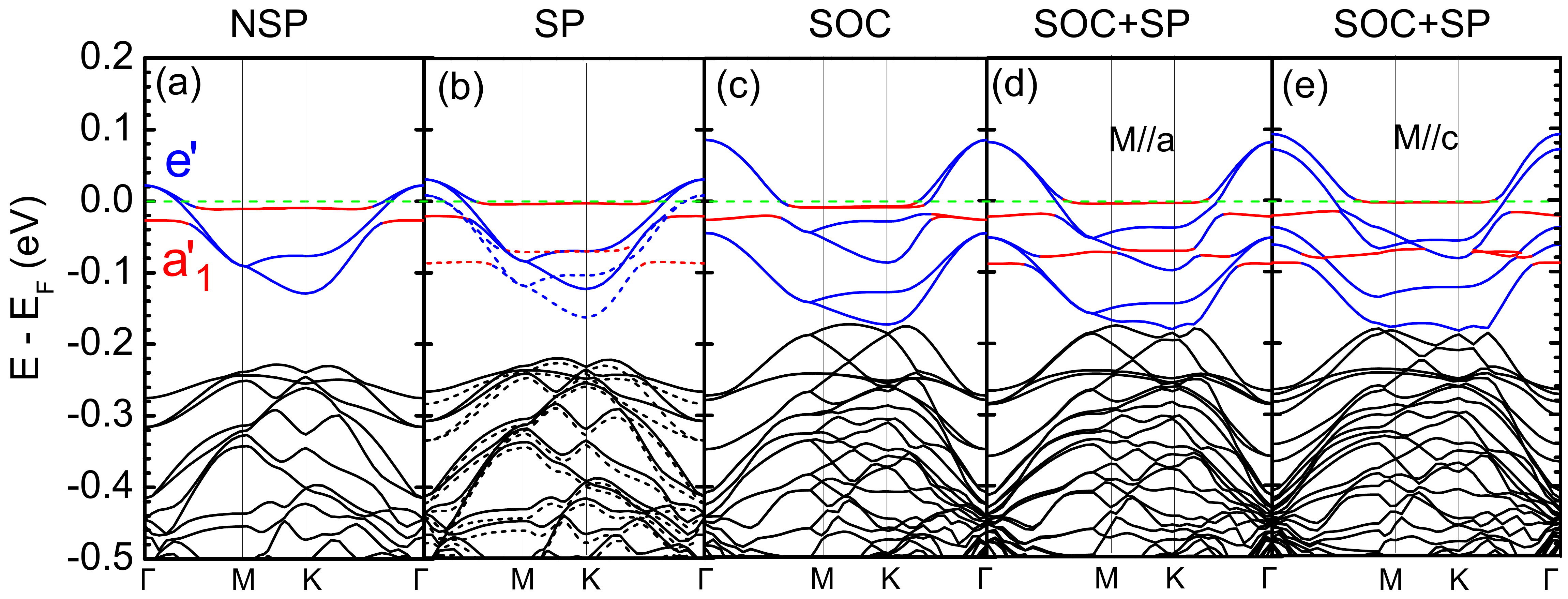}
\caption{The band structure of a single V dopant in a $6\times6$ supercell including relaxation. (a) non-spin polarized (NSP), (b) spin polarized (SP), (c) only spin orbit coupling (SOC). Including SOC and SP with the magnetization in-plane $({\bf M} \! \parallel \! a)$ (d) and perpendicular to the monolayer plane $({\bf M} \! \parallel \! c)$ (e). Because SOC does not mix the $e'$ ($m_l=\pm2$) and $a'_1$ ($m_l=0$) states strongly, we can continue to label these blue and red. In (b), the solid and dashed lines represent spin-up and spin-down states, respectively. The Fermi level shown as a horizontal green dashed line is chosen to be zero. }
\label{figS}
\end{figure*}

\subsection{Single ion anisotropy}
\label{ssec:SIA}

Microscopically, magnetic anisotropy arises when spin and orbital degrees of freedom are coupled by the spin-orbit interaction so that the total energy depends on the spatial orientation of the magnetic moment. According to the ``force theorem'' \cite{Mackintosh:80, Heine:80}, {\em changes} to the total energy, $\delta E$, that result from a small perturbation can be related to changes in the sum of the single-particle eigenstates of the Kohn-Sham equations \cite{Kohn:pr65} of DFT, $\delta E \sim \delta \sum_i^{\rm occ} \varepsilon_i$, which should not be iterated to self consistency. The force theorem has been applied to the calculation of the magnetic anisotropy energy (MAE) where the perturbation is the spin-orbit coupling (SOC) \cite{Daalderop:prb90a} and comparison with explicit total energy calculations yields essentially perfect agreement for Fe, Co and Ni \cite{Stiles:prb01}. The advantage of the force-theorem approach is that it allows the MAE to be directly related to (changes to) the electronic structure \cite{Daalderop:prb94, Daalderop:UMSI94} which are shown for the relaxed configuration of a single V atom in a $6\times6$ supercell of MoS$_2$ in \cref{figS}.

In the context of \cref{figJ}(e), we already discussed the exchange splitting of the $a'_1$ and $e'$ levels. In a $6\times6$ supercell, the increased band dispersion leads to a smaller exchange splitting. With (without) relaxation, these splittings averaged over the Brillouin zone are, respectively, 66 (64)~meV and  20 (22)~meV (\cref{figS}b). Because MX$_2$ monolayers do not have inversion symmetry, SOC leads to a substantial splitting of the spin degenerate states at K and K$'$ with $\{ d_{xy}, d_{x^2-y^2}\}$ character ($l=2, m = \pm2$) \cite{Zhu:prb11}. In \cref{figS}(c), we see that SOC splits the $e'$ level at $\Gamma$ by 130~meV while the $a'_1$ level with $d_{3z^2-r^2}$ character ($l=2, m=0$) is not affected. For a larger (12$\times$12) supercell, the effect of SOC on the bands shown in \cref{figJ}(d) is to split the upper $e'$ level so that it lies above the unaffected $a'_1$ level and accommodates the hole.

To understand the energy levels obtained with SOC and spin polarization (exchange splitting) in the single impurity limit, it is instructive to consider the model Hamiltonian
\begin{equation}
  H   = H_0 + \Delta {\bf m}.{\bf s}  + \xi {\bf l} \cdot {\bf s}
\end{equation}
where
$H_0$ is the spin-independent part of the Hamiltonian,
$\Delta {\bf m}$ is the exchange field that leads to an exchange splitting $\Delta$, and
$\bf m$ is a unit vector in the direction of the magnetization, $\bf m \equiv M/|M|$.
In the subspace of the $l=2, m_l=\pm 2$ orbitals
\begin{equation}
 \xi {\bf l} \cdot {\bf s}
 = \frac{\xi}{2}  \begin{pmatrix} l_z & l_- \\   l_+ & -l_z  \end{pmatrix}
 =                \begin{pmatrix} \xi & 0   \\   0   & -\xi  \end{pmatrix}
\end{equation}
where we use Hartree atomic units with $\hbar=1$. For ${\bf M} \! \parallel \! c$,
\begin{equation}
\Delta {\bf m}.{\bf s}
 =                \begin{pmatrix} \frac{\Delta}{2}   & 0   \\   0   & -\frac{\Delta}{2}   \end{pmatrix}.
\end{equation}
and the SOC Hamiltonian can be written as
\begin{equation}
H   = H_0 + \begin{pmatrix}
  \xi + \frac{\Delta}{2} &      0                  &     0                 &  0 \\
      0                  & -\xi - \frac{\Delta}{2} &     0                 &  0 \\
      0                  &      0                  & -\xi+\frac{\Delta}{2} &  0 \\
      0                  &      0                  &     0                 & \xi-\frac{\Delta}{2} \\
                        \end{pmatrix}
\end{equation}
For  ${\bf M} \! \parallel \! a$ we have
\begin{equation}
H   = H_0 + \begin{pmatrix}
      \xi        & \frac{\Delta}{2} &   0               &  0                \\
\frac{\Delta}{2} &      -\xi        &   0               &  0                \\
      0          &      0           &       -\xi        &  \frac{\Delta}{2} \\
      0          &      0           &  \frac{\Delta}{2} &        \xi        \\
                        \end{pmatrix}.
\end{equation}

\begin{figure}[t]
 \centering
\includegraphics[scale = 0.35]{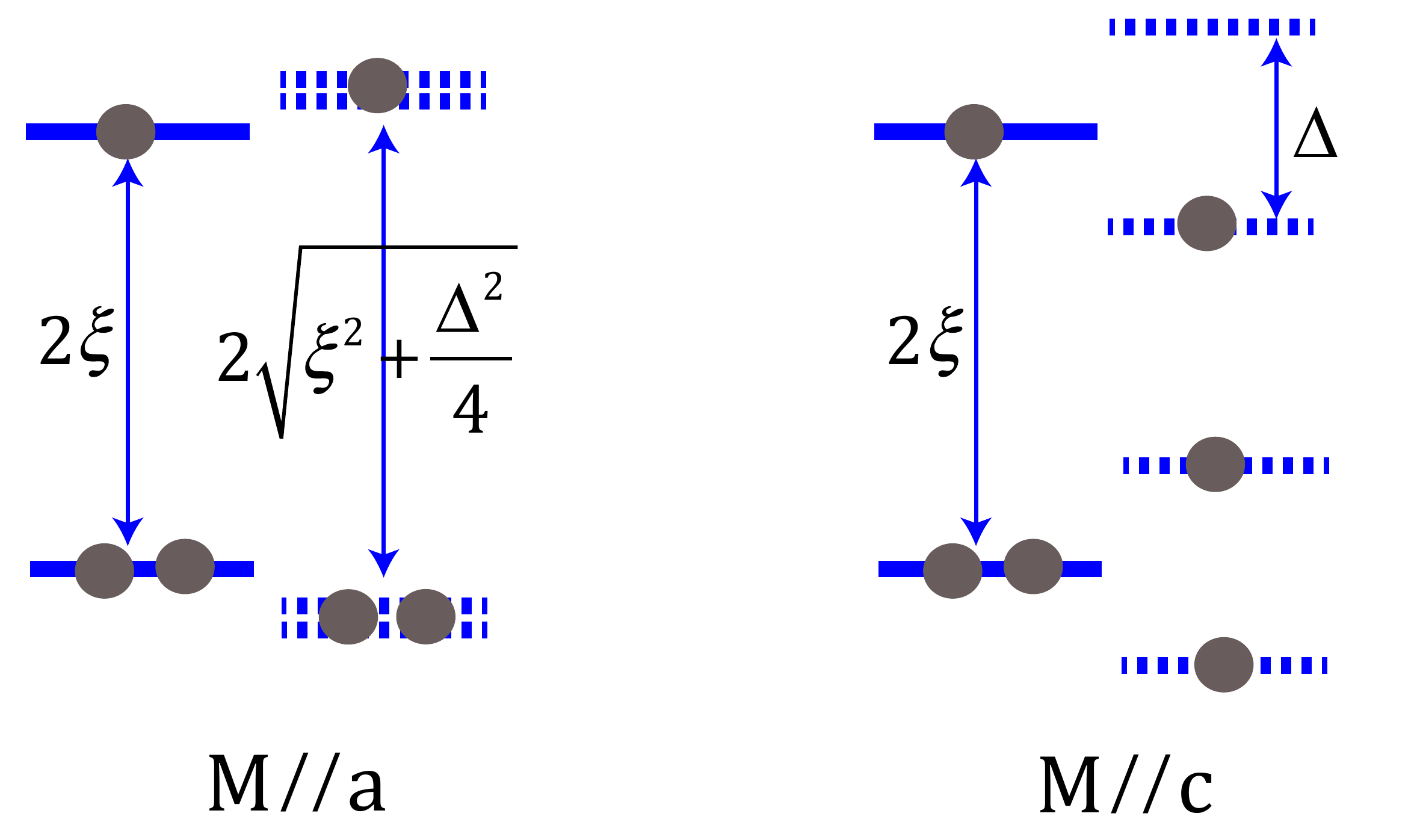}
\caption{Schematic of the $e'$ energy levels with spin orientation in-plane (${\bf M}\|a$) and out of plane (${\bf M}\|c$). The solid blue lines are spin degenerate.
 }
\label{figT}	
\end{figure}

Diagonalizing $H$ results in the energy level scheme sketched in \cref{figT}. The magnetic anisotropy energy is $E_{\rm MAE} = E_a-E_c $, where $E_a$ and $E_c$ are the total energies when ${\bf M} \! \parallel \! a$ and ${\bf M} \! \parallel \! c$, respectively. Using the force theorem, the energy change on including SOC is given by the change in the sum of single-particle eigenvalues. The reference energy (without SOC) cancels when the difference is taken for the two magnetization directions and, for occupancy with a single hole, $E_{\rm MAE}$ can be estimated to be
\begin{eqnarray}
\label{eqn:MAE}
E_{\rm MAE} && = E_a -E_c =   \xi + \frac{\Delta}{2} - \sqrt{\xi^2+ \frac{\Delta^2}{4}}
\end{eqnarray}
where we make use of the fact that the sum over all single particle eigenvalues is zero to express the sum over occupied states in terms of the sum over unoccupied states that is simply the energy of the hole. In this simple model, it is clear that for single acceptors the energy is lower when the magnetization is out of plane. In the limit that $\Delta \ll \xi$, $E_{\rm MAE} \sim \frac{\Delta}{2}(1-\frac{\Delta}{4\xi}) $.

\begin{figure}[t]
 \centering
\includegraphics[scale = 0.36]{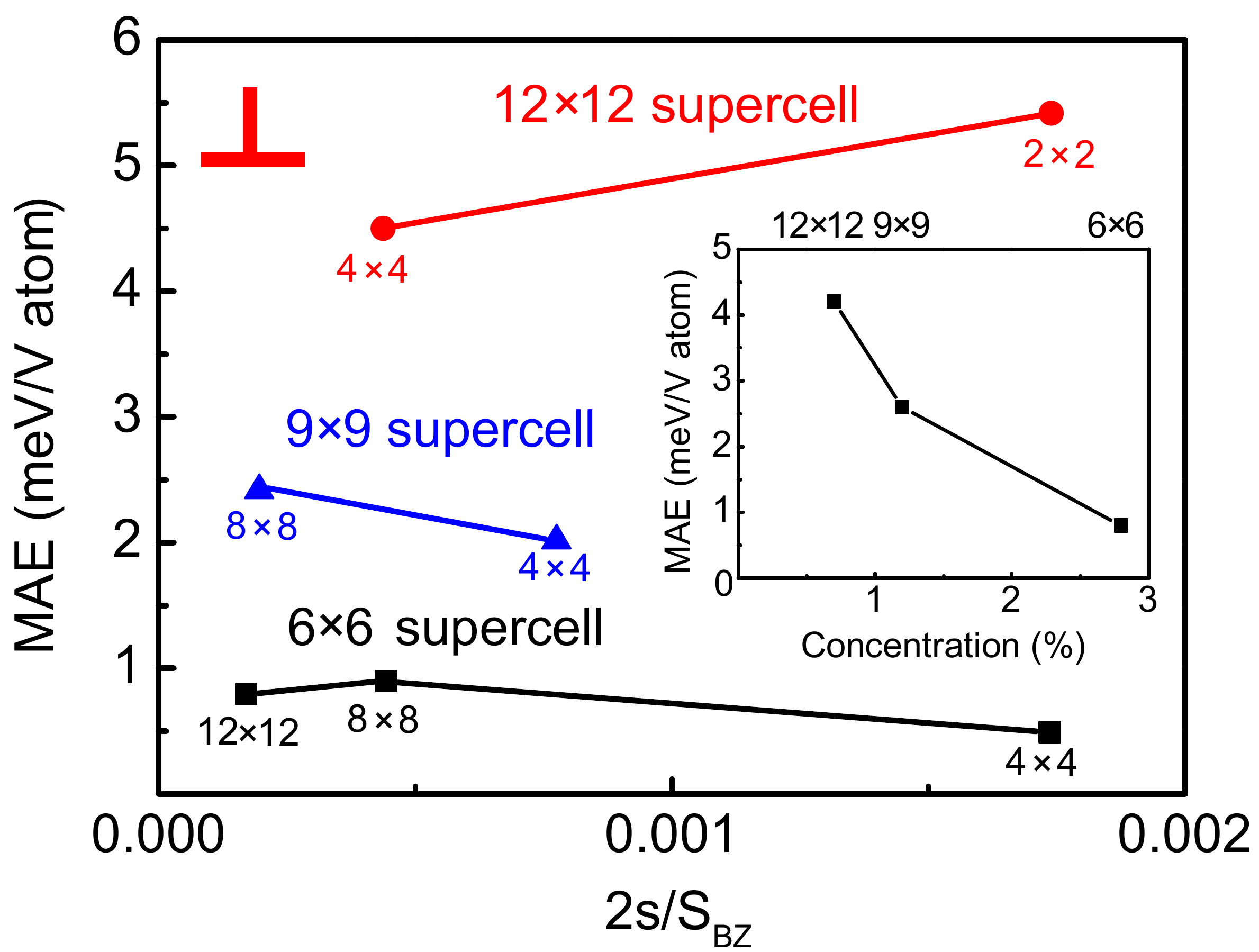}
\caption{\label{fig:MAE}(color online).
Convergence of the anisotropy energy of a relaxed substitutional V$_{\rm Mo}$ atom in a monolayer of MoS$_2$ for 6$\times$6 (black), 9$\times$9 (blue) and 12$\times$12 (red) supercells as a function of the area, $s$, of the triangular surface element used to perform the two-dimensional BZ integral, given as a fraction of the total area of the 2D BZ, $S_{\rm BZ}$, for a 1$\times$1 unit cell. The number of divisions of the reciprocal lattice vectors corresponding to each surface element is indicated for each supercell.
}
\label{figU}	
\end{figure}

To determine the  MAE using the {\sc vasp} code \footnote{The {\sc vasp} guide, https://cms.mpi.univie.ac.at/vasp/vasp.pdf}, we adopt a two step procedure. We first perform a well-converged self-consistent spin-polarized calculation for the minimum energy geometry without SOC. The output from that calculation is used as input to the second step where SOC is added and the Kohn-Sham equation is solved non self-consistently yielding a new eigenvalue spectrum, Fermi energy and wavefunctions from which a total energy can be determined; to use the force theorem, we will just make use of the eigenvalue-sum part of the total-energy output. When adding the SOC, an orientation for the exchange field (magnetization direction) needs to be chosen and this will yield an orientation dependent eigenvalue spectrum, Fermi energy etc. To determine the MAE, we need to perform two calculations with the magnetization chosen (i) perpendicular to the plane and (ii) in plane. The MAE will be expressed as the difference. To calculate the single particle eigenvalue sum for the electronic structure shown in \cref{figS} requires a careful BZ summation \cite{Daalderop:prb90a, Daalderop:prl92, Daalderop:prb94} for which we use the improved tetrahedron method \cite{Blochl:prb94a}. The results obtained using the force theorem for a 6$\times$6 supercell and 4, 8 and 12 divisions of the reciprocal lattice vectors are shown in \cref{figU} as a function of the BZ area element (2$s$) normalized to the area, $S_{\rm BZ}$, of the BZ for a 1$\times$1 primitive unit cell (black squares). An integral is defined as the limit where $s \rightarrow 0$ for an infinite number of sampling k points and from the figure we expect a value of $\sim 0.8 \pm 0.2\,$meV.

When the supercell size is increased and the dispersion of the $e'$ and $a_1'$ states becomes smaller, we might expect the BZ summation to converge faster but the situation is complicated by the near-degeneracy of the $e'$ and $a'_1$ bands. \cref{figU} includes results for 9$\times$9 and 12$\times$12 supercells indicating a strong increase in the size of the $E_{\rm MAE}$ in the single impurity limit. The strong dependence of the MAE on the supercell size can be understood in terms of the reduced dispersion of the impurity levels and the contribution to the MAE from  states near the Fermi level whose degeneracy is lifted when the magnetization direction is rotated from in-plane with ${\bf M} \! \parallel \! a$ to out-of-plane with ${\bf M} \! \parallel \! c$ as illustrated by \cref{figS}(d,e) and \cref{figT}. For a 12$\times$12 supercell, we can extrapolate the results obtained using a 2$\times$2 and 4$\times$4 k-point sampling to estimate a converged MAE of 4.2 meV. The exchange-splitting $\Delta$ is 15.2 meV for the $e'$ level and using this value of $\Delta$ and $2\xi=130 \,$meV in \eqref{eqn:MAE} yields a value of $E_{\rm MAE} \sim 7.2\,$meV that is still larger than the 4.2~meV estimate from the full calculation. We can extrapolate the results for the three sizes of supercell to $s = 0$ and then plot the results as a function of the inverse supercell size (inset) to estimate the SIA in the infinite supercell limit to be $4.5\pm 0.5\,$meV per V ion. This is much larger than the value reported for 2D CrI$_3$ \cite{Xu:npjcm18} that exhibits Ising behavior \cite{Huang:nat17}. The  dipole-dipole interactions that play an important role in determining whether or not the magnetization of thin magnetic layers and magnetic multilayers is in-plane or out-of-plane are orders of magnitude smaller in the present case and can be safely neglected. \cite{Daalderop:prb90a, Daalderop:prl92, Daalderop:prb94}

\subsection{Monte Carlo calculations}
\label{ssec:MC}

Very strong single-ion anisotropy combined with isotropic Heisenberg exchange results in Ising-like behaviour \cite{Leonel:jmmm06} which automatically gives a magnetically ordered phase at finite temperature \cite{Onsager:pr44, Yang:pr52}. We map the energy differences calculated between FM and AFM oriented spins onto an isotropic Heisenberg exchange interaction and then model the V-doped MoS$_2$ monolayer as an Ising spin system for which all odd moments disappear in zero field by symmetry. Monte Carlo calculations are used to determine the Curie temperature $T_C$ using Binder's cumulant method \cite{Binder:zfpb81, Landau:09} where the fourth order cumulant of the magnetization $\bf M$ simplifies to $U_4(T,L) = 1 - \langle M^4 \rangle / 3 \langle M^2 \rangle^2 $. As the system size $L \rightarrow \infty$, $U_4 \rightarrow 0$ for $T > T_C$ and $U_4  \rightarrow 2/3$ for $T < T_C$. For large enough lattice size, $U_4(T,L)$ curves for different values of $L$ cross as a function of temperature at a ``fixed point'' value $U^*$ and the location of the crossing fixed point is the critical point \cite{Landau:09}.

\begin{figure}[t]
\includegraphics[scale = 0.34]{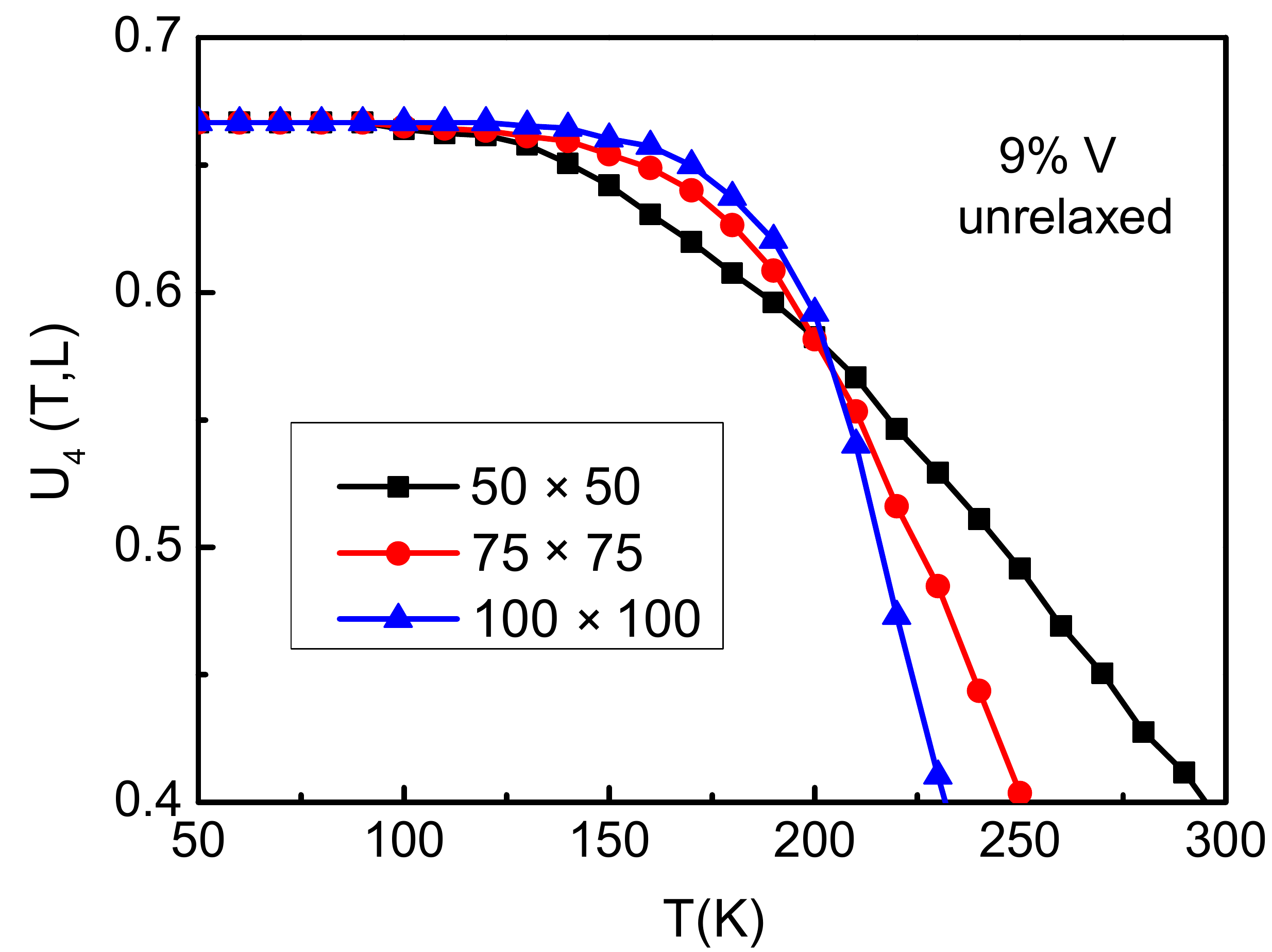}
\caption{Variation of the fourth order cumulant for three different supercell sizes $L= 50,\,75,\,100$ as a function of temperature.
}
\label{figV}
\end{figure}

\begin{figure}[t]
\includegraphics[scale = 0.35]{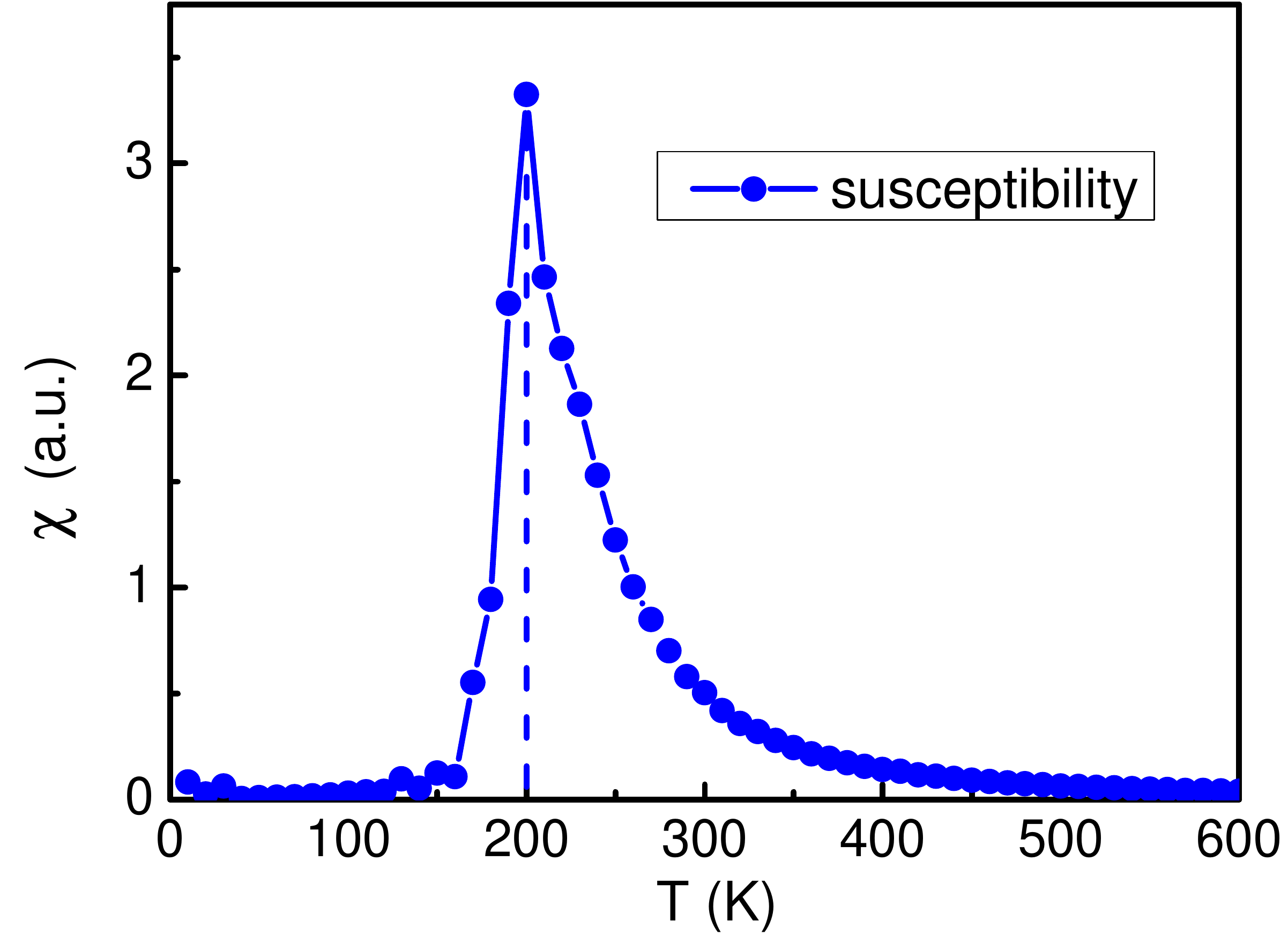}
\caption{Variation of the magnetic susceptibility as a function of temperature for 9$\%$ unrelaxed V dopant concentration with $L=100$.
}
\label{figW}
\end{figure}

At a given temperature and doping concentration, we establish thermodynamic equilibrium in $10^5$ Monte Carlo (MC) thermalization steps and then average over 48 different random dopant configurations to obtain $\langle M^2 \rangle$ and $\langle M^4 \rangle$. Three different lattice sizes with $L= 50,\,75,\,100$ are used to calculate $U_4$ as a function of the temperature $T$ with doping concentrations from 1\% to 11\%. An example of the results is shown in \cref{figV} where the fitting curve for the unrelaxed case in \cref{figL} is used to describe the exchange interactions for a doping concentration of 9\%. The temperature corresponding to the size independent universal fixed point $U^*$ where the $U_4(L,T)$ curves for different lattice sizes $L$ intersect yields an estimate for $T_C$. For the largest supercell size $L=100$, we calculated the magnetic susceptibility $\chi = [\langle M^2 \rangle-\langle |M| \rangle^2]/Nk_{b}T $ which diverges at the critical temperature in the thermodynamic limit \cite{Landau:09}. An example is shown in \cref{figW} in which the Curie temperature obtained from the position of the magnetic susceptibility peak is in good agreement with that obtained from the fourth order cumulant.

\subsection{Curie temperature}
\label{ssec:CT}

\begin{figure}[b]
\includegraphics[scale = 0.35]{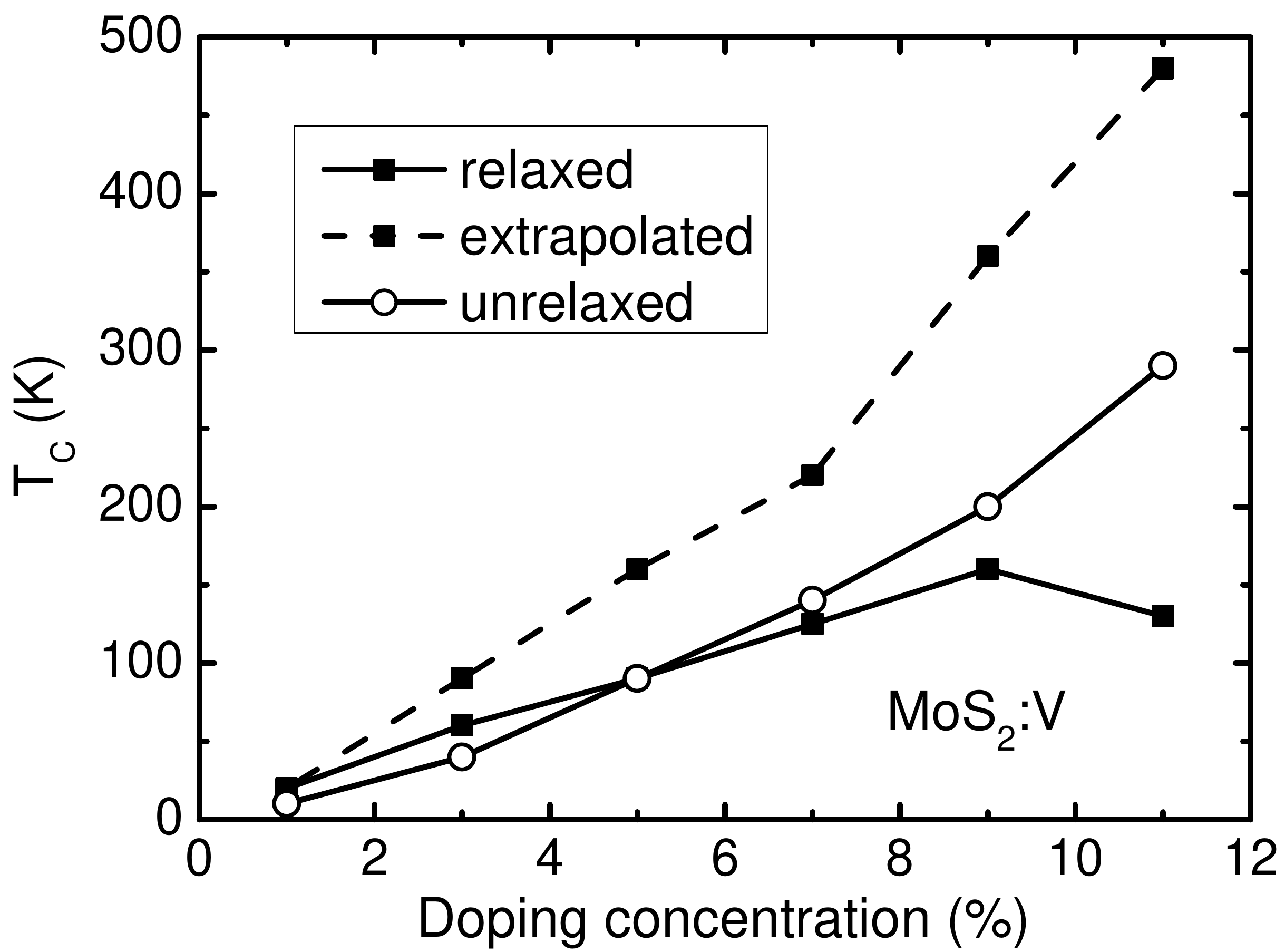}
\caption{Variation of the Curie temperature as a function of the doping concentration calculated using Binder's cumulant method and the exchange interactions shown in \cref{figL} for an MoS$_2$ monolayer doped with V. The dashed curve was calculated by extrapolating the exchange interaction for relaxed dopant pairs to separations shorter than the critical separations where quenching occurs.
}
\label{figX}
\end{figure}

\begin{table*}[t]
\caption{Computational studies of single acceptor dopants in MoS$_2$.
SC: Supercell.
XC: Exchange-Correlation functional.
MAE: Magnetic Anisotropy Energy.
US-PP: Ultrasoft pseudopotential \cite{Vanderbilt:prb90}.
NC-PP: Norm-conserving pseudopotential \cite{Troullier:prb91}.
PAW: Projector Augmented-Wave \cite{Blochl:prb94b}.
LDA: Local Density Approximation \cite{Perdew:prb81}.
CA:  Ceperley-Alder \cite{Ceperley:prl80}.
GGA: Generalized Gradient Approximation.
PBE: Perdew-Burke-Ernzerhof \cite{Perdew:prl96}.
HSE: Heyd-Scuseria-Ernzerhof \cite{Heyd:jcp03,*Heyd:jcp06}.
QE: {\sc quantum espresso} \cite{Giannozzi:jpcm09}.
{\sc vasp}: Vienna Ab-initio Simulation Package \cite{Kresse:prb96, Kresse:prb99}.
{\sc wien2k} \cite{Blaha:cpc02}.
{\sc siesta} \cite{Soler:jpcm02}.
}
\label{tabG}
\begin{ruledtabular}
\begin{tabular}{cclllcclccl}
  \multicolumn{2}{c}{SC size (\# atoms)}
& \multicolumn{1}{c}{}
& \multicolumn{1}{c}{}
& \multicolumn{1}{c}{}
& \multicolumn{1}{c}{}
& \multicolumn{1}{c}{Vac.}
& \multicolumn{1}{c}{}
& \multicolumn{1}{c}{}
& \multicolumn{1}{c}{Exchange}
& \multicolumn{1}{c}{}   \\
\cline{1-2}
  \multicolumn{1}{c}{Default}
& \multicolumn{1}{c}{Max/test}
& \multicolumn{1}{c}{Dopant}
& \multicolumn{1}{c}{Method}
& \multicolumn{1}{c}{XC}
& \multicolumn{1}{c}{U(eV)}
& \multicolumn{1}{c}{(\AA)}
& \multicolumn{1}{l}{Code}
& \multicolumn{1}{c}{MAE}
& \multicolumn{1}{c}{Int. $J(d)$}
& Reference \\
\hline
$4 \times 4$   &                & V, Nb, Ta & US-PP & GGA/PBE &  0  & 10 & QE     & No  & No  & Cheng PRB13 \cite{Cheng:prb13} \\
$4 \times 4$   &                & V         & PAW   & GGA/PBE &  0  & 12 & VASP   & No  & No  &   Yue PLA13 \cite{Yue:pla13} \\
$5 \times 5$   & $6 \times 6$   & Nb        & NC-PP & LDA/CA  &  0  & 15 & SIESTA & No  & No  & Dolui PRB13 \cite{Dolui:prb13} \\
               &                & Nb        & PAW   & HSE     &  0  &    & VASP   & No  & No  &  \hspace{1em}  \textquotedbl 
                                                                                                 \hspace{1.5em} \textquotedbl \\
$5 \times 5$   &                & V         & PAW   & GGA/PBE &  0  & 15 & VASP   & No  & No  &   Yun PCCP14 \cite{Yun:pccp14} \\
$8 \times 8$   &                & V         & PAW   & GGA/PBE & 5.5 &    & VASP   & No  & No  & Andriotis PRB14 \cite{Andriotis:prb14} \\
$5 \times 5$   & $7 \times 7$   & V, Nb, Ta & US-PP & GGA/PBE &  0  & 12 & QE     & No  & No  &    Lu NRL14 \cite{Lu:nrl14} \\
$4 \times 4$   &                & V         & PAW   & GGA/PBE &  0  & 15 & VASP   & No  & No  &  Miao JMS16 \cite{Miao:jms16} \\
$5 \times 5$   &                & V         & PAW   & GGA/PBE &  3  & 15 & VASP   & No  & Yes &   Fan NRL16 \cite{Fan:nrl16} \\
$6 \times 6$   &                & Cr, V     & PAW   & GGA/PBE &  0  & 20 & VASP   & No  & No  & Robertson ACSN16 \cite{Robertson:acsn16} \\
$4 \times 4$   & $4 \times 4$   & V, Cr     & FLAPW & GGA     & 2.5 & 15 & WIEN2K & No  & No  & Singh AM17 \cite{Singh:am17} \\
               & $8 \times 5$   & V         & PAW   & GGA/PBE &  0  & 16 & VASP   & No  & Yes &  Miao ASS18 \cite{Miao:ass18} \\
$6 \times 6$   &                & Nb, Ta    & PAW   & HSE     &  0  & 17 & VASP   & No  & No  &  Choi PRAP \cite{Choi:prap18} \\              
$3 \times 3$   & $4 \times 4$   & V         & US-PP & GGA     &  3  & 20 & QE     & No  & Yes & Mekonnen IJMPB18 \cite{Mekonnen:ijmpb18} \\
\hline
$12 \times 12$ & $15 \times 15$ & V, Nb, Ta & PAW   & LDA(GGA)& ---     & 20 & VASP   & Yes & Yes & This work \\
\end{tabular}
\end{ruledtabular}
\end{table*}


The ordering temperatures we calculate are shown in \cref{figX} for V doping concentrations $x$ in the range from 1\% to 11\%. Without relaxation, $T_C(x)$ increases monotonically with doping concentration and reaches room temperature for a concentration of $\sim 11$\%. With relaxation, the ferromagnetic exchange interaction is quenched for close dopant pairs and $T_C(x)$ exhibits  a maximum of $\sim 165\,$K for 9\% doping. If we extrapolate the exchange interaction for relaxed dopant pairs to separations smaller than the critical separation where quenching occurs (dashed line in \cref{figL}), the Curie temperature increases rapidly and monotonically with doping concentration and exceeds room temperature for dopant concentrations larger than 9\%. Because the maximum value of $T_C(x)$ we obtain would be higher but for the quenching of the magnetic moments of closely separated relaxed dopants, it becomes important to consider how to suppress the quenching to obtain higher Curie temperatures. This will be discussed in \cref{sec:Discussion}.

\section{Comparison with other work}
\label{sec:Comp}

\cref{tabG} summarizes earlier computational work on doping MoS$_2$ monolayers with V, Nb or Ta. Because much of it was concerned with doping rather than with magnetic ordering, no attempts were made to calculate the magnetic anisotropy. Most of the calculations were done using small supercells and the separation dependence of the exchange interaction was not studied systematically, if at all. We noted in Sec.~\ref{sec:CD} that the GGA positions the $\Gamma$-point VBM too high with respect to the K-point VBM by comparison with experiment \cite{Jin:prl13} and we therefore used the LDA that yields better agreement with experiment in this regard. Because most of the calculations referred to in \cref{tabG} were performed with the GGA exchange-correlation potential, we examine the effect of using the GGA rather than the LDA.

\begin{figure}[b]
\includegraphics[scale=0.35]{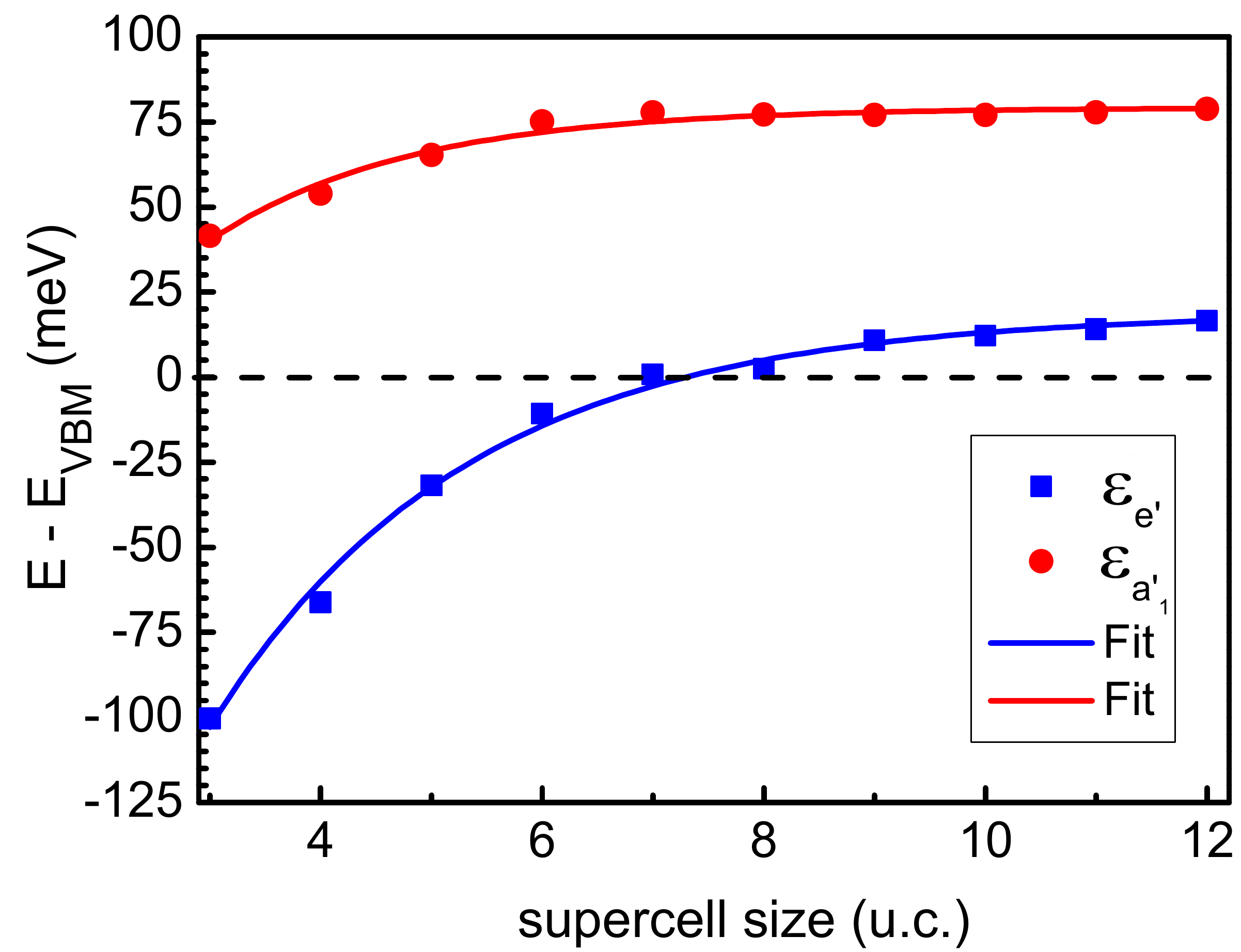}
\caption{Dependence on the supercell size $N$ of the $a'_1$ and $e'$ impurity levels induced by an unrelaxed substitutional vanadium atom V$_{\rm Mo}$ with respect to the valence band maximum in the GGA.
}
\label{figY}
\end{figure}

\subsubsection{GGA versus LDA}
\label{sssec:GGA}

If we use atomic configurations whose geometry was optimized using the LDA and repeat the electronic structure calculations using the GGA, we qualitatively reproduce the LDA results for the exchange interaction between V dopant atoms. The significant differences that will be documented below therefore have their origin in the only slightly different minimum-energy geometries predicted by the GGA.

Our starting point is an MoS$_2$ monolayer whose lattice constant $a_{\rm GGA}=3.185\,$\AA\ minimizes the GGA total energy, \cref{tab:A}. The GGA overestimate of the position of the $\Gamma$-point VBM with respect to the K/K$'$ VBM makes the $a'_1$ level much higher than the $e'$ levels as shown in \cref{figY} where these levels are plotted as a function of supercell size. The asymptotic values are $\sim \,$76 and $\sim 20 \,$meV for the $a'_1$ and $e'$ states, respectively compared to $\sim 62 \,$meV for both in the LDA case, \cref{figG}. Compared to the LDA estimate of a Bohr radius of 4.2\AA\ from the radial extent of the partial charge density in \cref{figE}, with the GGA we find a slightly larger value of 4.7\AA\ for the $a'_1$ state. By analogy with a hydrogen atom where the extent of the 1$s$ orbital increases greatly for the H$^-$ ion compared to the neutral atom, we attribute the slightly larger radial extent for the more strongly bound hole in the GGA case to its greater $a'_1$ (hole) occupancy. In the GGA, the doped system is fully spin-polarized for the smallest 3$\times$3 supercell we considered and the $a'_1$ exchange splitting of 142~meV is much larger than the 91~meV we found in the unrelaxed LDA case. (The LDA predicts an exchange splitting of $\Delta \varepsilon_{1s} = \varepsilon_{1s}^\uparrow -\varepsilon_{1s}^\downarrow = 0.35 \,$Rydberg for a hydrogen atom compared to 0.55~Rydberg for the GGA; see
\cref{sec:HA}).

\begin{figure}[t]
\includegraphics[scale=0.35]{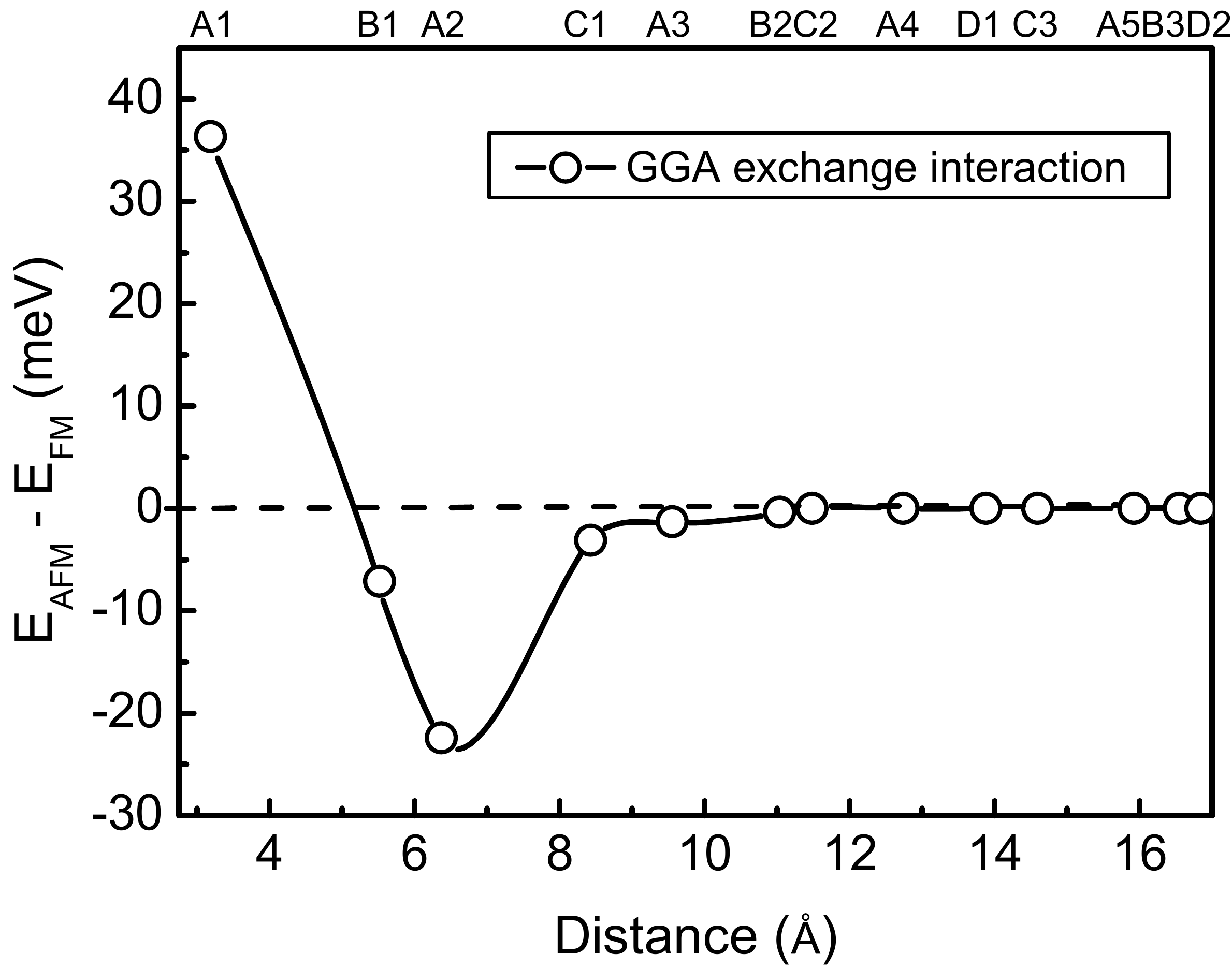}
\caption{The total energy difference between parallel and antiparallel aligned spins on the V dopants without relaxation within GGA.
}
\label{figZ}
\end{figure}

\cref{figZ} shows the exchange interaction calculated with the GGA for pairs of unrelaxed V$_{\rm Mo}$ dopants as a function of the distance between them. We observe an oscillatory behavior with FM coupling for neighbouring pairs (``A1'' configuration, see \cref{figC}) that switches to AFM for B1, A2, C1 and A3 configurations after which it is essentially zero reflecting the small effective Bohr radius of the $a'_1$ level ($a_0^* \sim 4.7\,$\AA). The relatively large separation in energy of the $a'_1$ and $e'$ levels means that for separations larger than $\sim 9$\AA, both holes are to be found in the anti-bonding $a^*$ level because the bonding-antibonding interaction is too weak to lead to hole occupancy of the $e^*$ level. Because both holes must of necessity occupy the $a^*$ level, their spins must be opposite and FM ordering is energetically unfavourable (though the energy difference is very small).

\begin{figure}[t]
\includegraphics[scale=0.31]{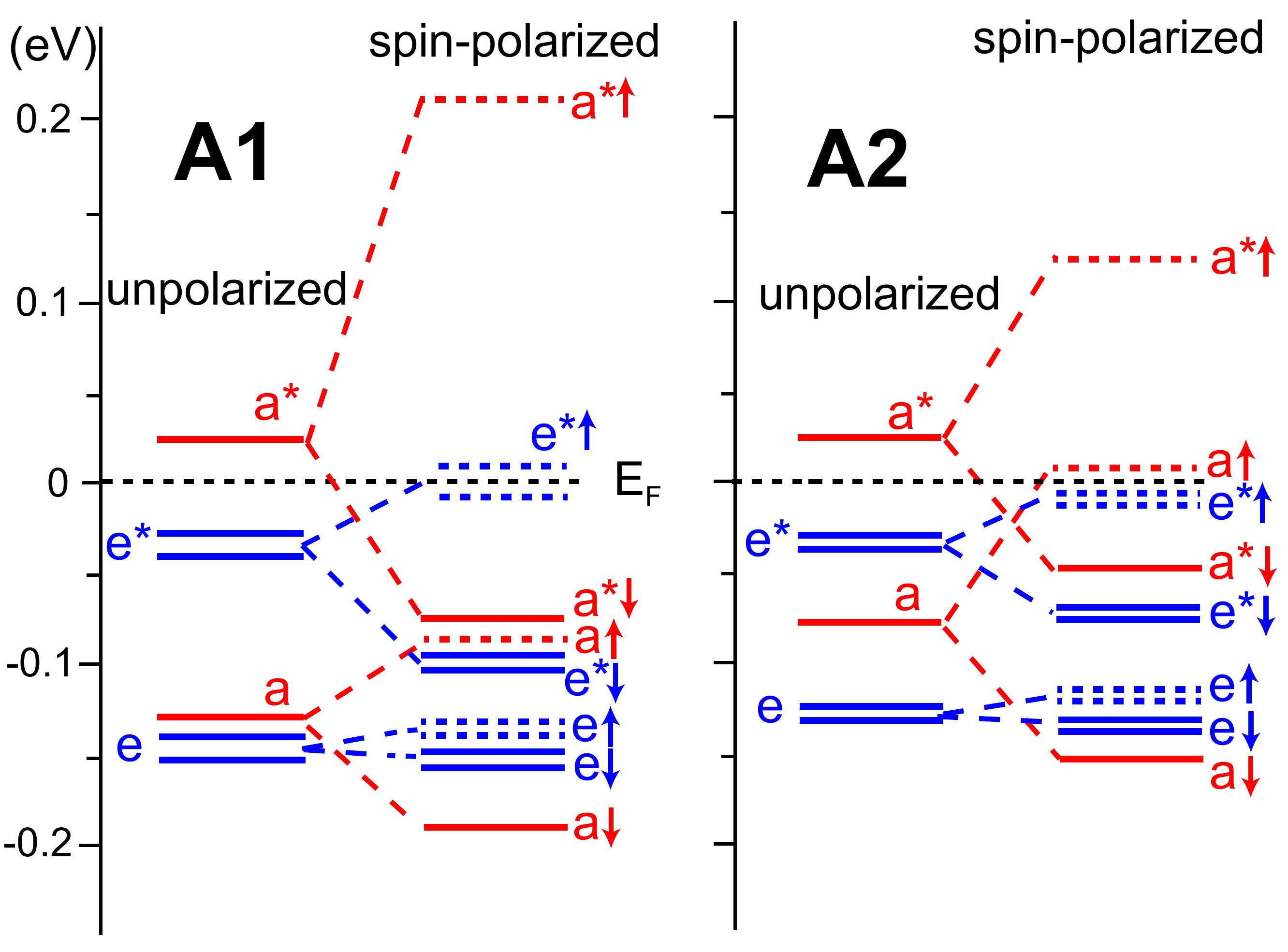}
\caption{Energy diagrams for unrelaxed dopant pairs in  A1 and A2 configurations (\cref{figC}) with and without spin polarization calculated in GGA. The levels are calculated from the appropriate weighted average of the $\Gamma$, K/K$'$ and M eigenvalues as in figures \ref{figM} and \ref{figP}. The Fermi level is indicated as a black dashed horizontal line. The spin up and spin down states are represented as dashed and solid lines, respectively.
}
\label{figgA}
\end{figure}

To understand the oscillation for small separations, we consider the electronic structures of the $a'_1$ and $e'$ derived $a$--$a^*$ and $e$--$e^*$ bonding and antibonding states for the A1 and A2 configurations in \cref{figgA} with and without spin polarization.
We begin with the A1 configuration. Compared to the LDA case that was shown in \cref{figM}(b), we see that the unpolarized $a$ and $a^*$ levels (lhs of \cref{figgA}) are higher than the $e$--$e^*$ levels reflecting the corresponding feature for a single substitutional dopant and that the $a$--$a^*$ bonding-antibonding interaction is larger because the Bohr radius of the $a'_1$ states is larger in the GGA.
On the rhs of the A1 panel, we see that the exchange splitting of the $a^*$ levels is approximately doubled to $\sim 300\,$meV for the A1 configuration compared to the single V$_{\rm Mo}$ case because of the overlapping hole densities; the same doubling occurs in the LSDA, where the exchange splitting of $92\,$meV for a single substitutional V is enhanced to $180\,$meV for neighbouring V pairs, see \cref{figO}. The exchange splitting is so large that the down-spin $a^*$ level moves below the (approximately doubly) degenerate up-spin $e^*$ level with the result that the hole can flip its spin and the Fermi level is pinned in the half-filled up-spin $e^*$ state. By having one hole in an $a^*$ state and the other in an $e^*$ state it is possible for their spins to be parallel and to simultaneously gain spin-polarization and bonding energy.

As the separation between the V$_{\rm Mo}$ atoms is increased, the exchange splitting and bonding interaction decrease rapidly for the $a$ states and a FM triplet state is only formed at the expense of having one hole occupying a bonding $a$ state (rhs of A2 panel); spin-polarization energy gain is offset by loss of bonding energy. Antiparallel alignment of the holes allowing gain of both spin-polarization and bonding energy becomes more favourable.

The effect of relaxation is to push the $a'_1$ level up in energy and increase the $a$--$a^*$ bonding-antibonding splitting so that the unpolarized electronic structure resembles that of \cref{figM}(e) with both holes in the $a^*$ level. For dopant separations less than $\sim 10$ \AA\ this leads to quenching of the magnetism. For separations larger than this, the $a'_1$ levels ($a_0^* \sim 4.7\,$\AA) interact only weakly with each other leading to very small energy differences $E_{\rm AFM}-E_{\rm FM}$ and negligible exchange interaction because the $e'$ related levels are too low in energy to be occupied.

As anticipated at the beginning of this section, we can trace the large difference between the GGA and LDA descriptions of the exchange interaction between pairs of V$_{\rm Mo}$ dopants to the 2\% difference between the LDA and GGA lattice constants. If we use the experimental lattice parameters listed in \cref{tab:A}, the differences between LDA and GGA disappear and we find that near-degenerate $e'$ and $a'_1$ impurity levels with binding energies of 56 meV in a direct band gap of 1.78 eV couple ferromagnetically for all separations. The different orbital character of the $e'$ and $a'_1$ levels means that they are very sensitive to the in-plane lattice constant $a$ and out-of-plane $d_{\rm SS}$, respectively, whose ratio determines their relative positions. From \cref{tab:A}, we see that the LDA ratio of 1.001 is much closer to the experimental value of 0.996 than the GGA ratio of 1.018 and argue that the LDA provides a more reasonable description of the relative position of the impurity levels.

When SOC is taken into consideration, the energy difference $\Delta_{\rm K\Gamma}$ between the K/K$'$ and $\Gamma$-point valence band maxima increases from 150 to 216 meV for LDA and from 12 to 88 meV for GGA, respectively, because of the large spin-orbit splitting at the K/K$'$ point, \cref{tabH}. The experimental value, 140~meV, is just in between making it unclear what will actually happen. To resolve this issue, experiment should focus on determining the position of the $a'_1$ level with respect to the top of the valence band in the single impurity limit.

Because the relative position of the impurity states determines the exchange interaction between dopant atoms, we have performed exploratory calculations to tune the relative positions of the K/K$'$ and $\Gamma$ valence band maxima and consequently of the $a'_1$ and $e'$ levels with strain. For GGA, a 1\% compressive strain ($\Delta a < 0$) is found to lower the $a'_1$ levels to be degenerate with the $e'$ levels and we find FM coupling for all separations.  2\% tensile strain ($\Delta a > 0$) lifts the $a'_1$ levels far above the $e'$ levels stabilizing the magnetic moment of single impurities but favouring singlet formation of impurity pairs and AFM coupling. So while tensile strain reduces the formation energies of Nb$_{\rm Mo}$ and Ta$_{\rm Mo}$ and facilitates $p$ doping of MoS$_2$ \cite{Choi:prap18}, it is detrimental for ferromagnetic ordering. 

\begin{table}[t]
\caption{Effect of SOC on the band gap $\Delta\varepsilon_g$ and valence band alignment for an MoS$_2$ monolayer in the LDA and GGA. $\Delta_{\rm K\Gamma}=\varepsilon_{\rm K}-\varepsilon_{\Gamma}$ is the energy difference between the valence band maximum (VBM) at the K and $\Gamma$ points.
}
\label{tabH}
\begin{ruledtabular}
\begin{tabular}{lllll}
    & \multicolumn{2}{c}{No SOC}     & \multicolumn{2}{c}{Including SOC} \\
\cline{2-3}        \cline{4-5}
    & $\Delta\varepsilon_g$(eV)
                & $\Delta_{\rm K\Gamma}$
                           & $\Delta\varepsilon_g$(eV)
                                        & $\Delta_{\rm K\Gamma}$  \\
\hline
GGA &   1.650   &  0.012   & 1.586      &  0.088  \\
LDA &   1.860   &  0.150   & 1.787      &  0.216  \\
Exp &           &          & 1.900$^a$  &  0.140$^b$  \\
  \end{tabular}
  \end{ruledtabular}
$^a$Ref.\onlinecite{Mak:prl10}
$^b$Ref.\onlinecite{Jin:prl13}
\end{table}

If the exchange interactions are so sensitive to the lattice constant and the ratio of $a$ to $d_{\rm SS}$, it might be useful to consider tuning this ratio by modifying $\Delta_{\rm K\Gamma}$ either using strain or by alloying, Mo(S/Se/Te)$_2$. In the  MoSe$_2$, MoTe$_2$ and WSe$_2$ systems, the $a'_1$ level lies so much lower than the $e'$ level that using GGA or LDA with their different lattice parameters has little effect; the coupling is dominated by the long range of the $e'$ levels.

\subsubsection{LDA + U}
\label{sssec:LDAU}

Two of the studies cited in \cref{tabG} use a finite value of $U$ to better describe onsite Coulomb repulsion between electrons in localized $d$ orbitals \cite{Andriotis:prb14, Singh:am17}. We find that LDA+U \cite{Dudarev:prb98} with modest values of $U$ makes the local magnetic moment more localized and enhances it.  The (more localized) $a'_1$ level is more sensitive to $U$ than the $e'$ level.

A small value of $U$ (less than 1 eV) increases the exchange splitting of the $a'_1$ level and increases the FM exchange interaction which would yield a larger $T_{\rm C}$ compared to calculations without $U$. A larger value of $U$ (larger than 3 eV) causes the hole to become even more localized and fully polarized even at very high concentrations (25\%). As a consequence, the exchange interaction decays more rapidly and the separation below which quenching occurs decreases because the bonding interaction decays more rapidly. Compared to LDA calculations, the Curie temperature would be lower at low dopant concentration but enhanced at high concentration.

In our LDA calculations, we find that the shallow vanadium 3$d$ orbitals hybridize strongly with S-3$p$ and Mo-4$d$ orbitals delocalizing the holes. We expect the Coulomb $U$ in our system to be small and with a small $U$, the Curie temperature should be enhanced. Our LDA results should thus represent a lower bound on the exchange interaction and ordering temperature.

\section{Discussion}
\label{sec:Discussion}

According to the Zener $p$--$d$ model used to interpret magnetic coupling in Ga(Mn)As dilute magnetic semiconductors \cite{Dietl:natm10}, holes in As ($p$) bonding states mediate the exchange interaction between strong local ($d^5$) magnetic moments on Mn$^{2+}$ dopant ions. In the present case, the magnetic moments that we find come from unpaired ($d^1$) spins in gap states that are only weakly bound by the Coulomb potential of the dopant ions, \cref{figB}. The large Bohr radii we find for these states, \cref{tabB}, allows them to overlap to form narrow bands and suggests that on-site Coulomb interactions may play a minor role and a model of itinerant ferromagnetism may be more appropriate than the various localized models used to study the Ga(Mn)As and related systems \cite{Jungwirth:rmp06, Sato:rmp10}. When the holes are in orbitally nondegenerate $a'_1$ levels, AFM coupling is favoured to satisfy Pauli's exclusion principle; when they are in the degenerate $e'$ levels, FM coupling is preferred to minimize the Coulomb interaction. For MoS$_2$, the (accidental) near-degeneracy of the $a'_1$ and $e'$ hole states leads to them competing to determine the magnetic properties whereby the strength of the exchange interaction is related to the exchange splitting of the impurity band and will be affected by the band dispersion for high doping concentrations.

In the low doping limit, the impurity states have no dispersion and are fully polarized. As the impurity concentration is increased, the impurity levels overlap to form narrow bands that broaden and eventually overlap the narrow Mo band that forms the top of the valence band. As seen in \cref{figH}, the impurity bandwidth increases exponentially with increasing doping concentration.  For the 9\% V dopant concentration for which we find $T_{\rm C}$ to be a maximum, the $e'$ impurity bandwidth is $\sim 400\,$meV. The $a'_1$ band is narrower, only about a third as wide. Both bandwidths exceed the 91~meV exchange splitting we find for single V impurities in \cref{figJ} that would imply partial quenching of the magnetic moments. For the ordered V dopants studied in \cref{tabD}, this quenching occurs as the concentration is increased above 3\% and is complete by 6\%. For itinerant electrons occupying narrow bands, it has been argued that the effective interaction predicted by the Stoner criterion will not be reduced by correlation effects or spin wave excitations \cite{Edwards:jpcm06}. In contrast to traditional dilute magnetic semiconductors with large local moments that do not contribute to the spin stiffness, the completely spin polarized carriers in narrow impurity bands lead to a large spin stiffness and develop ferromagnetic ordering by their mutual interaction.

\begin{figure}[t]
\includegraphics[scale=0.30]{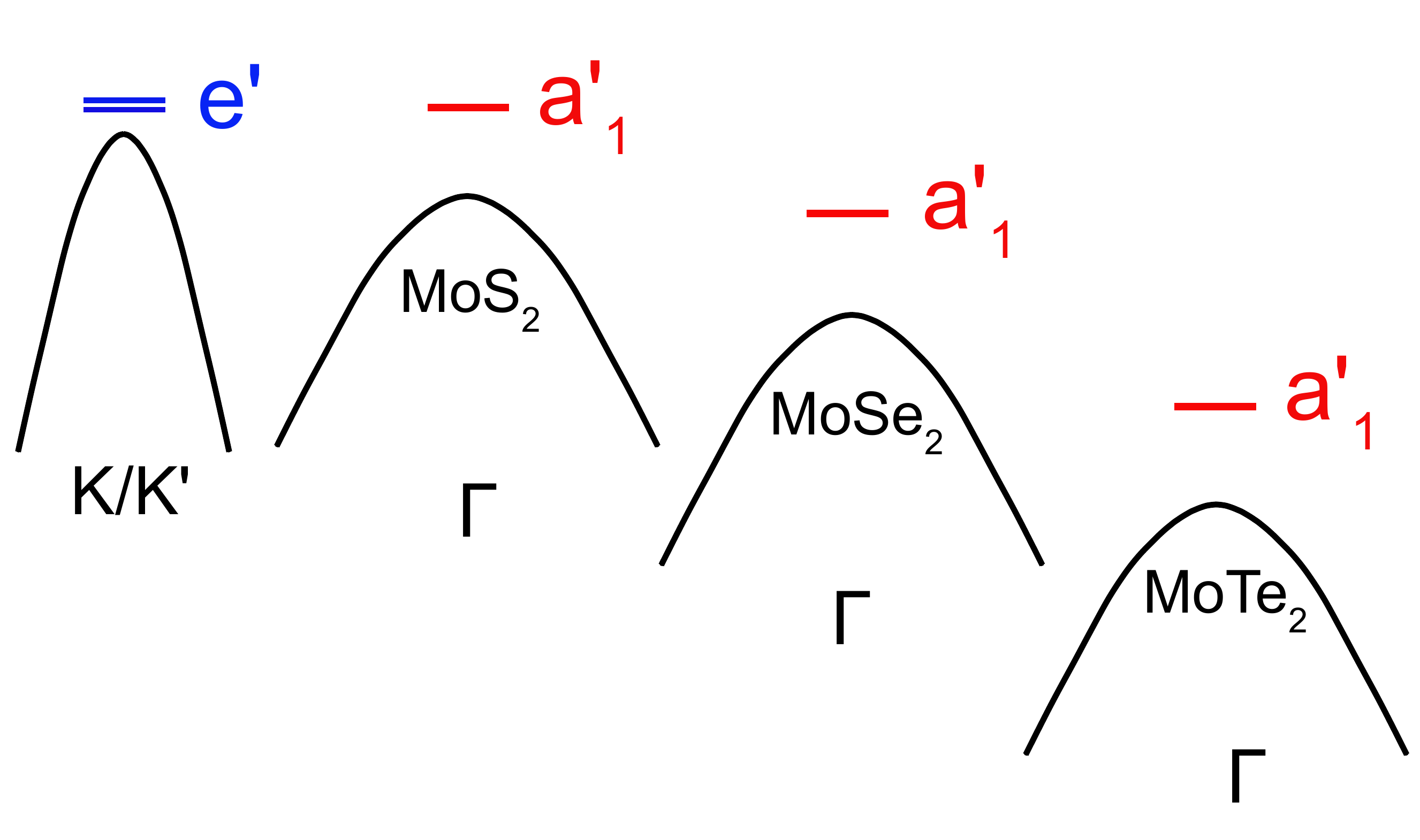}
\caption{Schematic of the relative position of VBM at K and Gamma points and corresponding impurity levels for MX2.
}
\label{figgB}
\end{figure}

The quenching of ferromagnetic pairing for close impurity pairs can be avoided by considering instead of MoS$_2$ as host semiconductor, MoSe$_2$ or MoTe$_2$ (WSe$_2$ or WTe$_2$) for which the $\Gamma$ point VBM drops with respect to the K/K$'$ VBM as S$\rightarrow$Se$\rightarrow$Te (sketched in \cref{figgB}). Preliminary calculations show that the $a'_1$ defect levels do indeed follow the $\Gamma$ point VBM leaving the holes in the orbitally degenerate $e'$ derived impurity bands. The increased lattice constant makes the $e'$ states more localized and enhances the spin polarization in MoSe$_2$ and MoTe$_2$. Very recently there have been reports of long-range and/or room temperature ferromagnetism occurring in V doped WSe$_2$ monolayers \cite{Yun:arXiv18}, in MoSe$_2$ and MoTe$_2$ \cite{Guguchia:sca18}, in V and Ta doped MoTe$_2$ \cite{Coelho:aem19, Yang:aem19} and  in MoS$_2$ \cite{Hu:acsami19} whereby the interaction with anion vacancies would appear to play an important role. These systems clearly warrant closer study.

We might expect double acceptors to have larger magnetic moments and exchange interactions. However, when MoS$_2$ is doped with Ti, Zr, Hf on the Mo site, the $a'_1$ level is lifted far above the $e'$ levels and accommodates both holes so such substitutional impurities are  nonmagnetic in the single impurity limit. Only when dopant pairs are sufficiently close ($\sim6$ \AA ) does strong $\pi$ bonding lift the $e^*$ antibonding level above the $a$ bonding level so all four holes occupy antibonding states. The two holes in the $e^*$ states can become ferromagnetic with a total magnetic moment of 2 $\mu_B$ but this does not represent an improvement on the single acceptor case. 

In MoSe$_2$ or MoTe$_2$ (WSe$_2$ or WTe$_2$) monolayers where the $\Gamma$ point VBM lies well below the K/K$'$ VBM, the two holes introduced by double acceptors occupy $e'$ states that can acccomodate four holes. Half-filling of these degenerate levels leads to a competition between Jahn-Teller distortion and exchange splitting. If the Jahn-Teller distortion is sufficiently strong, the magnetic moment will be quenched and we do not expect double acceptors to be magnetic for low doping concentrations. 

\section{Summary \& Conclusions}
\label{sec:Conclusions}

We have used {\em ab initio} calculations to explore the possibility of inducing ferromagnetism in an MoS$_2$ monolayer by substitutionally doping it with V, Nb or Ta on Mo sites. In the single impurity limit, the resulting repulsive Coulomb potential leads to a doubly degenerate $\{d_{xy}, d_{x^2-y^2}\}$ state with $e'$ symmetry bound to the K/K$'$ valence band maxima and a singly degenerate $a'_1$ state with $d_{3z^2-r^2}$ character bound to the slightly lower-lying $\Gamma$-point valence band maximum that are accidentally degenerate. The exchange interaction between two such hole states depends on whether the holes have $a'_1$ or $e'$ character, the former being quite localized, the latter quite extended in space. The magnetic moments of the spin $\frac{1}{2}$ acceptor states couple ferromagnetically at low concentrations but if the dopants are closer than the effective Bohr radius of the $d_{3z^2-r^2}$ orbital, the magnetic moments quench in order to profit from the bonding interaction. The details of the exchange interaction depend sensitively on the equilibrium structure of the undoped monolayer that in turn depends on the (approximate) exchange-correlation functional used. We argue that the LDA is preferable to the GGA because it describes the ordering of the K/K$'$ and $\Gamma$ valence band maxima better compared to experiment.

When spin-orbit coupling is included, we calculate a large magnetic anisotropy energy for acceptors with a preference for out of plane orientation and argue that this large single ion anisotropy justifies using an Ising spin model to study the ferromagnetic ordering. We estimate the ordering temperature by combining our (isotropic) separation-dependent exchange interactions with Monte Carlo calculations using Binder's cumulant method. For an MoS$_2$ monolayer doped with V (Nb or Ta), we estimate ferromagnetic Curie temperatures as a function of the dopant concentration and find a maximum $T_C$ of $\sim$170 K ($\sim$100 K) at around 9\% dopant concentration. At sufficiently high concentrations of impurity states, the acceptor states form bands and magnetism is quenched when the bandwidth exceeds a critical value; this critical value depends sensitively on the exchange-correlation functional used.

Although the maximum calculated $T_C$ is below room temperature, our work demonstrates that shallow impurities in MX$_2$ monolayers that bind weakly but have long range interactions are promising dopants to explore with a view to realizing room temperature ferromagnetism.


\acknowledgments
This work was financially supported by the ``Nederlandse Organisatie voor Wetenschappelijk Onderzoek'' (NWO) through the research programme of the former ``Stichting voor Fundamenteel Onderzoek der Materie,'' (NWO-I, formerly FOM) and through the use of supercomputer facilities of NWO ``Exacte Wetenschappen'' (Physical Sciences). Y. G. thanks the China Scholarship Council for financial support. N.G.\ is grateful to Dr.\ Supravat Dey for fruitful discussions. 

\appendix

\section{Sulphur reference atom}
\label{sec:SRA}

\begin{figure}[t]
\includegraphics[scale = 0.33]{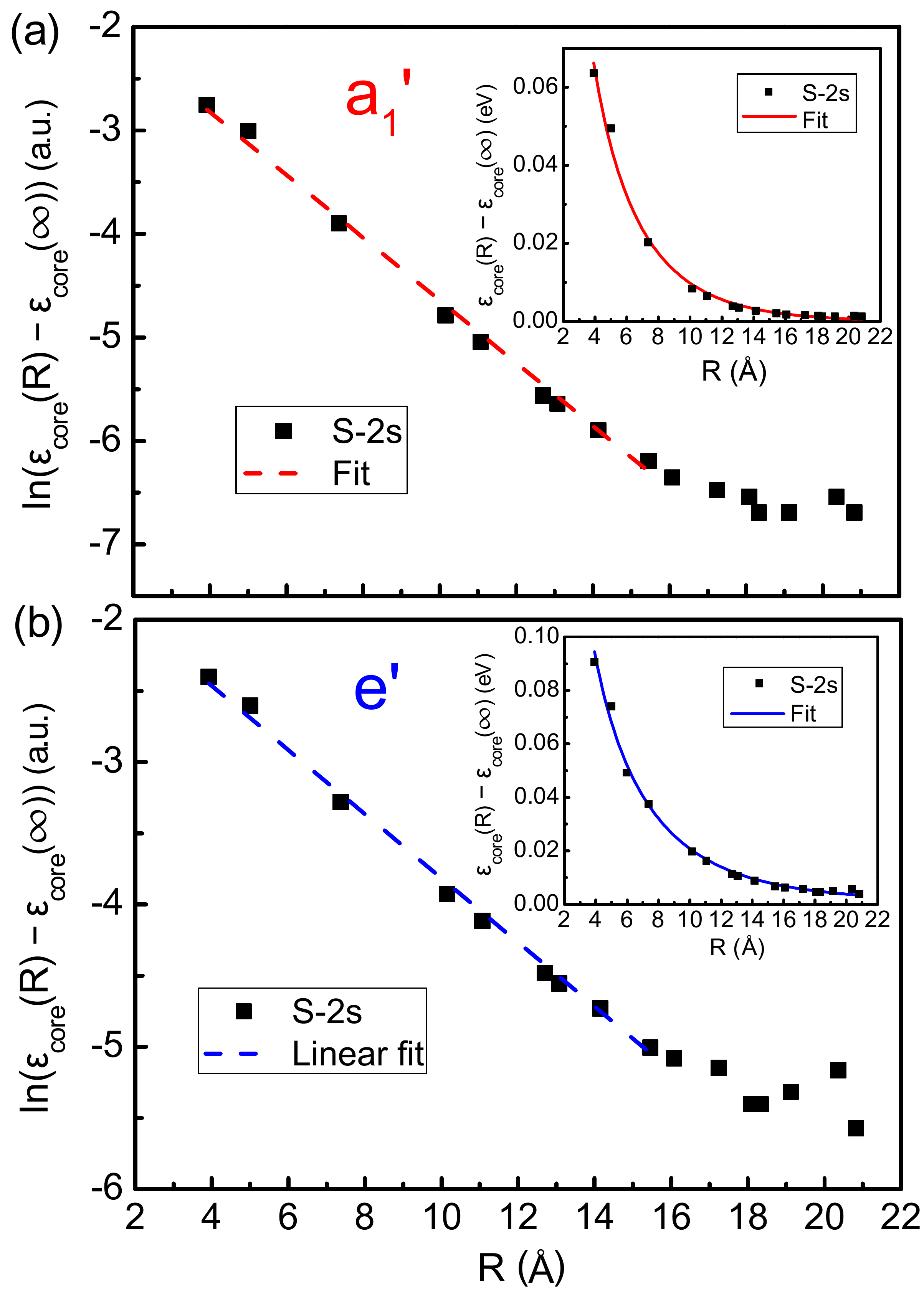}
\caption{Dependence of the S $2s$ semicore levels on the separation from the ${\rm V_{Mo}}$ dopant ion. The (screened) Coulomb potential of the V dopant is screened by the $a'_1$ hole (upper panel) respectively by the $e'$ hole (lower panel). The asymptotic value $\varepsilon_{\rm core}(\infty)$ was determined by fitting the calculated data points in the insets to an exponential wave function and using this fit (red curves) to extrapolate to $R=\infty$.
}
\label{figgC}
\end{figure}

The impurity potential of a vanadium acceptor screened by a hole in the $a'_1$ and $e'$ states as felt by S atoms is plotted in the insets to \cref{figgC}. The effective Bohr radii of these impurity states determined by fitting to Eq.~\eqref{eq:shb} are 8.8~\AA\ for screening by the $e'$ state and 6.5~\AA\ for screening by the $a'_1$ state. For the $e'$ level, this effective Bohr radius is consistent with the other estimates summarized in \cref{tabB}. For the $a'_1$ level, however, the value of 6.5~\AA\ yielded by the S atom probes is larger than the value of 5.4~\AA\ yielded by using Mo atoms as probes. We already saw more scatter in the estimate of the $a'_1$ radius and in view of its very small value and the importance of the central cell potential and local screening effects on this length scale, it is not surprising to see this type of variation measured by probes at different radial distances.

\section{Hydrogen atom in LDA/GGA}
\label{sec:HA}

\begin{table}[t]
\caption{Comparison of total energies ($E$) and Kohn-Sham (KS) eigenvalues ($\varepsilon$) for a hydrogen atom as described by the LDA and the LSDA, the non-spin polarized (NSP)-GGA and spin-polarized (SP)-GGA in Rydberg units (13.606~eV).}
\label{tabI}
\begin{ruledtabular}
\begin{tabular}{ldddd}
    & \multicolumn{2}{c}{LDA}      & \multicolumn{2}{c}{GGA} \\
\cline{2-3}        \cline{4-5}
                              & \moc{NSP} & \moc{SP} & \moc{NSP} & \moc{SP}  \\
\hline
$E$                           &   -0.89   &  -0.97   & -0.92     &  -0.99  \\
$\varepsilon_{1s}^\uparrow$   &   -0.46   &  -0.18   & -0.47     &   0.00  \\
$\varepsilon_{1s}^\downarrow$ &   -0.46   &  -0.53   & -0.47     &  -0.55  \\
$\Delta \varepsilon_{1s}$     &    0.00   &   0.35   &  0.00     &   0.55  \\
\end{tabular}
\end{ruledtabular}
\end{table}

In the local density approximation (LDA), the total energy of a neutral hydrogen atom is not $-1$~Rydberg but its absolute value is about 10\% smaller, $\sim -0.89$~Rydberg and the Kohn-Sham eigenvalue for the 1$s$ state is $\varepsilon_{1s}=-0.46$ \cite{Gunnarsson:prb74}. It has been shown that the discrepancy can be substantially reduced by using the spin-polarized (SP) version of the LDA, the local spin density approximation (LSDA) \cite{Gunnarsson:prb74}. Using the Perdew-Zunger \cite{Perdew:prb81} parameterization of L(S)DA, we obtain total energies ($E$) and KS eigenvalues ($\varepsilon$) of $E=-0.89$~Ry, $\varepsilon= -0.46 $Ry (LDA) and $E=-0.97$ Ry (LSDA) and an exchange splitting of the Kohn-Sham $1s$ eigenvalues of -0.35 Ry.
If instead of the L(S)DA, we use the Perdew-Burke-Ernzerhof \cite{Perdew:prl96} GGA, we obtain energies of  -0.92~Ry (GGA) and -0.99 Ry (SP-GGA) and an exchange splitting of -0.55 Ry.

%

\end{document}